\documentclass{aa}  

\usepackage{float}
\usepackage{epstopdf}
\usepackage{booktabs}
\usepackage{longtable}
\usepackage{multirow}
\usepackage{textcomp}
\usepackage{graphicx}
\usepackage{array}
\newcolumntype{H}{>{\setbox0=\hbox\bgroup}c<{\egroup}@{}}
\usepackage{txfonts}
\begin{document}

   \title{Radio halos in a mass-selected sample of 75 galaxy clusters}
   \subtitle{Paper I - Sample selection and data analysis}

\authorrunning{V. Cuciti et al.}
\author{V. Cuciti\inst{1},
          R. Cassano\inst{2},
          G. Brunetti \inst{2}, 
          D. Dallacasa\inst{3,2},
          R.~J. van Weeren\inst{4},
          S. Giacintucci\inst{9},
          A. Bonafede\inst{3,2},
          F. de Gasperin\inst{1},
          S. Ettori\inst{7,8},
          R. Kale\inst{5},
          G.~W. Pratt\inst{6},
          T. Venturi\inst{2}
          }

\institute{Hamburger Sternwarte, Universit\"at Hamburg, Gojenbergsweg 112, 21029, Hamburg, Germany,
\email{vcuciti@hs.uni-hamburg.de} \and INAF-Istituto di Radioastronomia, via P. Gobetti 101, 40129 Bologna, Italy 
\and Dipartimento di Fisica e Astronomia, Universit\`{a} di Bologna, via P. Gobetti 93/2, 40129 Bologna, Italy
\and Leiden Observatory, Leiden University, PO Box 9513, 2300 RA Leiden, The Netherlands
\and National Centre for Radio Astrophysics, Tata Institute of Fundamental Research       Savitribai Phule Pune University Campus, Pune 411 007 Maharashtra, INDIA
\and AIM, CEA, CNRS, Université Paris-Saclay, Université Paris Diderot, Sorbonne Paris Cité, F-91191 Gif-sur-Yvette, France
\and INAF, Osservatorio di Astrofisica e Scienza dello Spazio, via Pietro Gobetti 93/3, 40129 Bologna, Italy 
\and
INFN, Sezione di Bologna, viale Berti Pichat 6/2, I-40127 Bologna
\and Naval Research Laboratory, 4555 Overlook Avenue SW, Code 7213, Washington, DC 20375, USA}

   \date{Received --; accepted --}

 
  \abstract
   {Radio halos are synchrotron diffuse sources at the centre of a fraction of galaxy clusters. The study of large samples of clusters with adequate radio and X-ray data is necessary to investigate the origin of radio halos and their connection with the cluster dynamics and formation history.}
   {The aim of this paper is to compile a well-selected sample of galaxy clusters with deep radio observations to perform an unbiased statistical study of the properties of radio halos.}
   {We selected 75 clusters with $M\geq 6\times 10^{14} M_\odot$ at $z=0.08-0.33$ from the Planck Sunyaev-Zel’dovich catalogue. Clusters without suitable radio data were observed with the Giant Metrewave Radio Telescope (GMRT) and/or the Jansky Very Large Array (JVLA) to complete the information about the possible presence of diffuse emission. We used archival \textit{Chandra} X-ray data to derive information on the clusters' dynamical states.}
   {This observational campaign led to the detection of several cluster-scale diffuse radio sources and candidates that deserve future follow-up observations. Here we summarise their properties and add information resulting from our new observations. For the clusters where we did not detect any hint of diffuse emission, we derived new upper limits to their diffuse flux.}
   {We have built the largest mass-selected ($>80$\% complete in mass) sample of galaxy clusters with deep radio observations available to date. The statistical analysis of the sample, which includes the connection between radio halos and cluster mergers, the radio power -- mass correlation, and the occurrence of radio halos as a function of the cluster mass, will be presented in paper II.}
   \keywords{Galaxies: clusters: general --
                Galaxies: clusters: intracluster medium --
                Radiation mechanisms: non-thermal
               }

   \maketitle

\section{Introduction}

Clusters of galaxies occupy an exclusive position in the cosmic hierarchy since they are the most massive gravitationally bound structures in the Universe. They form and grow at the intersection of cosmic filaments, where smaller systems are channelled by the gravitational field that is dominated by dark matter. Mergers between clusters are the most energetic events in the Universe. Most of this energy contributes to heat the intracluster medium (ICM) up to the observed temperature ($10^7-10^8$ K). At the same time, a fraction of this energy, is channelled into the acceleration of particles and amplification of magnetic fields in the ICM by complex mechanisms, presumably invoking turbulence and shocks operating in plasma with unique properties \citep{brunettijones14}. This creates diffuse cluster-scale synchrotron emission, which has been observed in a growing number of clusters. Depending on their size and location, these sources are classified as radio relics or radio halos \citep[see][for a review]{vanweeren19}. Radio relics are elongated, often arc-shaped, sources located at the periphery of dynamically disturbed clusters. They are considered as tracers of merger-driven shocks propagating through the ICM \citep[e.g.][]{ensslin98, markevitch05, kang12, pinzke13}. Moreover, some relaxed clusters host mini halos, confined within the core, generally on scales $<0.2\times R_{500}$ \citep{giacintucci08, giacintucci17, giacintucci19}. The mechanisms responsible for the formation of mini halos are still a matter of debate, possibilities include electrons re-acceleration by turbulence generated in the core by several mechanisms \citep{gitti02, zuhone13} and secondary particles generated by hadronic collisions in the ICM \citep[e.g.][]{zuhone15, pfrommer04, jacob17}.

This work is mainly focused on radio halos, which are centrally located sources whose emission is roughly coincident with the X-ray emission of the host clusters. In the current theoretical scenario, radio halos form via the turbulent re-acceleration of electrons in the ICM \citep{brunetti01, petrosian01, brunettilazarian07, brunettilazarian11, brunettilazarian16, pinzke17}. The basic idea is that such turbulence is injected into the ICM during merging events. As a consequence, a strong connection between the properties of radio halos and the cluster mass and dynamical state is expected. 
Radio halos should be preferentially found in massive, merging clusters, should be rare in small and less disturbed systems, and should absent in relaxed clusters. Less energetic merger events are expected to form radio halos with very steep spectra ($\alpha<-1.5$, with $S(\nu)\propto \nu^{\alpha}$), the so-called ultra steep spectrum radio halos \citep[USSRHs, e.g.][]{brunetti08nature}. 

The study of the statistical properties of radio halos in galaxy clusters is a powerful tool to investigate the connection and evolution of these sources with the cluster dynamics and formation history and to test theoretical models for their formation. Pioneering studies using Arecibo, the NVSS and the WENNS surveys \citep{hanisch82,andernach86,giovannini99,kempner01} revealed that radio halos are not ubiquitous in galaxy clusters and that their occurrence increases with increasing the X-ray luminosity of the host clusters \citep{liang00}, although the role of selection biases due to the sensitivity limit of the used surveys was unclear \citep{kempner01, rudnick06}. In this respect, an important step forwards has been achieved with the Giant Meterwave Radio Telescope (GMRT) radio halo survey \citep{venturi07,venturi08} and its extension \citep{kale13,kale15}. That work led to the discovery of the so-called radio `bimodality' of galaxy clusters. Indeed radio halos are found in merging systems and follow the correlation between the radio power and the X-ray luminosity of the host cluster, while relaxed clusters without radio halos lie well below that correlation \citep{brunetti09, cassano10}. All these studies were based on the selection of the most X-ray luminous clusters, while the key parameter for the formation of radio halos is the cluster mass. The advent of clusters surveys via the Sunyaev-Zel'dovich (SZ) effect offers the opportunity to compile nearly mass selected samples of clusters with high levels of completeness \citep{basu12,cassano13,cuciti15,knowles19}, owing to the tight correlation between the SZ effect and the cluster mass \citep{motl05, nagai06}. The first results based on SZ selected samples of clusters suggested that the fraction of radio halos is larger with respect to X-ray selected samples \citep{sommerbasu14}. \citet{cassano13} showed that clusters are bimodal behaviour also in the radio luminosity--mass diagram.

With the aim of performing the first unbiased census of radio halos in a mass-selected sample of galaxy clusters, we selected 75 massive clusters from the Planck SZ catalogue \citep{planck14}. The first results on the occurrence of radio halos, based on a sub-sample of clusters that had available radio information, were presented in \citet{cuciti15}. We showed that the fraction of radio halos drops in low mass clusters, in line with turbulent re-acceleration models. However, that result could be affected by the incompleteness of the radio information for the total sample. Therefore, we carried out an observational campaign with the GMRT and the Jansky Very Large Array (JVLA) to complete the information about the possible presence of diffuse emission for all the clusters of the sample. In this paper we present the results of these new observations and we summarise the properties of the total sample, both from the radio and the X-ray points of view. 

In Section \ref{Sec:sample} we present the selection of the sample, in Section \ref{Sec:radio} we describe the procedures adopted to reduce the radio data and in Section \ref{Sec:results} we show the results of the radio data analysis. We derive upper limits to the radio emission of clusters without radio halos in Section \ref{Sec:UL}. In Section \ref{Sec:profile}, we derive the surface brightness radial profile of radio halos. The analysis of the X-ray data is described in Section \ref{Sec:X-ray analysis} and the dynamical properties of the clusters are discussed in Section \ref{Sec:dynamics}. In Section \ref{Sec:conclusions} we summarise the work and give our conclusions. We perform the statistical analysis of the radio and X-ray properties of the cluster of this sample in Paper II.

Throughout this paper we assume a $\Lambda$CDM cosmology with $H_0=70$ km s$^{-1}$Mpc$^{-1}$, $\Omega_\Lambda=0.7$ and $\Omega_m=0.3$.

\section{Sample selection}
\label{Sec:sample}
In \citet{cuciti15}, we selected a sample of massive objects from the Planck SZ cluster catalogue \citep{planck14}. The selection criteria are discussed in \citet{cuciti15} and are summarised below.

At redshift $0.08<z<0.2$ we adopted $M_{500}\footnote{$M_{500}$ is the mass enclosed in a sphere with radius $R_{500}$, which is defined as the radius within which the mean
mass over-density of the cluster is 500 times the cosmic critical density at the cluster redshift}\geq5.7\times10^{14}M_\odot$
and we selected clusters observed in the NVSS \citep[$\delta>-40^{\circ}$,][]{condon98}.
The lower redshift limit ($z>0.08$) is driven by the fact that radio interferometers suffer from the lack of sampling at short baselines, resulting in decreased sensitivity to emission on large spatial scales, such as the typical scales of radio halos. Moveover, the largest angular scale detectable with the JVLA at 1.5GHz (C and D configuration) is 970 arcsec, meaning that at redshift $z<0.08$ only scales smaller than 1.5 Mpc can be recovered.

At redshift $0.2<z<0.33$ we selected clusters with $M_{500}\geq6\times10^{14} \,M_\odot$ and we adopted a declination limit $\delta>-31^\circ$ and $|b|\geq 20^\circ$ ($|b|$ is the galactic latitude), which coincides with that of the GMRT radio halo Survey \citep{venturi07,venturi08,kale13,kale15}, in order to maximise the availability of information in the literature. The upper redshift limit ($z<0.33$) is mainly related to the insufficient completeness of the PSZ1 catalogue at higher redshift for these masses.

\begin{figure*}
 \centering
 \includegraphics[width=15cm]{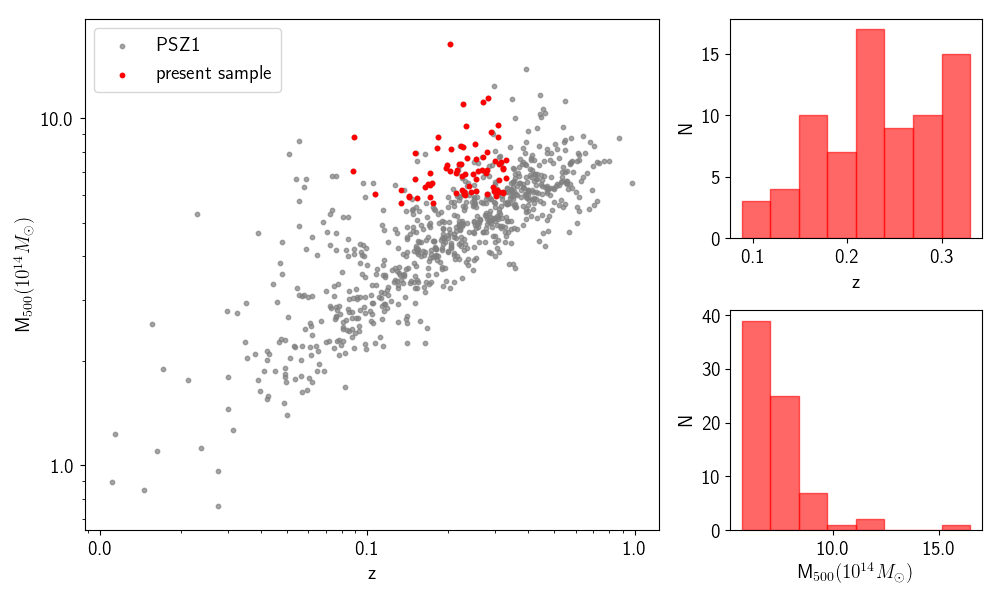}

\caption{Mass and redshift distribution of the clusters of the sample. \textit{Left}: Distribution of the clusters of the Planck SZ cluster catalogue in the $M_{500}-z$ diagram. Clusters belonging to the sample presented here are marked in red. \textit{Top right}: Redshift distribution of the clusters of our sample distribution. \textit{Bottom right}: Mass distribution of the clusters of our sample.}
 \label{Fig:M_z}
\end{figure*}

We adopted a slightly different cut in mass in the two redshift bins to increase the statistics and, at the same time, assure about the same mass completeness of the sample in both redshift ranges \citep{planck14, cuciti15}. Indeed, in the selected mass ranges, the completeness of the Planck catalogue is $\sim 90$\% at $z<0.2$ and $\sim 80$\% at $z>0.2$, thus we estimated a completeness of our sample of $\sim 83$\%\footnote{Estimated as $\frac{0.9\times21+0.8\times54}{75}$.}.
Our sample consists of 75 clusters (21 at $z<0.2$ and 54 at $z>0.2$), whose properties are listed in Table \ref{tab:completesample} and the mass and redshift distributions are shown in Fig. \ref{Fig:M_z}. The median redshift of the sample is 0.23. The redshift distribution is rather uniform, except for the very low-redshift tail, where the volume of the Universe is too small to host such massive objects. As expected, the mass distribution is peaked around the mass cut of the sample and then it declines with increasing mass \citep[e.g.][]{press74}. The median value for the mass is $\sim7\times10^{14} \,M_\odot$. Only a few clusters with $M_{500}\gtrsim9\times10^{14}M_\odot$ are present in the Universe at the redshifts considered here. 

\longtab{
\begin{center}
\begin{small}
\renewcommand{\arraystretch}{1.5}
\begin{longtable}{l c c c c c c c}
\caption{Total sample clusters properties} \\ 
\label{tab:completesample} \\
 
\hline      
\hline
Cluster name  	&	RA 	&	DEC 	& 	z	&			$M_{500}$			  &	$\rm{R_{500}}$ & Radio	&		$P_{1.4 \mathrm{GHz}}$ \\
             	& 		&			&		& ($10^{14}$ M$_{\odot}$) & (kpc) 		       & info	& ($10^{24}$ W Hz$^{-1}$)	       \\  
\hline
A1437		    &  12 00 22.3   &   $+$03 20 33.9 	&   0.134 & $5.68_{0.39}^{0.38}$   &  1200 & no RH*	      & -              \\
A2345               &  21 27 06.8   &   $-$12 07 56.0 	&   0.176 & $5.71_{0.49}^{0.46}$   &  1190 & UL$^{4}$	      & $<0.38$        \\
A2104		    &  15 40 08.2   &   $-$03 18 23.0 	&   0.153 & $5.91_{0.60}^{0.57}$   &  1200 & UL*		      & $<0.24$        \\
Zwcl 2120.1+2256    &  21 22 27.1   & 	$+$23 11 50.3 	&   0.143 & $5.91_{0.34}^{0.33}$   &  1200 & RH(c)*             & -              \\
RXC J0616.3-2156    &  06 16 22.8   & 	$-$21 56 43.4 	&   0.171 & $5.93_{0.45}^{0.43}$   &  1200 & UL*		      & $<0.25$        \\
A1413 		    &  11 55 18.9   &   $+$23 24 31.0 	&   0.143 & $5.98_{0.40}^{0.38}$   &  1220 & MH$^5$	      & -              \\
A1576 		    &  12 37 59.0   &   $+$63 11 26.0 	&   0.302 & $5.98_{0.50}^{0.48}$   &  1160 & UL$^6$	      & $<0.64$        \\
A2697 		    &  00 03 11.8   &   $-$06 05 10.0 	&   0.232 & $6.01_{0.61}^{0.58}$   &  1190 & UL$^2$	      & $<0.41$        \\
Z5247 		    &  12 33 56.1   &   $+$09 50 28.0 	&   0.229 & $6.04_{0.59}^{0.56}$   &  1190 & RH(c)$^{7}$ 	      & -              \\
Zwcl 0104.9+5350    &  01 07 54.0   & 	$+$54 06 00.0 	&   0.107 & $6.06_{0.43}^{0.41}$   &  1240 & RH$^8$	      & $1.62\pm0.15$  \\
RXC J0142.0+2131    &  01 42 02.6   & 	$+$21 31 19.0 	&   0.280 & $6.07_{0.83}^{0.77}$   &  1170 & RH$^{40,US(c)}$    & $0.42\pm0.01$  \\
A1423 		    &  11 57 22.5   &   $+$33 39 18.0 	&   0.214 & $6.09_{0.51}^{0.49}$   &  1200 & UL$^2$	      & $<0.38$        \\
ZwCl 1028.8+1419    &  10 31 28.2   &  	$+$14 03 34.0   &   0.310 & $6.11_{0.69}^{0.65}$   &  1160 & no RH*	      & -              \\
A3041     	    &  02 41 22.1   & 	$-$28 38 13.0   &   0.230 & $6.12_{0.57}^{0.54}$   &  1190 & RH(c)*	      & -              \\
RXC J2051.1+0216    &  20 51 08.0   &  	$+$02 15 55.0 	&   0.320 & $6.13_{0.74}^{0.69}$   &  1150 & UL* 		      & $<0.73$        \\
A2472 		    &  22 41 50.6   &   $+$17 31 43.0   &   0.310 & $6.15_{0.78}^{0.72}$   &  1160 & UL*		      & $<1.0$         \\
A2895 		    &  01 18 11.1   & 	$-$26 58 23.0   &   0.230 & $6.15_{0.55}^{0.52}$   &  1190 & UL* 		      & $<0.5$         \\
RXC J1314.4-2515    &  13 14 28.0   & 	$-$25 15 41.0 	&   0.244 & $6.15_{0.73}^{0.69}$   &  1190 & RH$^{1,US(c)}$     & $0.68\pm0.24$  \\
A2537 		    &  23 08 23.2   &   $-$02 11 31.0 	&   0.297 & $6.17_{0.66}^{0.62}$   &  1170 & UL$^2$	      & $<0.45$        \\
A68 		    &  00 37 05.3   &   $+$09 09 11.0 	&   0.255 & $6.19_{0.68}^{0.64}$   &  1190 & UL$^7$ 	      & $<0.42$        \\
A56        	    &  00 33 50.4   & 	$-$07 47 28.0   &   0.300 & $6.20_{0.73}^{0.69}$   &  1170 & UL*		      & $<1.2$         \\
A1682 		    &  13 06 49.7   &   $+$46 32 59.0	&   0.226 & $6.20_{0.46}^{0.45}$   &  1200 & RH(c)$^2$	      & -              \\
A1132 		    &  10 58 19.6   &   $+$56 46 56.0	&   0.134 & $6.23_{0.31}^{0.31}$   &  1240 & RH$^{27,US}$	      & $0.16\pm0.08$  \\ 
RXJ1720.1+2638 	    &  17 20 10.1   & 	$+$26 37 29.5	&   0.164 & $6.34_{0.40}^{0.38}$   &  1240 & MH+USSRH$^{9,40}$    & -              \\
A781 	            &  09 20 23.2   &   $+$30 26 15.0 	&   0.295 & $6.35_{0.61}^{0.58}$   &  1180 & UL$^2$	      & $<0.36$        \\
A384       	    &  02 48 13.9   & 	$-$02 16 32.0   &   0.240 & $6.38_{0.61}^{0.58}$   &  1210 & UL*		      & $<0.74$        \\
A2218  		    &  16 35 51.6   &   $+$66 12 39.0 	&   0.171 & $6.41_{0.26}^{0.26}$   &  1240 & RH$^{3}$	      & $0.44\pm0.10$  \\   
A3411  		    &  08 41 55.6   &   $-$17 29 35.7 	&   0.169 & $6.48_{0.38}^{0.37}$   &  1250 & RH$^{10}$	      & $0.27\pm0.1$   \\   
Zwcl 0634.1+4750    &  06 38 02.5   & 	$+$47 47 23.8 	&   0.174 & $6.52_{0.45}^{0.43}$   &  1250 & RH$^{25}$ 	      & $0.31\pm0.02$  \\   
RXC J1322.8+3138    &  13 22 48.8   &   $+$31 39 17.0  	&   0.310 & $6.63_{0.62}^{0.59}$   &  1190 & no RH*	      & -              \\
A3888 		    &  22 34 26.8   &   $-$37 44 19.1 	&   0.151 & $6.67_{0.34}^{0.33}$   &  1260 & RH$^{*,33}$	      & $1.90\pm0.20$  \\   
A3088 		    &  03 07 04.1   &   $-$28 40 14.0 	&   0.254 & $6.71_{0.58}^{0.55}$   &  1220 & UL$^2$	      & $<0.43$        \\
A220  	            &  01 37 19.5   & 	$+$07 56 16.0   &   0.330 & $6.74_{0.92}^{0.85}$   &  1190 & UL* 		      & $<0.97$        \\
A2667 		    &  23 51 40.7   &   $-$26 05 01.0 	&   0.226 & $6.81_{0.49}^{0.47}$   &  1240 & MH$^{28}$	      & -              \\
A521 		    &  04 54 09.1   &   $-$10 14 19.0 	&   0.248 & $6.90_{0.64}^{0.61}$   &  1240 & RH$^{11,US}$	      & $1.45\pm0.13$  \\  
A2355    	    &  21 35 22.5   &   $+$01 23 26.0  	&   0.230 & $6.92_{0.51}^{0.49}$   &  1240 & UL* 		      & $<0.83$        \\
A2631 		    &  23 37 40.6   &   $+$00 16 36.0 	&   0.278 & $6.97_{0.62}^{0.58}$   &  1230 & UL$^2$	      & $<0.41$        \\
A1914 		    &  14 26 03.0   &   $+$37 49 32.0 	&   0.171 & $6.97_{0.36}^{0.35}$   &  1280 & no RH$^{35}$	      & -              \\    
RXC J1504.1-0248    &  15 04 07.7   &	$-$02 48 18.0 	&   0.215 & $6.98_{0.60}^{0.57}$   &  1330 & MH$^{13}$	      & -              \\
A1733    	    &  13 27 03.7   &  	$+$02 12 15.0  	&   0.260 & $7.05_{0.65}^{0.62}$   &  1240 & UL*		      & $<0.53$        \\
A520 		    &  04 54 19.0   &   $+$02 56 49.0 	&   0.203 & $7.06_{0.58}^{0.56}$   &  1270 & RH$^{14}$	      & $2.45\pm0.18$  \\    
A478 		    &  04 13 20.7   &   $+$10 28 35.0 	&   0.088 & $7.06_{0.36}^{0.35}$   &  1320 & MH$^{15,40}$	      & -              \\
A773 	            &  09 17 59.4   &   $+$51 42 23.0 	&   0.217 & $7.08_{0.46}^{0.44}$   &  1260 & RH$^{14}$	      & $1.48\pm0.16$  \\     
PSZ1 G019.12+3123   &  16 36 29.4   &   $+$03 08 51.0   &   0.280 & $7.08_{0.67}^{0.63}$   &  1230 & UL*		      & $<0.63$        \\
PSZ1 G139.61+24.20  &  06 22 13.9   &  	$+$74 41 39.0   &   0.270 & $7.09_{0.58}^{0.56}$   &  1210 & MH+USSRH$^{28,29}$      & -              \\
A1351 		    &  11 42 30.8   &   $+$58 32 20.0 	&   0.322 & $7.14_{0.53}^{0.51}$   &  1210 & RH$^{16}$	      & $9.30\pm1.5$   \\     
A115 		    &  00 55 59.5   &   $+$26 19 14.0 	&   0.197 & $7.21_{0.52}^{0.50}$   &  1280 & no RH$^{14}$	      & -              \\
A402 		    &  02 57 41.1   & 	$-$22 09 18.0 	&   0.320 & $7.20_{0.68}^{0.65}$   &  1220 & MH (c)$^{*}$       & -              \\
A1451 		    &  12 03 16.2   &   $-$21 32 12.7 	&   0.199 & $7.32_{0.48}^{0.47}$   &  1290 & RH$^{25}$	      & $0.64\pm0.07$  \\   
RXCJ 0510.7-0801    &  05 10 47.9   & 	$-$08 01 06 	&   0.220 & $7.36_{0.63}^{0.61}$   &  1280 & no RH$^{7}$	      & -              \\
PSZ1 G205.07-6294   &  02 46 27.5   & 	$-$20 32 5.29 	&   0.310 & $7.37_{0.66}^{0.63}$   &  1240 & no RH$^p$	      & -              \\
A2261 		    &  17 22 17.1   &   $+$32 08 02.0 	&   0.224 & $7.39_{0.46}^{0.45}$   &  1280 & RH$^{30,34,US(c)}$ & $0.68\pm0.07$  \\   
RXCJ2003.5-2323     &  20 03 30.4   & 	$-$23 23 05.0 	&   0.317 & $7.48_{0.67}^{0.64}$   &  1240 & RH$^1$ 	      & $10.71\pm1.73$ \\    
A2552 		    &  23 11 26.9   &   $+$03 35 19.0 	&   0.300 & $7.53_{0.62}^{0.59}$   &  1250 & RH(c)$^7$ 	      & -              \\
MACS J2135-010      &  21 35 12.1   &  	$-$01 02 58.0   &   0.330 & $7.57_{0.64}^{0.61}$   &  1240 & UL*		      & $<1.17$        \\
A3444 		    &  10 23 50.8   &   $-$27 15 31.0 	&   0.254 & $7.62_{0.56}^{0.53}$   &  1270 & MH$^{7,28}$	      & -              \\
S780 		    &  14 59 29.3   &   $-$18 11 13.0 	&   0.236 & $7.71_{0.63}^{0.60}$   &  1290 & MH$^{7,28}$	      & -              \\
A1443         	    &  12 01 27.7   &   $+$23 05 18.0 	&   0.270 & $7.74_{0.56}^{0.54}$   &  1270 & RH$^{31}$ 	      & $0.91\pm0.04$  \\ 
A2204 		    &  16 32 45.7   &   $+$05 34 43.0 	&   0.151 & $7.96_{0.38}^{0.37}$   &  1340 & MH$^{15}$	      & -              \\
A1758a 		    &  13 32 32.1   &   $+$50 30 37.0 	&   0.280 & $7.99_{0.46}^{0.44}$   &  1280 & RH$^{17}$	      & $5.75\pm0.98$  \\     
A209		    &  01 31 53.0   &   $-$13 36 34.0 	&   0.206 & $8.17_{0.44}^{0.43}$   &  1330 & RH$^1$	      & $1.99\pm0.21$  \\     
A665 		    &  08 30 45.2   &   $+$65 52 55.0 	&   0.182 & $8.23_{0.40}^{0.39}$   &  1340 & RH$^3$	      & $2.51\pm0.21$  \\     
A1763 		    &  13 35 17.2   &   $+$40 59 58.0 	&   0.228 & $8.29_{0.41}^{0.40}$   &  1320 & no RH$^2$	      & -              \\
RXC J1514.9-1523    &  15 14 58.0   & 	$-$15 23 10.0 	&   0.223 & $8.34_{0.55}^{0.53}$   &  1330 & RH$^{18, US(c)}$   & $2.39\pm0.70$  \\      
A1835 		    &  14 01 02.3   &   $+$02 52 48.0 	&   0.253 & $8.46_{0.57}^{0.55}$   &  1320 & MH$^{19}$	      & -              \\
A2142 		    &  15 58 16.1   &   $+$27 13 29.0 	&   0.089 & $8.81_{0.29}^{0.29}$   &  1420 & RH$^{24,32,US(c)}$ & $0.19\pm0.06$  \\
A1689 		    &  13 11 29.5   &   $-$01 20 17.0 	&   0.183 & $8.86_{0.45}^{0.44}$   &  1380 & RH$^{20}$	      & $0.95\pm0.28$  \\     
A1300 		    &  11 31 56.3   &   $-$19 55 37.0 	&   0.308 & $8.83_{0.62}^{0.59}$   &  1310 & RH$^{21,US(c)}$    & $3.80\pm1.43$  \\      
A2813  		    &  00 43 27.4   & 	$-$20 37 27.0   &   0.290 & $9.16_{0.55}^{0.53}$   &  1340 & UL*		      & $<1.4$         \\
A2390 		    &  21 53 34.6   &   $+$17 40 11.0 	&   0.234 & $9.48_{0.42}^{0.41}$   &  1380 & no RH$^{34}$	      & -  	       \\   
A2744 		    &  00 14 18.8   &   $-$30 23 00.0 	&   0.307 & $9.56_{0.51}^{0.49}$   &  1350 & RH$^{14}$	      & $17.40\pm0.90$ \\   
A2219 		    &  16 40 21.1   &   $+$46 41 16.0 	&   0.228 & $11.01_{0.37}^{0.36}$  &  1450 & RH$^{12}$	      & $5.63\pm0.80$  \\    
PSZ1 G171.96-40.64  &  03 12 57.4   & 	$+$08 22 10.0 	&   0.270 & $11.13_{0.58}^{0.56}$  &  1440 & RH$^{22,US(c)}$    & $4.90\pm1.35$  \\      
A697 		    &  08 42 53.3   &   $+$36 20 12.0 	&   0.282 & $11.48_{0.47}^{0.46}$  &  1190 & RH$^{2,US}$	      & $1.51\pm0.14$  \\      
A2163 		    &  16 15 46.9   &   $-$06 08 45.0 	&   0.203 & $16.44_{0.41}^{0.40}$  &  1680 & RH$^{23}$	      & $22.90\pm1.16$ \\   
\hline
\hline 

\end{longtable}
\end{small}
\end{center}
\tablefoot{RH = Radio Halo, MH = Mini Halo, UL = Upper Limit, US = Ultra Steep, c = candidate. $P_{1.4 \mathrm{GHz}}$ = $k$-corrected radio power at 1.4 GHz. * This paper,
$^1$\citet{venturi07}, $^2$\citet{venturi08}, $^3$\citet{giovannini00}, $^{4}$\citet{bonafede17} $^5$\citet{govoni09}, $^6$\citet{kale13}, $^7$\citet{kale15}, $^8$\citet{vanweeren11}, $^9$\citet{giacintucci14_1720}, $^{10}$\citet{vanweeren13}, $^{11}$\citet{brunetti08nature}, $^{12}$\citet{bacchi03}, $^{13}$\citet{giacintucci11_1504}, $^{14}$\citet{govoni01}, $^{15}$\citet{giacintucci14}, $^{16}$\citet{giacintucci09}, $^{17}$\citet{giovannini06}, $^{18}$\citet{giacintucci11}, $^{19}$\citet{murgia09}, $^{20}$\citet{vacca11}, $^{21}$\citet{reid99}, $^{22}$\citet{giacintucci13}, $^{23}$\citet{feretti01}, $^{24}$\citet{farnsworth13}, $^{25}$\citet{cuciti18}, $^{26}$\citet{cassano13}, $^{27}$\citet{wilber18}, $^{28}$\citet{giacintucci17}, $^{29}$\citet{savini18}, $^{30}$\citet{sommer17}, $^{31}$\citet{bonafede15}, $^{32}$\citet{venturi17}, $^{33}$\citet{shakouri16}, $^{34}$ \citet{savini19}, $^{35}$\citet{mandal19}, $^p$Ferrari et al. (private communication). }
}

\section{Radio data analysis}
\label{Sec:radio}
The presence of diffuse radio emission has been already studied with deep radio observations in the literature for 55, out of the 75 clusters of the sample. For three of them, which are known to host diffuse emission, we obtained observations at different frequencies. Overall, in this paper we present new radio observations of the 23 galaxy clusters listed in Table \ref{tab:data}. These observations were carried out with the GMRT and/or the JVLA. In particular, we analysed GMRT 610 MHz observations of 11 clusters, GMRT 330 MHz observations of eight clusters and JVLA 1.5 GHz observations of 15 clusters. The details about the radio data analysed here are given in Table \ref{tab:data}. The frequency coverage of the sample is heterogeneous, however, this is currently a necessary compromise in order to build a complete sample that is large enough to perform a solid statistical analysis. This limit will be overcome thanks to ongoing surveys with, for example, LOFAR \citep{shimwell19} and MeerKAT \citep{knowles17}. 
We describe the main steps of the data reduction in the following subsections.

\begin{table*}
\begin{center}
\caption{Summary of the radio data analysis}
\begin{small}
\begin{tabular}{l c c c c c c c}
\midrule
\midrule
Name		&	telescope	& project code&	$\nu$	&	$\Delta t$	&	beam		&	rms	&	detection\\	
		&			&	&	(MHz)	&	(min)		&	($''\times ''$)	&	(mJy/beam)	&	\\
\midrule
A56		&	JVLA C		& 14B-190&	1500		&	40			&	$13.8\times10.4$&	0.080	&	UL\\
\midrule
A2813	&	JVLA C	& 14B-190	&	1500		&	40			&	$11.4\times9.5$&		0.035	&	UL\\
\midrule
A2895	&	JVLA C	& 14B-190	&	1500		&	40			&	$14.6\times9.0$&		0.040	&	UL\\
\midrule
A3041	&	JVLA C	& 14B-190	&	1500		&	40			&	$18.0\times8.6$&		0.035	&	candidate RH\\
\midrule
A220		&	JVLA C	& 14B-190	&	1500		&	40			&	$11.6\times9.8$&	0.045	&	UL\\
\midrule
\multirow{2}*{A384}& JVLA C	& 14B-190	&	1500		&	40  &	$13.5\times10.4$&0.035	&	UL\\
&	GMRT		&26\_021 &	610	 &	220	&	$5.9\times4.8$	&	0.050	&	--\\
\midrule
Zwcl1028.8+1419 &	GMRT	&27\_025	&	610	&	330	&		$5.3\times4.8$	&	0.056	&	no UL\\
\midrule
RXC J1322.8+3138 &	GMRT	&27\_025	&	610	&	150	&		$5.7\times4.4$	&	0.060	&	no UL\\
\midrule
A1733			 &	GMRT	& 27\_025	&	610	&	250	&		$7.9\times5.0$	&	0.060	&	UL\\
\midrule
PSZ1 G019.12+3123 &	GMRT	& 26\_021	&	610	&	250	&		$5.0\times3.8$	&	0.035	&	UL\\
\midrule
MACS J2135-010	 &	GMRT	&30\_019	&	610	&	300	&		$7.8\times5.8$	&	0.080	&	UL\\
\midrule
\multirow{2}*{A2355}& JVLA C	& 14B-190	&	1500		&	40  &	$11.6\times10.8$&0.040	&	UL\\
&	GMRT		&30\_019&	610	& 	300	&	$8.3\times6.1$	&	0.130	&	--\\
\midrule
\multirow{2}*{RXC J2051.1+0216}& JVLA C	& 14B-190	&	1500		&	40  &	$14.1\times11.5$&0.050	&	UL\\
&	GMRT		& 26\_021&	610	 &	200	&	$6.0\times4.8$	&	0.100	&	--\\
\midrule
A2472		&	JVLA C	& 14B-190	&	1500		&	40			&	$10.5\times10.1$&	0.040	&	UL\\
\midrule
\multirow{3}*{PSZ1 G139.61+2420}& JVLA C	& 14B-190	&	1500		&	40  &	$10.3\times5.1$&0.030	&	\multirow{3}*{MH}\\
&	GMRT$^a$		&27\_025+28\_077&	610	 &	300+300	&	$6.0\times5.0$	&	0.030	\\
\midrule
\multirow{4}*{A1443}& JVLA C+D	&13A-268	&	1500		&	60+90  &	$15.0\times12.0$&0.020	&	\multirow{4}*{RH}\\
&	GMRT		&27\_025&	610	 &	150	&	$5.6\times4.2$	&	0.050	\\
&	GMRT$^b$		&23\_020&	330	 &	270	&	$8.6\times7.3$	&	0.060	\\
\midrule
\multirow{2}*{RXC J0510.7-0801}& GMRT	&23\_004	&	610	&	330  &	$5.4\times4.8$&0.200	&	no UL\\
&	GMRT	&23\_004	&	240	&	330	&	$15.7\times13.1$	&	1.200	&	no UL\\
\midrule
\multirow{2}*{A402} &	GMRT \large{*}	&22\_021	&	330	&	400	&		$13.4\times8.6$	&	0.100	&	\multirow{2}*{candidate MH}\\
&	GMRT	 \large{*}&25\_018	&	330	&	380	&	$57.0\times42.0$	&	1.000	\\
\midrule
A1437			 &	GMRT	&29\_001	&	330	&	330	&		$9.0\times7.4$	&	0.400	&	no UL\\
\midrule
A2104			 &	GMRT &	05VKK01	&	330	&	350	&		$13.3\times10.4$	&	0.130	&	UL\\
\midrule
\multirow{2}*{Zwcl2120.1+2256}& JVLA D		&15B-035	&1500		&	40  &	$33.0\times30.3$&	0.075	&	\multirow{2}*{candidate RH}\\
&	GMRT	 \large{*}&23\_046	&	330	&	250	&	$10.0\times9.2$	&	0.100	\\
\midrule
RXC J0616.3-2156		&	JVLA DnC& 	15B-035	&	1500		&	40			&	$55.4\times19.9$&	0.080	&	UL\\
\midrule
A3888			 &	GMRT &28\_066		&	330	&	260	&		$14.6\times8.3$	&	0.300	&	RH\\
\midrule
\midrule
\end{tabular}
\tablefoot{$\Delta t$ = time on source; UL= Upper limit; RH = radio halo; MH= mini halo; no UL = no detection and no UL available; {\large*} = processed with SPAM \citep{intema09, intema14, intema17}; a) \citet{savini18}, b) \citet{bonafede15} . }
\label{tab:data}
\end{small}
\end{center}
\end{table*}

\subsection{GMRT data analysis}
\label{Sec:GMRT_data}
The GMRT observations listed in Table \ref{tab:data} were carried out using an observing bandwidth of 32 MHz subdivided into 256 channels\footnote{Only A2104 was observed with the old GMRT setup, namely with the simultaneous observation in two bands, the upper side band and the lower side band, each 16 MHz wide.}. We reduced these observations with the Astronomical Image Processing System (AIPS) or with the Common Astronomy Software Applications (CASA). Regardless of the software we used, the calibration procedure is essentially the same and it is outlined in the following.
The flux density scale was set according to \citet{scaifeheald12}. The bandpass was corrected using the flux density calibrators. As a first step we obtained amplitude and gain corrections for the primary calibrators in few central channels free of radio-frequency interference (RFI); these solutions were applied before determining the bandpass in order to remove possible time variations of the gains during the observation. Once we applied the bandpass, gain solutions for all the calibrator sources on the full range of channel were determined and transferred to the target source. Automatic removal of RFI was performed either with the CASA task \texttt{flagdata} or with the AIPS task \texttt{RFLAG}. Further manual editing of the data was performed. The central channels were averaged to a smaller number of channels each $1-2$ MHz wide to reduce the size of the dataset without introducing significant bandwidth smearing within the primary beam. A number of phase-only self-calibration rounds were carried out on the target field to reduce residual phase variations. A final amplitude and phase self-calibration was applied. Wide field imaging was implemented to account for the non-coplanarity of the baselines. In particular, we used the wprojection algorithm \citep{cornwell05, cornwell08} in CASA, while in AIPS we subdivided the field of view in tens of facets (the exact number of facets depending on the frequency, the resolution and the presence of bright sources). Facets were imaged separately, with a different phase centre, and then recombined. In CASA, wide band imaging (\texttt{mode=mfs}, \texttt{nterms=2}) was also used to consider the combination of the sources spectral index and the frequency dependency of the primary beam attenuation. To deal with the bright sources in the field of view that typically reduce the dynamic range of the image we adopted the so called `peeling' technique. Specifically, we obtained direction-dependent amplitude and phase solutions for those sources and then subtracted them out from the \textit{uv}-data. 
We did not add the `peeled' sources back into the final data, however, being typically far from the pointing centre, they are outside the portions of images shown in the paper.

We used the `Briggs' weighting scheme \citep{briggs95} with \texttt{robust=0} throughout the self-calibration\footnote{ we used \texttt{robust=0} both in AIPS and CASA, although we are aware that the definition of the robust parameter is slightly different in the two softwares.} and we produced final high-resolution images whose properties are listed in Table \ref{tab:data}. The images of the three clusters marked with an asterisks in the column `telescope' in Table \ref{tab:data}, were severely affected by artefacts due to bright sources in the field or residual RFI. To improve their quality, we processed those datasets with the Source Peeling and Atmospheric Modelling (SPAM) pipeline, which is extensively described in \citet{intema09}, \citet{intema14}, and \citet{intema17}. 

We subtracted all the compact sources from the \textit{uv}-data. First we made high-resolution images excluding the baselines sensitive to the emission on scales larger than $\sim250$ kpc (\texttt{uvrange}$<2-3$ klambda depending on the cluster redshift). We subtracted the clean components of the sources detected in the high-resolution images and we used the new dataset to produced low-resolution images. These low-resolution images are more sensitive to the extended low surface brightness emission and thus are suitable to evaluate the presence of sources such as radio halos or relics. Images were corrected for the primary beam response. The uncertainty on the flux scale is estimated to be 10\% \citep[e.g.][]{chandra04}.

\subsection{JVLA data analysis}

We performed the data reduction, both calibration and imaging, of the JVLA datasets with CASA. The total bandwidth, from 1 to 2 GHz, is divided into 16 spectral windows, each with 64 channels of 2 MHz in width. In this paper we use 1.5 GHz as the reference frequency for JVLA observations.

As a first step, the data were Hanning smoothed. We applied the pre-determined antenna position offset and elevation-dependent gain tables. The flux density scale was set according to \citet{perleybutler13}. We determined amplitude and phase solutions for the flux calibrators in the ten central channels of each spectral window in order to remove possible time variations during the calibrator observation. These solutions were pre-applied to find the delay terms and to correct for the bandpass response. We obtained the complex gain solutions for the calibrator sources on the full bandwidth pre-applying the bandpass and delay solutions. Finally, we applied all the calibration tables to the target fields. Automatic RFI flagging was applied to the target fields using the CASA task \texttt{flagdata}. To reduce the size of the dataset, we averaged the 48 central channels of each spectral window to six channels and we averaged in time with a time bin of 15 sec.\\
We ran several rounds of phase-only self-calibration on each target field and a final amplitude and phase self-calibration to end up the process. The wprojection algorithm was used to take into account the non-coplanar nature of the array. Wide band imaging is crucial when dealing with the 1 GHz bandwidth of the JVLA; therefore, we used three Taylor terms (\texttt{nterms=3}) to take the frequency dependence of the brightness distribution into consideration.
The imaging process involves the use of clean masks that have been made with the \texttt{PyBDSF} package \citep{pybdsm}. For the self-calibration we used the `Briggs' weighting scheme with \texttt{robust=0} and we made final high-resolution images whose properties are reported reported in Table \ref{tab:data}. Then, in order to highlight the possible diffuse emission, we subtracted all the discrete sources with the same technique described in Section \ref{Sec:GMRT_data} and we produced low-resolution images, using higher values for the \texttt{robust} parameter and/or tapering down the long baselines.

Images were corrected for the primary beam attenuation. The absolute flux scale uncertainties are assumed to be within 2.5\% \citep{perleybutler13}.

\section{Detection of diffuse emission}
\label{Sec:results}

In the next Sections, we present the three clusters hosting diffuse emission whose discovery has been already reported in dedicated papers. We summarise the properties of these sources and add complementary information that we obtained with our new data. In Section \ref{Sec:candidate} we discuss clusters hosting candidate diffuse emission. In the following, the errors reported for the diffuse sources take into account the uncertainty associated with the source subtraction, when applicable \citep[see e.g.][]{cassano13}.

\subsection{Abell 3888}

\begin{figure*}
\centering
\includegraphics[scale=0.45]{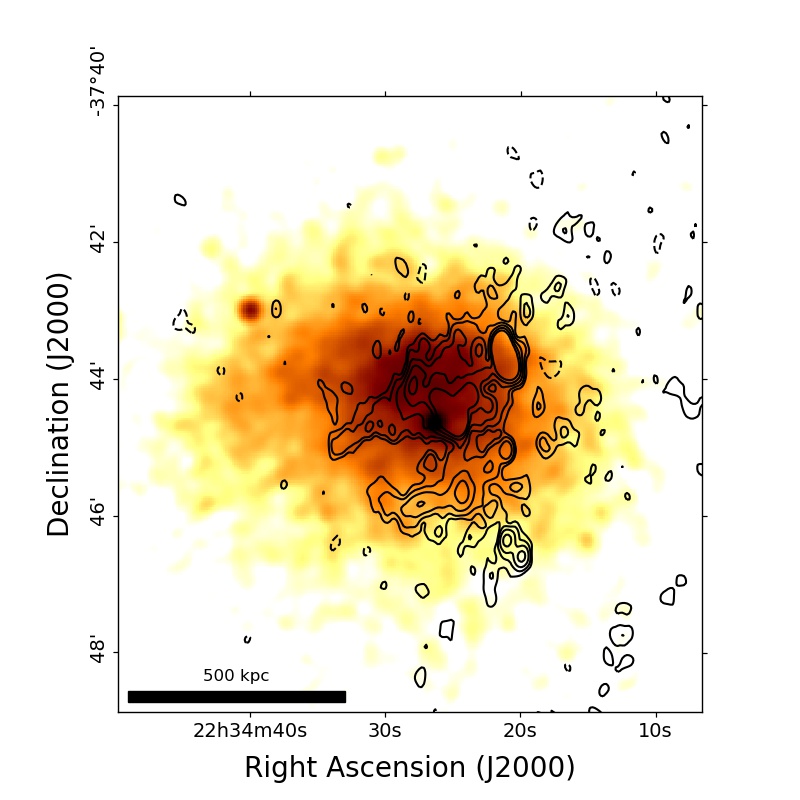}
\includegraphics[scale=0.45]{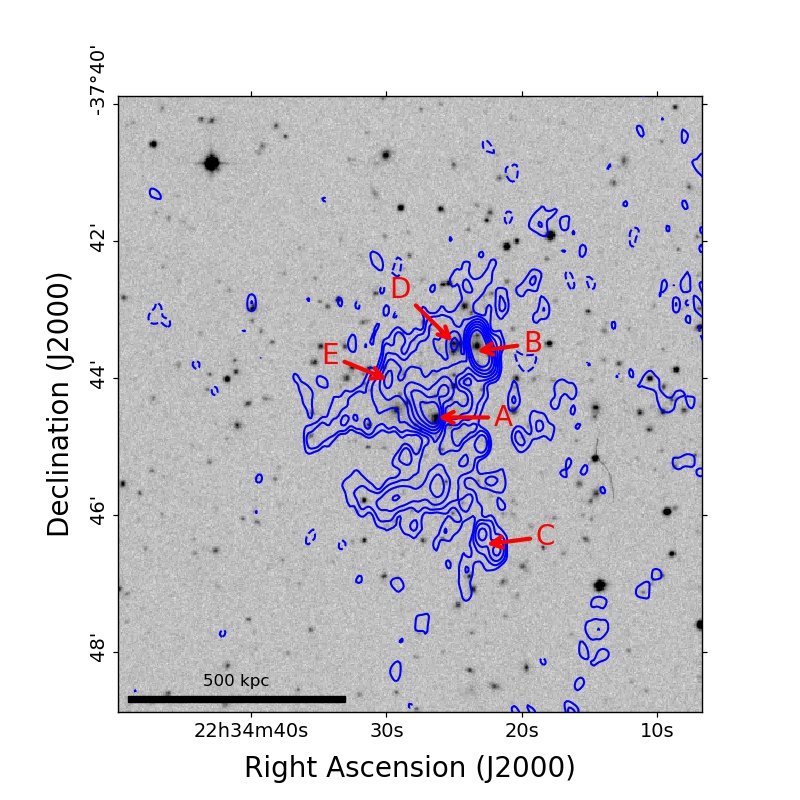}
\caption{Images of the cluster A3888. \textit{Left}: X-ray \textit{Chandra} image (colours) with GMRT 330 MHz contours superimposed. Contours start from 0.9 mJy/beam and are spaced by a factor of two. The -0.9 mJy/beam contour is dashed. \textit{Right}: Same contours as in the left panel superimposed on the optical DSS image. Labels mark the position of discrete radio sources (A to D) and E indicates a patch of diffuse emission with no optical counterpart that we consider to be part of the halo.}
\label{Fig:A3888_330_chandra_opt}
\end{figure*}

A3888 has a mass $M_{500}=6.67\times10^{14}M_\odot$ and it is at $z=0.151$. The dynamical state of A3888 has been debated in the literature \citep{pratt09, bohringer10, chon12, weissmann13, haarsma10}. Spectroscopic observations of the member galaxies reveal that they are distributed in two subgroups, suggesting that a merger is ongoing \citep{shakouri16opt}. Our dynamical analysis, based on the morphological parameters, confirms that A3888 is a merging cluster (Section \ref{Sec:dynamics}). A radio halo in A3888 was discovered with ATCA observations in the frequency range $1.25-2.55$ GHz \citep{shakouri16}. The bright radio halo is also detected in our GMRT 330 MHz image (Fig. \ref{Fig:A3888_330_chandra_opt}) and its morphology resembles the one described by \citet{shakouri16}. The radio emission of the cluster is complex and characterised by the presence of many bright sources embedded in the radio halo emission. As \citet{shakouri16} pointed out, source A\footnote{Source A is actually the blending of two head tail radio galaxies, clearly resolved in the ATCA high-resolution image \citep{shakouri16}.} and B are head tail member galaxies, while source C is a background radio galaxy. Being diffuse themselves, the subtraction of the head tail galaxies from the visibilities is very difficult. In addition to the compact sources detected in \citet{shakouri16}, we detect another patch of emission (labelled E in Fig. \ref{Fig:A3888_330_chandra_opt}, right panel) located north-east of source A. The superposition between the radio contours of A3888 and the optical DSS image in shown in Fig. \ref{Fig:A3888_330_chandra_opt}. While source A, B and D have clear optical counterparts, the brightest part of E does not have a corresponding galaxy, thus it could be a peak of the radio halo emission.
We measured a total flux density of $\sim1.29$ Jy inside the contours shown in Fig.\ref{Fig:A3888_330_chandra_opt}. In order to derive the radio halo flux density we estimated the contribution of the sources embedded in the diffuse emission (except for the patch E) and we subtracted their flux density from the total emission. We obtained a radio halo flux density $S_{330 \mathrm{MHz}}=380\pm60$ mJy. 
The LAS of the radio halo, measured from the 3$-\sigma$ contours, is $190''\times160''$ corresponding to a LLS of $500~\mathrm{kpc}\times420$ kpc. 

We note that the comparison between the radio halo flux density at 330 and 1400 MHz would give a very steep spectrum ($\alpha<-1.8$). However, the contribution of the sources embedded in the radio halo is not properly addressed at either frequency and further analysis is necessary to investigate the spectral properties of this radio halo.

\subsection{Abell 1443}
\begin{figure}
\centering
\includegraphics[scale=0.4]{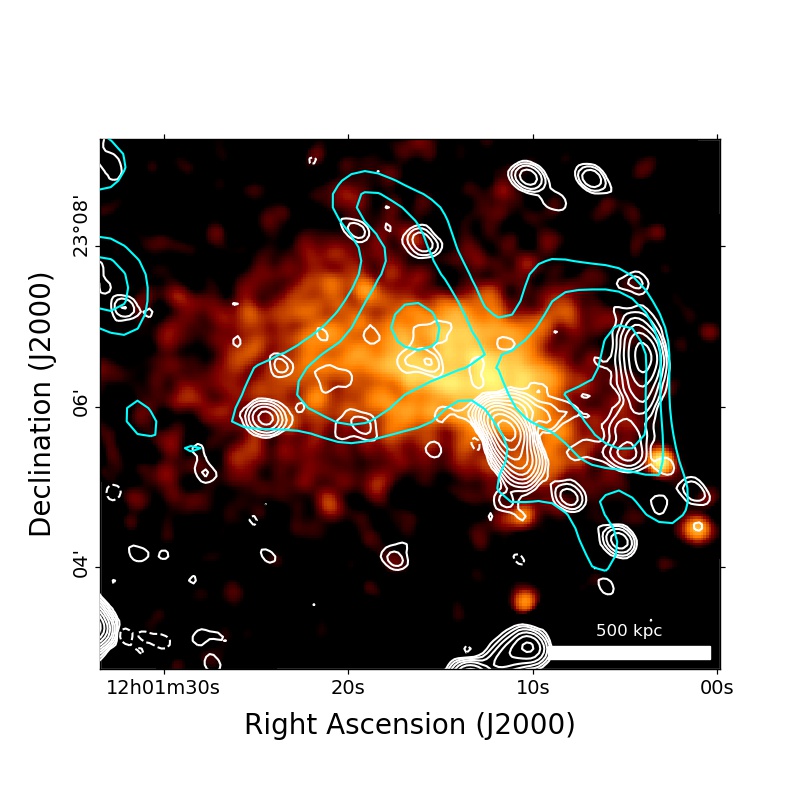}
\caption{X-ray Chandra image of A1443 with JVLA high-resolution (white) and low-resolution (cyan) contours overlaid. Contours start at 3$-\sigma$ and are spaced by a factor of two. 1 $-\sigma$ rms noise is 0.02 mJy/beam with beam=15$''\times12''$ (white contours) and it is 0.08 mJy/beam with beam=55$''\times54''$ (cyan contours).}
\label{Fig:A1443}
\end{figure}

A1443 is a massive ($M_{500}=7.74\times10^{14}M_\odot$) cluster at redshift $z=0.27$. A radio halo in A1443 has been discovered with GMRT 330 MHz \citep{bonafede15}. In addition, the authors detected a peculiar extended source, named `$\Gamma$-shaped' source, and a candidate radio relic on the western side of the cluster. While the `$\Gamma$-shaped' source and the candidate relic are well detected in our GMRT 610 MHz image (not shown here), the radio halo is only marginally visible. We reduced archival JVLA 1.5 GHz observations of A1443, the contours are shown in Fig. \ref{Fig:A1443} (white) and are overlaid on the X-ray Chandra image of the cluster. Also at 1.5 GHz, the radio halo is only visible as patches of diffuse emission. We subtracted the compact sources from the {\it uv}-data and we re-imaged this field at very low-resolution ($\sim$1 arcmin) to increase the possibility of imaging the diffuse emission. The low-resolution contours are shown in cyan in Fig. \ref{Fig:A1443}. Although the western part is most likely associated with the residuals of the two extended sources that are particularly challenging to subtract, we are confident that the central part of the diffuse emission belongs to the radio halo.
We estimated the flux density of the halo in a region that does not include the two extended sources on the west. The radio halo flux density is $S_{1.5 \mathrm{GHz}}=3.5\pm0.10$ mJy, corresponding to $P_{1.4 \mathrm{GHz}}=(9.1\pm0.25)\times10^{23}$ W/Hz. A1443 has been recently observed with LOFAR at 144 MHz as part of LoTSS \citep{shimwell19}. 
A multi-frequency study including the reanalysis of the 330 MHz GMRT data and the combination with LOFAR data is ongoing and will be presented in a future paper (Cuciti et al., in prep.).

\subsection{PSZG139.61+2420}
PSZG139.61+2420 (PSZG139, hereafter) is a low-entropy cool core cluster \citep{giacintucci17} with some trace of dynamical disturbance. Indeed, the X-ray morphology is slightly elongated and the centroid shift parameter has intermediate value between merging and non-merging systems. The mass of PSZG139 is $M_{500}=7.09\times10^{14}M_\odot$ and the redshift is $z=0.27$. Two GMRT 610 MHz observations are available for this cluster. Their combination has been presented in \citet{savini18} and \citet{giacintucci19}. We detected three faint discrete sources at the cluster centre, which blend with a diffuse component in the low-resolution image. We classified such a diffuse component as a mini halo \citep{giacintucci19}. 

PSZG139 has been also observed with LOFAR at 144 MHz. In the LOFAR image the mini halo is surrounded by a larger-scale diffuse component with an estimated spectral index steeper than $\alpha=-1.7$ \citep{savini18}. The coexistence of a mini halo in the cool-core with a larger scale, steep spectrum emission, may be the consequence of a minor merger. Diffuse emission with similar properties has bee recently found also in the galaxy cluster RXC J1720.1+2638 \citep{savini19}.

\begin{figure}
\centering
\includegraphics[scale=0.35]{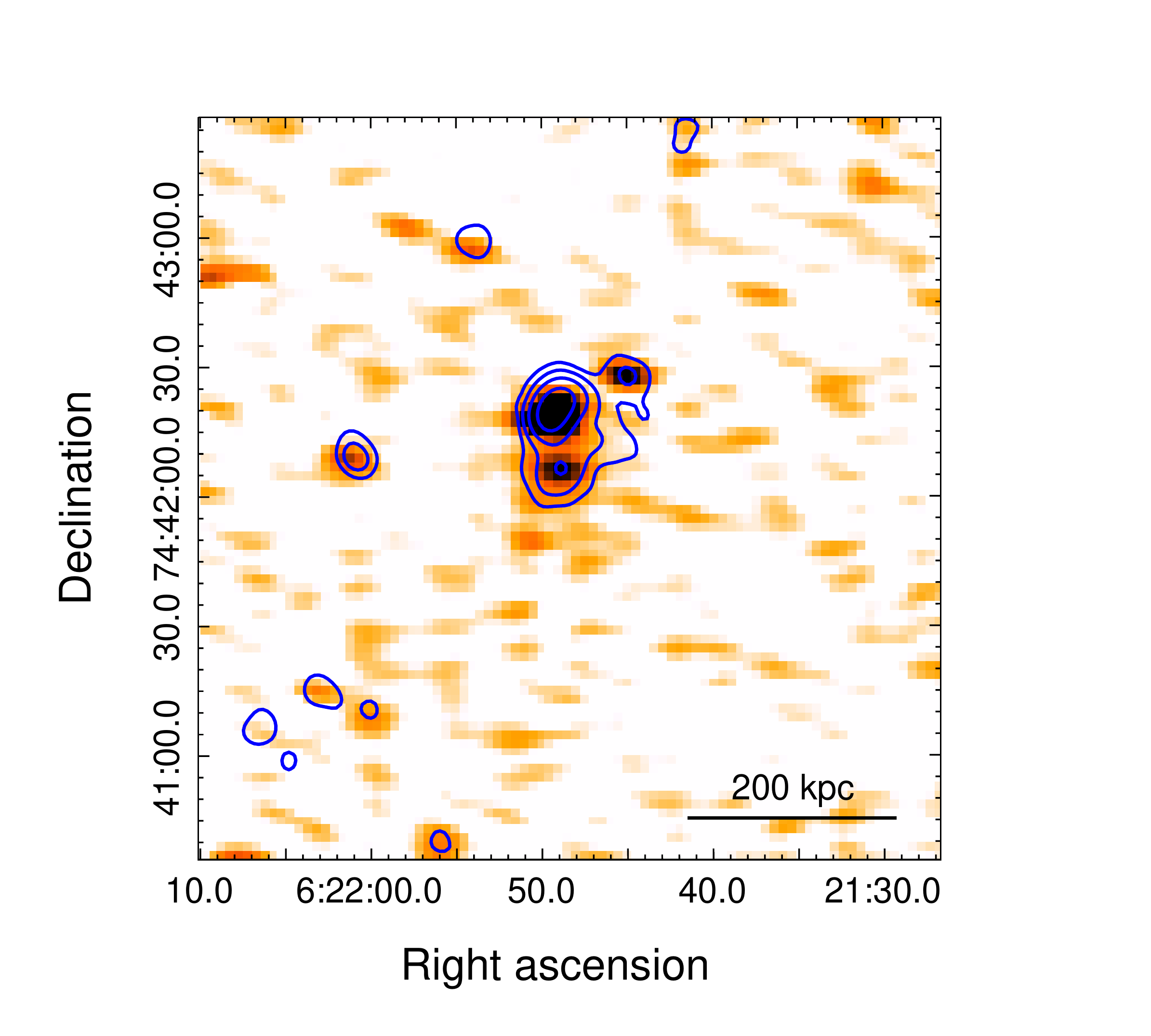}
\caption{JVLA C array $10.3''\times5.1''$ resolution image of the center of PSZG139.61+2420 with GMRT 610 MHz contours overlaid. Contours are $(\pm3,6,9,12...)\times \sigma_{rms}$ with $\sigma_{rms}=0.03$ mJy/beam and beam=$7''\times7''$ \citep{giacintucci17}.}
\label{Fig:P139_C_GMRT}
 \end{figure}

In Fig. \ref{Fig:P139_C_GMRT} we present the JVLA 1.5 GHz C array high-resolution image of the central region of PSZG139 compared to the low-resolution GMRT 610 MHz contours \citep[from][]{giacintucci19}. 
At a similar resolution, the mini halo appears less extended towards the west at high frequencies. We subtracted the discrete sources from the \textit{uv}-data and we estimated the flux density of the mini halo in the same region used for the GMRT image (3$-\sigma$ contours shown in Fig. \ref{Fig:P139_C_GMRT}). We obtained a flux density of the mini halo $S_{1.5 \mathrm{GHz}}=0.60\pm 0.05$ mJy, in agreement with the value measured with GMRT 1.28 GHz observations by \citet{giacintucci19}.

\subsection{Candidate diffuse emission}
\label{Sec:candidate}
\subsubsection{Zwcl2120.1+2256}

\begin{figure*}
\centering
\includegraphics[scale=0.43,trim={0cm 2.5cm 0cm 0cm},clip]{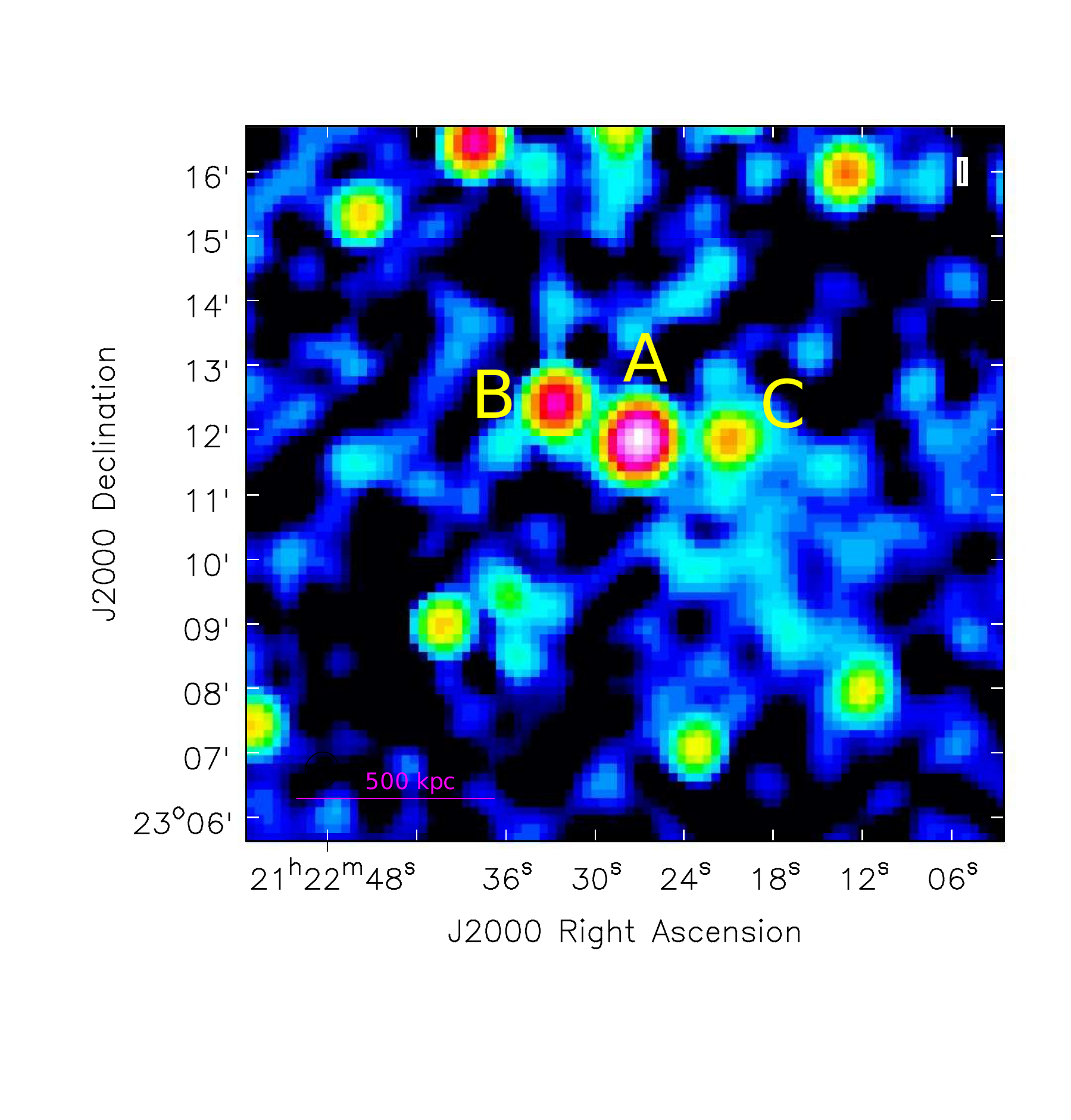}
\hspace{1cm}
\includegraphics[scale=0.47,trim={0cm 0cm 0cm 0cm},clip]{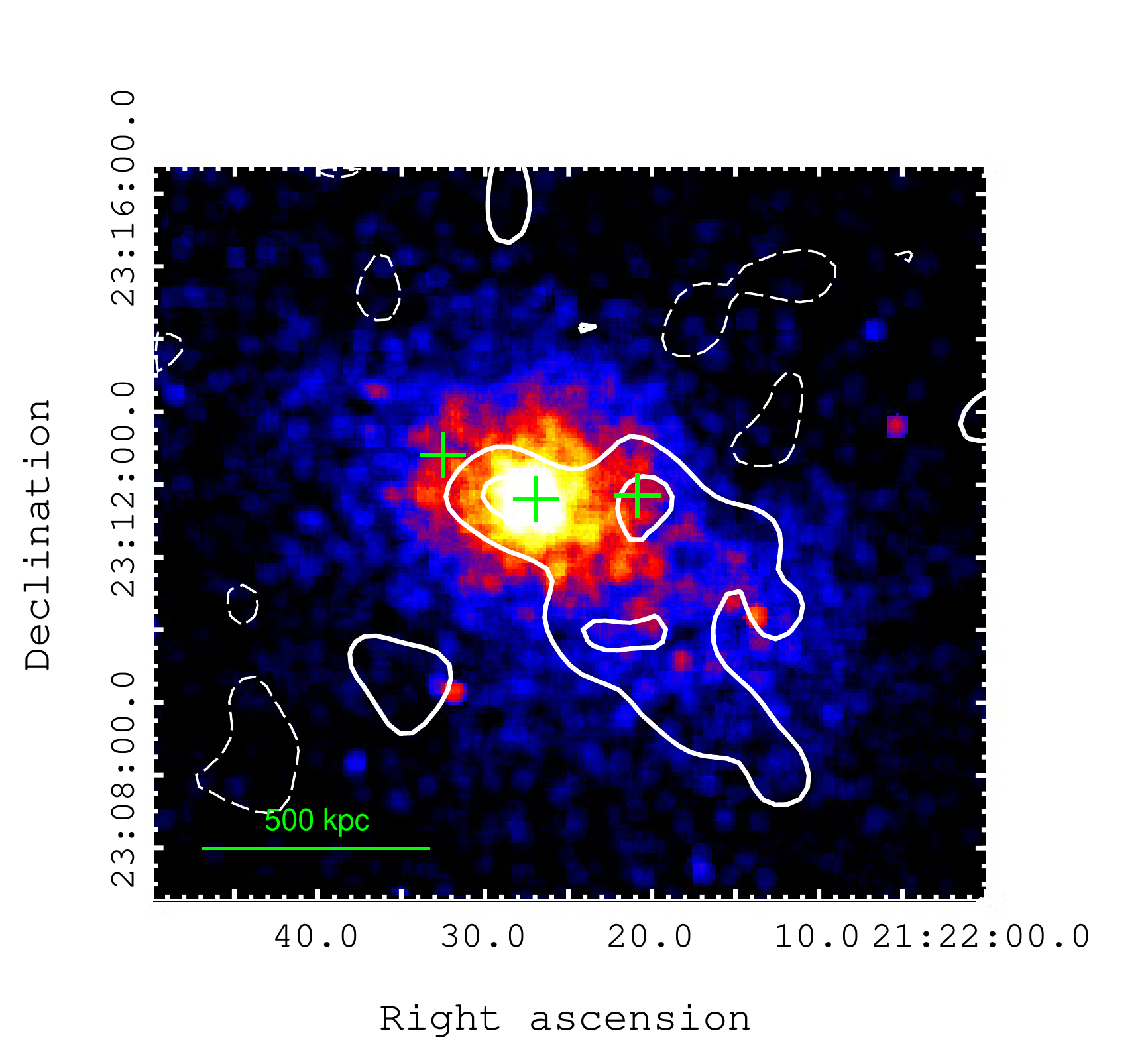}
\caption{JVLA image of the cluster Z2120. \textit{Left}: JVLA D array image. The resolution is $33''\times30''$ and the rms noise is 0.06 mJy/beam. Labels mark the discrete sources in the cluster field. \textit{Right}: Low-resolution ($55''\times54''$) JVLA contours after the compact sources subtraction overlaid over the X-ray \textit{Chandra} image. Contours are drawn at (3,6...)$\times \sigma$, with $\sigma=0.1$ mJy/beam. The 2$-\sigma$ negative contour is dashed. The green crosses mark the position of the three sources at the cluster centre, which have been subtracted. }              
\label{Fig:Z2120_D}
\end{figure*}

Zwcl2120-1+2256 (Z2120, hereafter) is at $z=0.143$ and it is one of the less massive clusters in our sample, with $M_{500}=5.91\times10^{14}M_\odot$. The information available in the literature about this cluster is rather sparse. According to the morphological analysis of the X-ray surface brightness distribution (Section \ref{Sec:dynamics}), we find that Z2120 sits in the intermediate region between merging and relaxed clusters. In fact, the X-ray emission of this cluster is fairly peaked at the centre, but there is a low-surface brightness `tail' extending to the south-west, suggestive ongoing dynamical activity (Fig. \ref{Fig:Z2120_D}, right panel).

The JVLA D array image of Z2120 is shown in Fig. \ref{Fig:Z2120_D} (left panel). In addition to the three compact sources (labelled A, B and C) in the cluster central region, we detected some faint diffuse emission extending towards south-west. The source-subtracted low-resolution image (Fig. \ref{Fig:Z2120_D}, right panel) shows some faint residual emission elongated in the NE-SW direction which, interestingly, follows the X-ray emission of the cluster, especially in the southern area. We note that, while the radio emission on top of the peak of the X-ray emission may be partly due to some residuals from the subtracted sources, the emission coincident with the low-surface brightness south-west X-ray tail is not affected by subtraction. The residual flux density measured within the 3$-\sigma$ contours of Fig. \ref{Fig:Z2120_D} (right panel), considering also the emission in the central region of the cluster, is $S_{1.5GHz}\sim7.7$ mJy. We classify this emission as a candidate radio halo. 

\begin{figure*}
\centering
\includegraphics[scale=0.45]{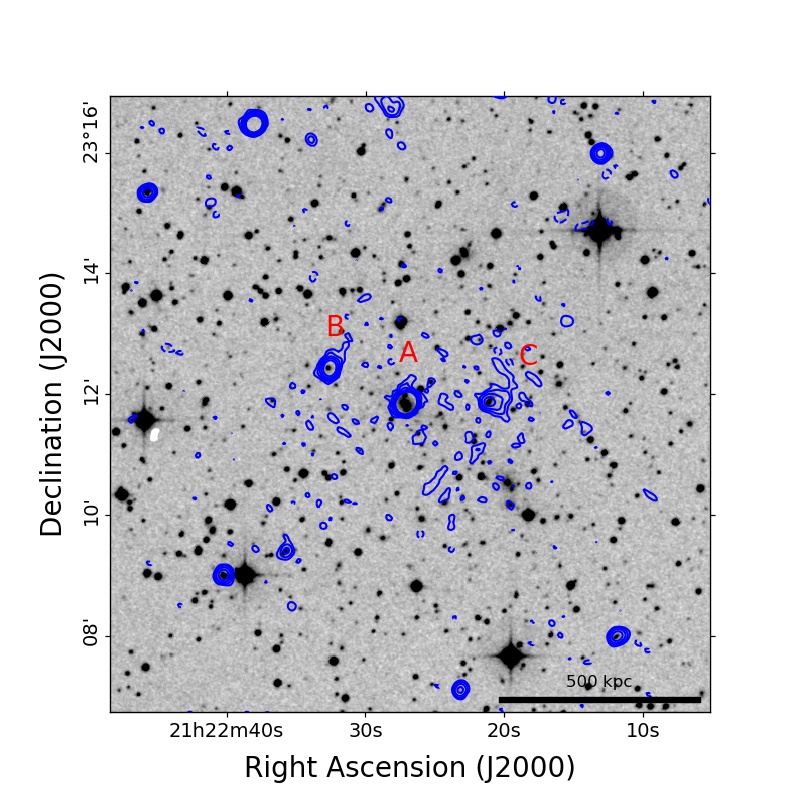}
\includegraphics[scale=0.45]{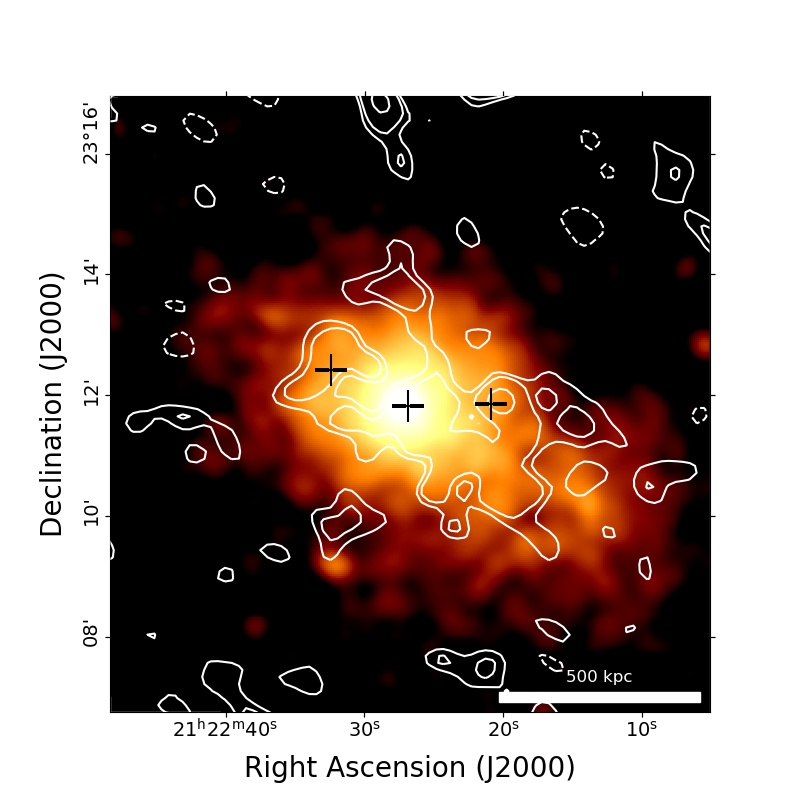}
\caption{GMRT images of the cluster Z2120. \textit{Left}: DSS optical image with GMRT high-resolution contours overlaid. Contours start at 0.1 mJy/beam and are spaced by a factor of two. The first negative contour is dashed. Labels mark the position of discrete sources. \textit{Right}: GMRT 330 MHz low-resolution ($32.7''\times28''$) contours after the subtraction of compact sources superimposed on the \textit{Chandra} X-ray image. Contours are (2,3,6...)$\times \sigma$, with $\sigma=0.3$ mJy/beam. The position of the discrete sources is marked with a black cross.}
\label{Fig:Z2120_330}
\end{figure*}

We reduced an archival GMRT 330 MHz observation of Z2120 (P.I. C. Jones) with the SPAM pipeline and then we imaged the processed data with CASA. Only some patches of diffuse emission are visible in the cluster central region on the high-resolution image ($\sim10''$, Fig. \ref{Fig:Z2120_330}, left panel). In the source-subtracted low-resolution image (Fig. \ref{Fig:Z2120_330}, right panel) a residual emission of $\sim46$ mJy is detected at low significance level. Remarkably, this emission is spatially coincident with the one detected at higher frequency, thus supporting the idea of a low surface brightness emission associated with the perturbed ICM. Given the low signal to noise ratio of the detection at both frequencies it is difficult to obtain a solid spectral information.


\subsubsection{Abell 3041}

\begin{figure*}
\centering
\includegraphics[scale=0.45]{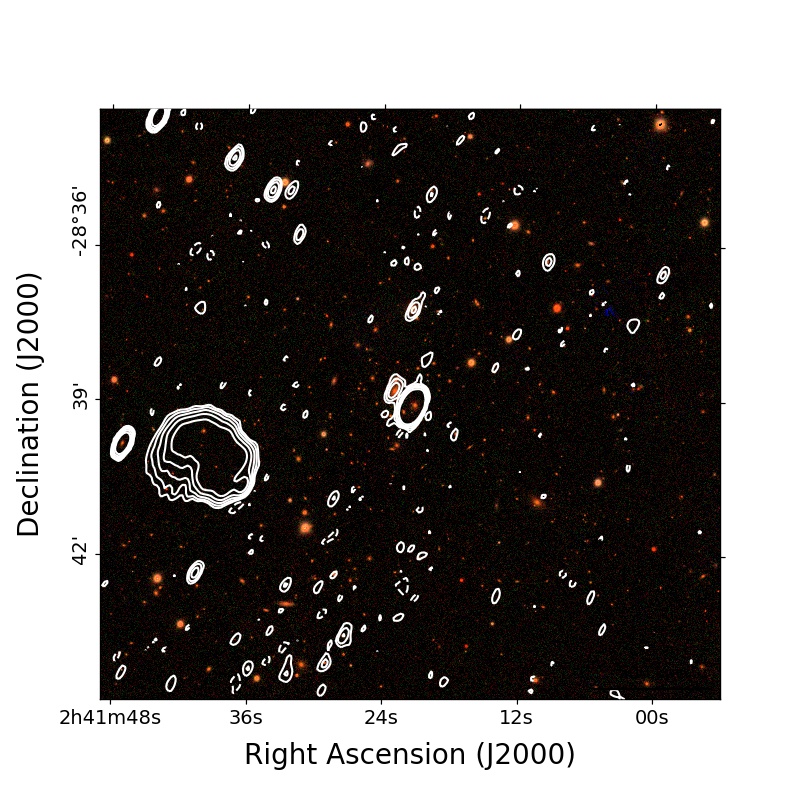}
\includegraphics[scale=0.45]{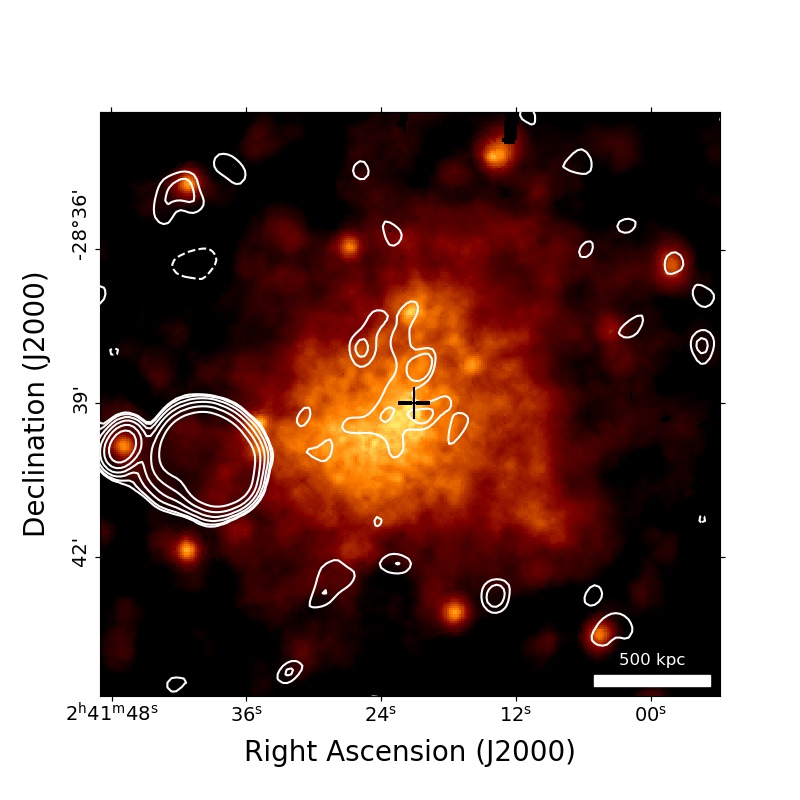}
\caption{Images of the cluster A3041. \textit{Left}: High-resolution JVLA C array contours (red) superimposed on the PanSTARRS optical image. Contours start at 0.1 mJy/beam and are spaced by a factor of two. The first negative contour is dashed. \textit{Right}: X-ray \textit{XMM-Newton} image with low-resolution ($35.8''\times25.5''$) JVLA C array contours overlaid. Contours are ($\pm2,3,6...)\times\sigma_{rms}$ with $\sigma_{rms}=0.08$ mJy/beam. The discrete sources in the field are subtracted, except for the region of the FRII radio galaxy. The black cross marks the position of the central brightest radio source.}
\label{Fig:A3041}
\end{figure*}

Abell 3041 is at redshift $z=0.23$ with mass $M_{500}=6.12\times10^{14}M_\odot$. An X-ray \textit{Chandra} observation of A3041 is available in the archive. Although it is very shallow (exposure time 9 ks), we processed and used it to derive the morphological parameters, which place A3041 in the merging region of the morphological diagrams (Section \ref{Sec:dynamics}). 
In Fig. \ref{Fig:A3041} (left panel) we show a deeper \textit{XMM-Newton} observation that better highlights the disturbed morphology of the cluster.

The cluster hosts two central radio sources, which are blended at the resolution of our C array observation ($\sim 10''$). Their overall flux density is $105\pm 3$ mJy. Both have optical counterparts in the DSS optical image (Fig. \ref{Fig:A3041}, left panel). There is no available redshift for the brightest one, while the fainter one is associated with a member galaxy \citep{colless03}. On the eastern side of the cluster there is an FRII radio galaxy extending over $\sim5.8$ arcmin. The nucleus is located $\sim 5.5$ arcmin from the centre of A3041, and no spectroscopic redshift is available for this source. The host galaxy is detected in the 2 MASS catalogue \citep{2mass}, with a $K$ magnitude of 15.62. Using the $K-z$ relation by \citet{willott03} we estimated that the redshift of the FRII galaxy is $z=0.42\pm0.16$, suggesting that it may be a background giant radio galaxy extending over almost 2 Mpc.

We subtracted all the discrete sources from the dataset, except for the FRII galaxy. We paid special attention to the subtraction of the two central sources. In particular, we did not adopt the usual approach described in Section \ref{Sec:GMRT_data}, but we subtracted a model made using the whole \textit{uv}-range. In this way, we made sure that, if there is some faint extended radio emission associated with the central sources, it is subtracted from the data that we then use to produce the low-resolution image, shown in Fig. \ref{Fig:A3041} (right panel). A residual emission of $\sim$ 4 mJy is present at the cluster X-ray peak, mostly detected at the 2$-\sigma$ level only. The spatial coincidence between the thermal and non-thermal emission may suggest that these residuals belong to a cluster diffuse radio source. However, the detection is marginal and we are aware that even a small calibration error around the bright central source might leave some residuals showing up in the low-resolution image. We thus consider A3041 as a case of candidate diffuse emission. 

\subsubsection{Abell 402}   

Abell 402 is a massive ($M_{500}=7.21\times10^{14}M_\odot$) cluster of galaxies at $z=0.32$. Its temperature within $R_{2500}$ measured on the [0.7--7] keV band excluding the central 70 kpc region is $8.0^{+1.1}_{-0.9}$ keV \citep{cavagnolo08,giacintucci17}. Although the X-ray morphology of the cluster is fairly regular and peaked at the center, the central entropy floor of A402 is relatively high \citep[$K_0=156\pm25$ keV/cm$^2$,][]{cavagnolo09,giacintucci17} suggesting that it does not possess a cool core and some sort of dynamical activity may be taking place in this cluster.

\begin{figure*}
\centering
\includegraphics[scale=0.45]{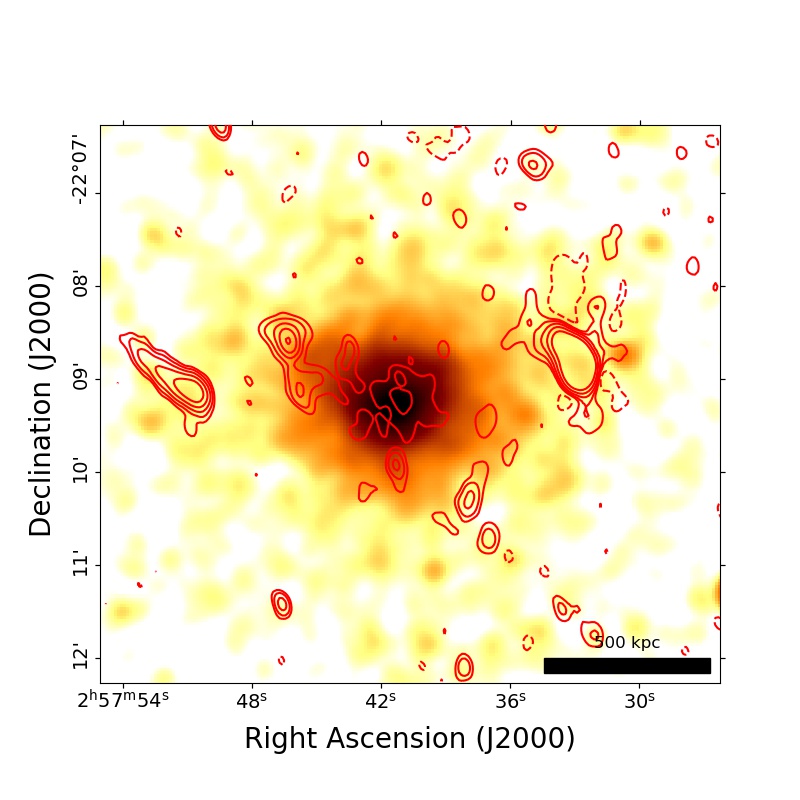}
\includegraphics[scale=0.45]{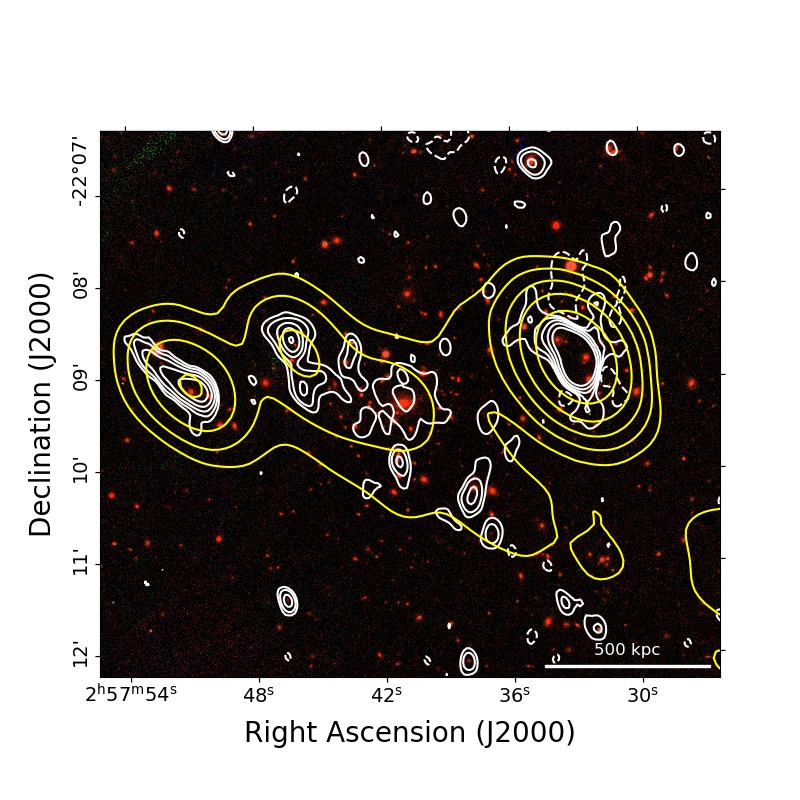}
\caption{Images of the cluster A402. \textit{Left}: GMRT 330 MHz contours superimposed on the X-ray \textit{Chandra} image of the cluster. Contours start at 0.35 mJy/beam and are spaced by a factor of two. The first negative contour is dashed. The beam is $13.4''\times8.6''$ and the rms noise of the radio image is $\sim0.1$ mJy/beam. \textit{Right}: PanSTARRS optical image of the field of A402 with the same contours of the left panel (white) plus the low-resolution GMRT 330 MHz contours (yellow). Contours start at 3 mJy/beam and are spaced by a factor of two. The rms noise of the low-resolution image is $\sim1$ mJy/beam with beam=$57''\times42''$. }
\label{Fig:A402}
\end{figure*} 

There are two archival GMRT 330 MHz observations available for this cluster (Obs No: 6153, P.I.: G. Macario and Obs No: 6837, P.I.:A. Bonafede). Both have been processed with the SPAM pipeline and both show evidence of diffuse emission at the cluster center. From the combination of the two images, \citet{giacintucci17} classified the diffuse emission in A402 as a candidate radio halo. The images of A402 are shown in Fig. \ref{Fig:A402}. On the left panel we show the GMRT 330 MHz high-resolution contours from Obs No 6153. There are several diffuse sources in the cluster field. The most interesting one, for our purposes, is the central one, which is co-spatial with the core of the cluster, as it is typical for radio mini halos \citep{mazzotta08,giacintucci14}. The flux density of this source, measured inside the 3$-\sigma$ contours shown in Fig. \ref{Fig:A402} (left panel) is $\sim$12 mJy and its LAS in the east-west direction is $\sim80''$ corresponding to $\sim370$ kpc.

On the right panel of Fig. \ref{Fig:A402} we also show the low-resolution contours of the same portion of the sky (derived from Obs No 6837), where some diffuse emission seems to be present on a larger scale. However, the image has a fairly low sensitivity (rms noise $\sim1$ mJy/beam with beam=$57''\times42''$) and the discrete sources in the cluster field (which have not been subtracted out) may largely contribute to the apparently extended emission. We fitted the surface brightness profile of the diffuse source in A402 with the technique discussed in Section \ref{Sec:profile} and we found a central surface brightness $I_0 = 9.1 \pm 1.1~\mu$Jy/arcsec$^2$ and an $e$-folding radius $r_e = 84.8 \pm 13.9$ kpc. These values would place A402 among mini halos in the $I_0-r_e$ diagram from \citet{murgia09}.
To be cautious we consider the diffuse emission in A402 as a candidate mini halo\footnote{While this paper was in preparation, \citet{giovannini20} classified the diffuse emission in A402 as a radio halo, using a short JVLA observation (25 minutes in C array and 10 minutes in D array), however the concerns about the possible contamination from the discrete sources apply also in this case.}.

\subsection{Radio diffuse emission in the full sample}

The observational campaign carried out during this work has enabled the completion of the radio information of our sample. In summary, among the 75 clusters presented in Table \ref{tab:completesample}, there are: 28 ($\sim37\%$) radio halos, ten of which are USSRHs or candidate USSRHs; seven ($\sim10\%$) radio relics, five of which in clusters with radio halos (already counted above); 11 ($\sim15\%$) mini halos, two of which show steep spectrum emission on a larger scale; five candidate radio halos and one candidate mini halo ($\sim8\%$); 31 ($\sim41\%$) clusters without central diffuse emission (two of which host relics, already counted above).

\section{Non-detections and upper limits}
\label{Sec:UL}
Among the 31 clusters of the sample without radio halos, 17 have been analysed in this paper (Table \ref{tab:UL}). High-resolution images of fields of $\sim10'\times10'$ centred on the cluster centre are shown in Appendix \ref{app:UL}. To extract quantitative information from the non-detection of radio halos in it is crucial to derive meaningful upper limits to the diffuse flux density of those clusters. Due to the bad quality of four of the datasets listed in Table \ref{tab:UL}, in the following, we will derive upper limits for 13 clusters. The issues related to these four clusters will be briefly discussed at the end of this section. We adopted the method of injecting mock radio halos in the \textit{uv}-datasets. The injection technique was introduced in the GMRT Radio Halo Survey \citep{brunetti07,venturi07,venturi08,kale13,kale15} and it has been used in the literature since then \citep{dallacasa09,russell11,bonafede17,johnston-hollitt17,cuciti18}. 

In this work, following \citet{bonafede17}, we modelled the radio halo brightness profile with an exponential law in the form:
\begin{equation}
\label{eq:model_halo}
I(r)=I_0 e^{-\frac{r}{r_e}},
\end{equation}
\noindent where $I_0$ is the central surface brightness and $r_e$ is the $e$-folding radius \citep{orru07,murgia09}. 
In order to inject Megaparsec scale radio halos, as reference we used $r_e=500~\mathrm{kpc}/2.6=192$ kpc, where 2.6 is the median value of the quantity $R_H/r_e$ for the radio halos studied by both \citet{murgia09} and \citet{cassano07} \citep{bonafede17}. In particular, $R_H$ is calculated, in \citet{cassano07}, as $R_H=\sqrt{R_{min}\times R_{max}}$, where $R_{min}$ and $R_{max}$ are the minimum and maximum radii of the 3$-\sigma$ surface brightness isocontours.

\begin{figure*}
\centering
\includegraphics[scale=0.33,trim={1cm 0cm 0cm 0cm},clip]{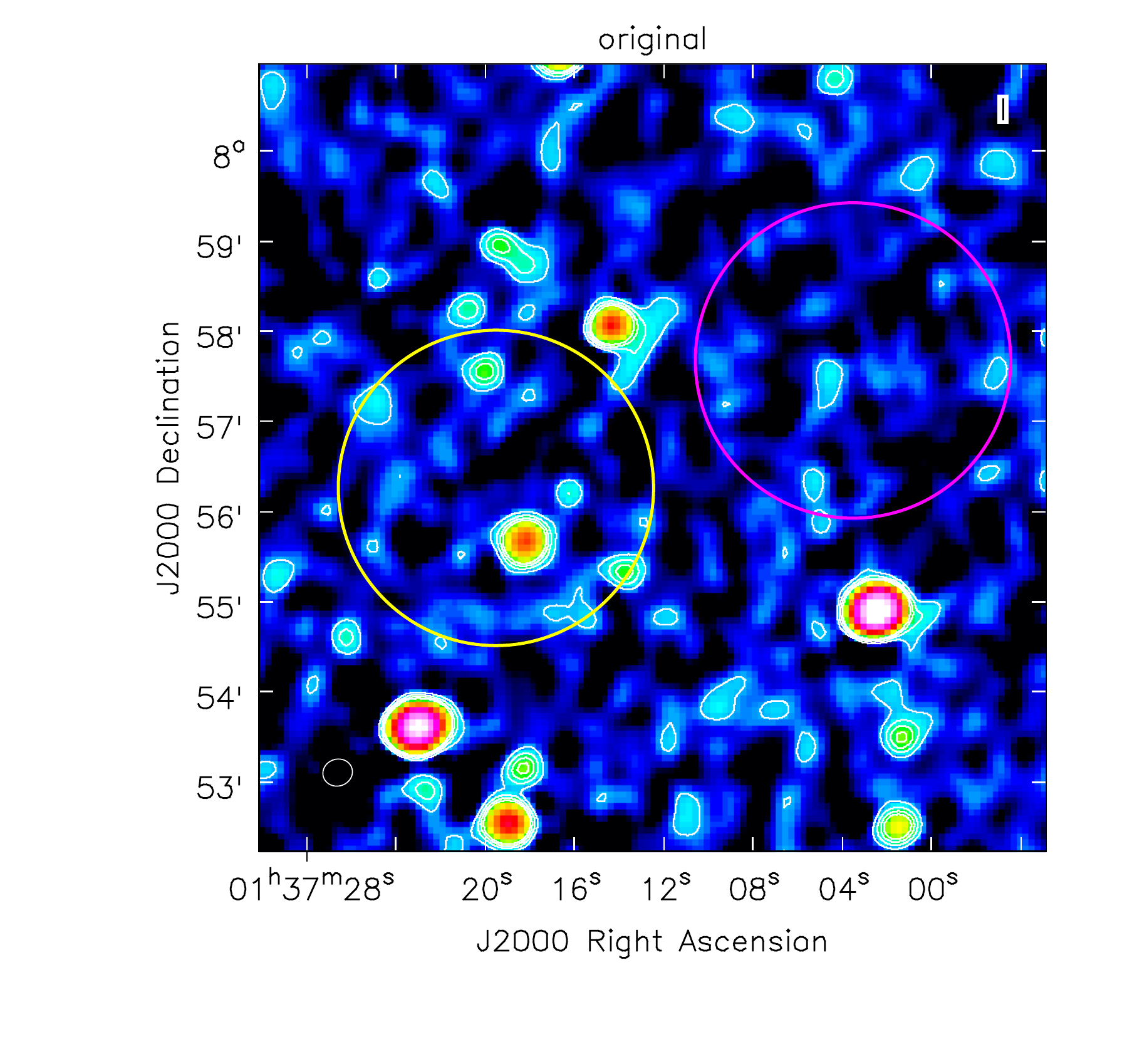}
\includegraphics[scale=0.33,trim={1cm 0cm 0cm 
0cm},clip]{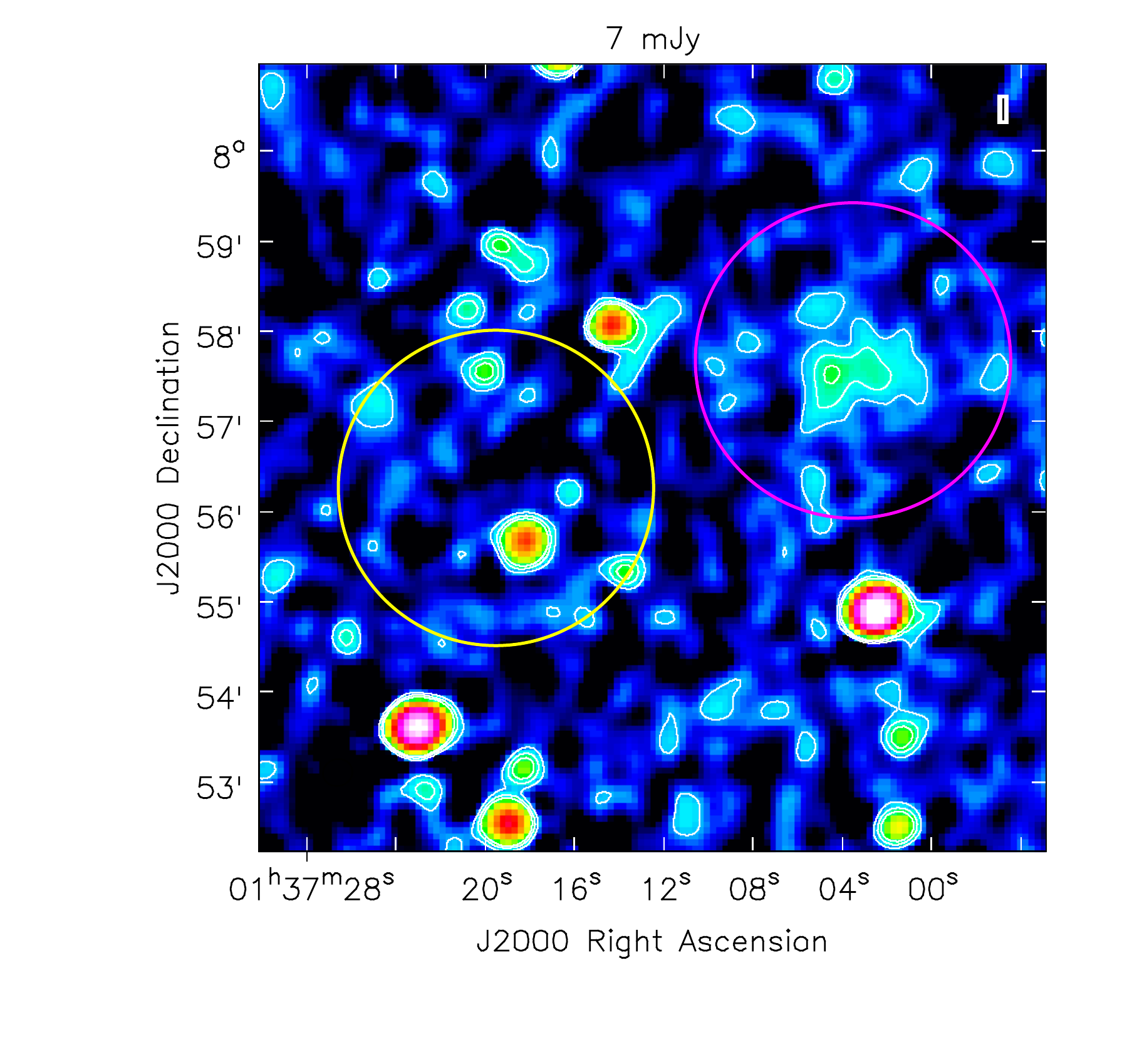}
\includegraphics[scale=0.33,trim={1cm 0cm 0cm 
0cm},clip]{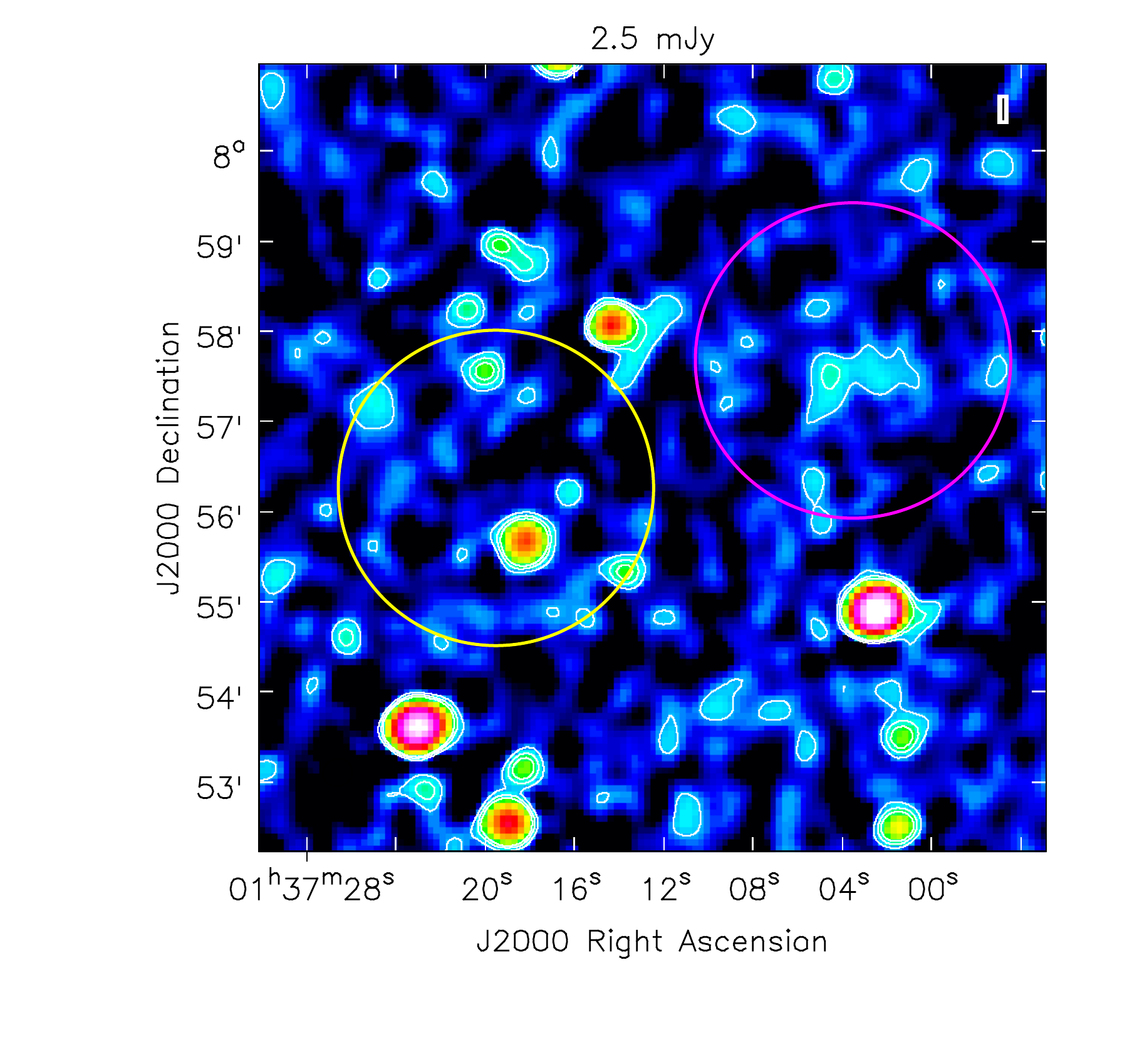}
\caption{Injection of mock radio halos in the cluster A220. \textit{Left}: Original image. \textit{Middle}: Injected radio halo with flux density of 7 mJy. \textit{Right}: Injected radio halo with flux density of 2.5 mJy. Contours start at 2$-\sigma$ rms noise. The two circles have radii of $2.6\times r_e =500$ kpc and are centred on the cluster centre (yellow) and on the modelled radio halo (magenta). We consider 2.5 mJy to be the upper limit to the diffuse emission of A220.}
\label{Fig:ul}
\end{figure*}

For each cluster, we chose a region in the image, close to the pointing center and void of sources and clear noise pattern and we created a set of mock radio halos with different integrated flux densities, centred on that region. 
We added the modelled radio halos to the datasets and, for each modified dataset, we followed the same procedures described in the previous sections to produce images optimised for the detection of the extended emission. For a given cluster mass, we started injecting a mock radio halo that would lie on the radio power--mass correlation from \citet{cassano13} and we reduced the injected flux density until the injected radio halo appeared just as some positive residuals leading to the `suspect' of diffuse emission. More quantitatively, we stop when the largest linear scale ($2-\sigma$ contours) of the recovered halo is $\sim2\times r_e$, implying that only $\sim30\%$ of the injected flux has been recovered, in line with the approach used in \citet{bonafede17} and \citet{cuciti18}.
The injected flux density corresponding to this marginal detection can be considered as the upper limit for that particular cluster. We did not inject the mock radio halo at the cluster centre because the possible presence of some weak residual emission in the cluster field may favour the detection of the mock radio halo, biasing our upper limit towards lower values. Furthermore, faint cluster radio galaxies below the detection limit of our observations may contribute to a positive plateau in the cluster field \citep[e.g.][]{farnsworth13, cuciti15}. An example of the injection procedure is shown in Fig. \ref{Fig:ul} for the cluster A220: the original image is in the upper left panel and a series of fake radio halos with decreasing flux densities is in the other panels. As a sanity check, we fitted the surface brightness radial profile, with the technique described in Section \ref{Sec:profile}, of an injected mock halo. In particular, we focused on A2104 and we injected a mock halo with flux density 70 mJy and $r_e=150$ kpc ($I_0=4.3~\mu$Jy/arcsec$^2$). We obtained $r_e=147\pm15$ kpc and $I_0=3.4\pm0.4~\mu$Jy/arcsec$^2$ as best fit parameters. We repeated this test with flux densities in the range 50-70 mJy and $r_e$ in the range 100-192 kpc, obtaining a discrepancy of $\sim15\%$ at most with respect to the injected values.

We applied this procedure to the 17 clusters listed in Table \ref{tab:UL}. We scaled the upper limit to 1.4 GHz, assuming a spectral index $\alpha=-1.3$, for clusters with GMRT 330 or 610 MHz data only (Table \ref{tab:UL}). In general, the possibility to place deep upper limits is related to the quality of the images in terms of sensitivity, density of the inner \textit{uv}-coverage and presence of bright sources with residual calibration errors. 
In line with the results from \citet{venturi08}, we found that the presence of radio halos with flux densities $\gtrsim10$ mJy can be safely established with GMRT 610 MHz observations, while injected radio halos of 5-10 mJy appear in the images as positive residuals that would lead to the suspect of diffuse emission and can be thus considered as upper limits. The values of the upper limits obtained with the JVLA range from 2 to 5 mJy. 

Unfortunately, we were not able to derive reliable upper limits for four clusters: RXC J0510.7-0801, RXC J1322.8+3138, A1437 and Zwcl1028.8+1419. The GMRT 610 MHz and 240 MHz images of RXC J0510.7-0801 are affected by the presence of a strong radio source in the field, whose sidelobes cross the cluster field even after several runs of self-calibration and peeling \citep[][see also Fig. \ref{Fig:R0510}]{kale15}. The field of RXC J1322.8+3138 is dominated by an FRII radio galaxy extending over $\sim11.5'$ close to the cluster region (Fig. \ref{Fig:R1322}). This source limited the possibility of producing sensitive low-resolution images. A lot of editing was needed for the datasets of A1437 and Zwcl1028.8+1419, especially at the short baselines; this, combined with the relatively high noise level of the images, did not allow a useful upper limit to be derived. We observed Zwcl1028.8+1419 with LOFAR and RXC J1322.8+3138 has been observed with LOFAR as part of LoTSS. We carried out a preliminary analysis of these observations and we anticipate that they do not suggest the presence of clear diffuse emission on the cluster scale.

\section{Surface brightness radial profile}
\label{Sec:profile}

\begin{figure*}
   \centering
   \includegraphics[height=7.5cm]{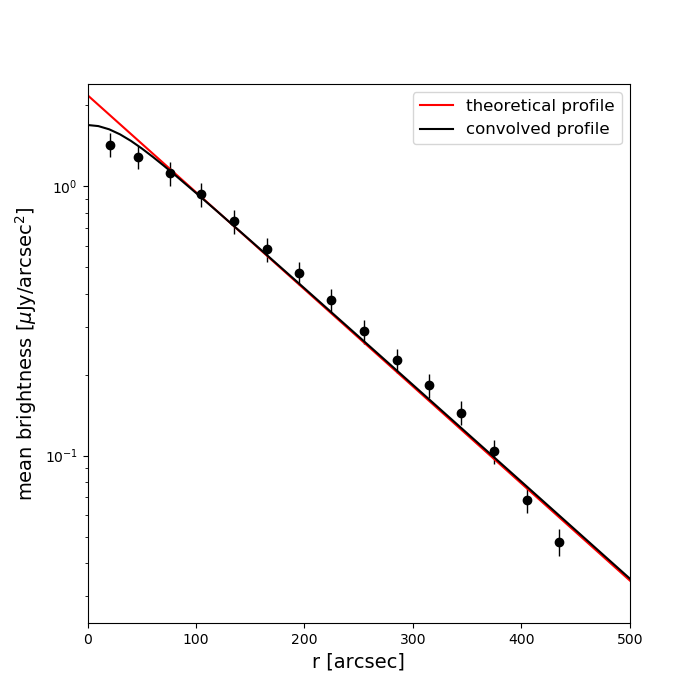}
   \includegraphics[height=7.5cm]{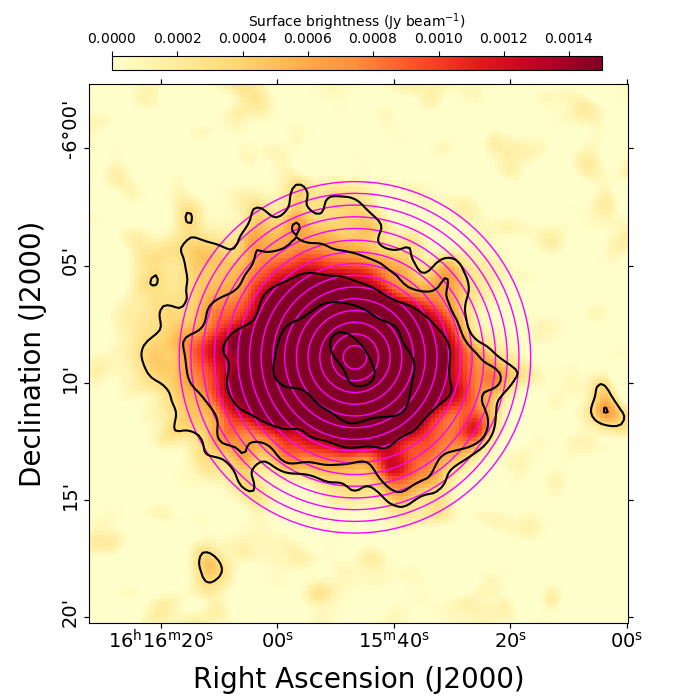}
   
 \caption{Radial surface brightness profile of the radio halo in A2163. \textit{Left}: Data points represents the averaged surface brightness measured in the annuli shown in the right panel. The red curve is the theoretical exponential profile, the black curve is the profile convolved with the beam of the image. \textit{Right}: VLA 1.4 GHz image from where the profile on the left panel has been extracted. The magenta annuli represent the regions where the average surface brightness was measured. Contours start at 3$-\sigma$ and are spaced by a factor of two. The 1$-\sigma$ rms noise of the image is 0.1 mJy/beam with beam=$60''\times60''$.}
         \label{Fig:profile}
   \end{figure*}
   
Following the approach described in \citet{murgia09}, we derived the azimuthally averaged surface brightness radial profile for the confirmed radio halos in our sample, and we fitted them with an exponential law in the form (\ref{eq:model_halo}) in order to derive the central surface brightness, $I_0$, and the $e$-folding radius, $r_e$. To do that, we used low-resolution source-subtracted images, convolved with a Gaussian circular beam. In case the image contains diffuse sources that cannot be properly subtracted we masked them (see e.g. the case of A2744 in Fig. \ref{Fig:profile_A2744}, where the radio relic has been masked) and we did not consider the masked pixels when calculating the surface brightness. We averaged the radio brightness in concentric annuli, centred on the peak of the image (Fig. \ref{Fig:profile}, right panel). The width of the annuli was chosen to be half of the FWHM of the beam of the image. We considered only annuli with an average surface brightness profile higher than two times the rms noise of the image. We note that the estimated fitting parameters differs only marginally ($<5\%$) if we consider only annuli with an average surface brightness profile higher than three times the rms noise of the image. We first generated a two--dimensional model using eq. (\ref{eq:model_halo}) with the same size and pixel size of the radio image and we convolved it with a Gaussian with the FWHM equal to the beam of the image. Then, we azimuthally averaged the exponential model with the same set of annuli used for the radio halo. The one-dimensional surface brightness profile of the two-dimensional exponential model (convolved with the beam) is our fitting model (see e.g. the black line in Fig. \ref{Fig:profile}), which takes into account the resolution of the image and the uncertainties associated with the sampling of the radial profile. We show the radial profile of the radio halo in A2163 with the best fit model in Fig. \ref{Fig:profile}, the others are shown in Appendix \ref{app:profile}.

\begin{table}
\begin{center}
\begin{footnotesize}
\caption{Radio halo brightness profiles }
\begin{tabular}{lccc}
\\
\hline
name & $r_e$  & $I_0$ & $\chi^2$ \\
     & (kpc)  & ($\mu$Jy/arcsec$^2$)&\\
\hline\\
  A773 & 203$\pm$ 14 & 0.42$\pm$ 0.05 & 0.16 \\
  A665 & 209$\pm$ 14 & 0.89 $\pm$ 0.08 & 1.10\\
  A209 & 225 $\pm$ 10 & 0.58 $\pm$ 0.05 & 1.88 \\
  A2163 & 402 $\pm$ 10 & 2.18 $\pm$ 0.11 & 1.16 \\
  A2218 & 82 $\pm$ 5 & 1.27 $\pm$ 0.15 & 1.97\\
  A2744 & 261 $\pm$ 8 & 3.02 $\pm$ 0.02 & 0.55\\
  Z0634 & 116 $\pm$ 14 & 0.70 $\pm$ 0.10 & 2.14 \\
  A2219 & 246 $\pm$ 12 & 1.31 $\pm$ 0.10 & 0.41 \\
  A1758 & 191 $\pm$ 13 & 1.71 $\pm$ 0.18 & 3.26 \\
  A697 & 184 $\pm$ 7 & 4.24 $\pm$ 0.34 & 8.02\\ 
  RXC J1314.4-2515 & 60 $\pm$ 4 & 7.45 $\pm$ 0.89 & 1.43 \\
  A521 & 187 $\pm$ 6 & 8.49 $\pm$ 0.57 & 2.21\\
  PSZ1 G171.96-40.64 & 253 $\pm$ 15 & 1.35 $\pm$ 0.11 & 2.23\\
  RXC J0142.0+2131 & 147 $\pm$ 18 & 4.74 $\pm$ 0.63 & 0.49\\
  
\hline
\label{Tab:profile_emissivity}
\end{tabular}
\end{footnotesize}
\end{center}
\end{table}

This method is based on the assumption that radio halos have a central peak and then the brightness decreases with increasing distance from the centre. This assumption is valid for most of the radio halos in our sample, however, ten radio halos (Z0104, R2003, A520 and A1351, A3411, R1514, A1300, A1132, A2142, A1443) clearly have multiple peaks, We did not include these radio halos in the analysis of the radial profiles. Moreover, we excluded four radio halos for which we do not have suitable source subtracted images (A1451, A3888, A2261, A1689). Thus, we derived and fitted the radial surface brightness profile of the 14 radio halos listed in Table \ref{Tab:profile_emissivity}. These radial profiles, together with the set of annuli used to derive them and the best fit model are shown in appendix \ref{app:profile}. In Table \ref{Tab:profile_emissivity} we summarise the best fit parameters, $I_0$ and $r_e$ and the reduced $\chi^2$ of the fit. We will use this information to derive the emissivity of radio halos in paper II.

\begin{table}
\begin{center}
\begin{scriptsize}
\caption{Clusters without detected extended emission}
\begin{tabular}{lcccc}
\\
cluster name  &rms&beam& UL* & $P_{1.4GHz}$** \\
&(mJy/beam)& ($''\times''$)&(mJy)&($10^{23}$ W/Hz)\\

\hline
\hline
\\
\textbf{JVLA L-band}& & & \\

A56$^{(C)}$ & 0.09 & $19.8\times19.2
$ &	4&	12\\
A2813$^{(C)}$& 0.055 & $20.5\times18.1$ &	5&	14\\
A2895$^{(C)}$& 0.055 & $22.4\times17.9$ &	3&	5\\
A220$^{(C)}$ & 0.055 & $27.6\times26.3$ &	2.5&9.7\\
A384$^{(C)}$& 0.045 & $28.3\times26.0$ &	4&	7.4\\
A2472$^{(C)}$& 0.06 & $18.1\times17.4$ &	3&	10\\
A2355$^{(C)}$& 0.08 & $28.2\times24.2$ &	5&	8.3\\
RXC J2051.1+0216$^{(C)}$& 0.04& $19.8\times18.3$& 2& 7.3\\
RXC J0616.3-2156$^{(D)}$& 0.08 & $78.3\times52.8$
 &	3&	2.5\\
\hline\\
\textbf{GMRT 610 MHz}& & & \\

Zwcl1028.8+1419& 0.25 & $18.6\times16.0$ &	--&	--\\
RXC J1322.8+3138& 0.35 & $19.2\times17.8$ &	--&	--\\
A1733& 0.34 & $34.7\times27.0$ &	7&	5.3\\
PSZ1 G019.12+3123& 0.08 & $21.5\times18.0$ &	7&	6.3\\
MACS J2135-010& 0.2 & $21.6\times17.2$ &	10& 11.7\\
RXC J0510.7-0801& 0.2 & $5.4\times4.8$ &	--&--\\
\hline\\

\textbf{GMRT 330 MHz}& & & \\

A1437& 0.27 & $19.4\times16.5$ &	--&--\\
A2104&  0.25 & $20.6\times19.6$ &	25&2.4\\

\hline
\hline
\label{tab:UL}
\end{tabular}
\tablefoot{\textit{Top panel}: Upper limits (UL) derived with JVLA L-band (C) C array or (D) D array observations. \textit{Middle panel}: UL derived with GMRT 610 MHz observations. \textit{Bottom panel}: UL derived with GMRT 330 MHz observations. *Flux density of the UL measured at the observing frequency. **Radio power of the UL at 1.4 GHz. For UL derived at 330 or 610 MHz we assumed a spectral index $\alpha=-1.3$.}
\end{scriptsize}
\end{center}
\end{table}

\section{X-ray data analysis} 
\label{Sec:X-ray analysis}
In this Section we describe the analysis of the dynamical properties of the clusters of our sample. Among the 75 clusters of the sample, 63 have archival X-ray \textit{Chandra} data. 54 of them already have literature information on their dynamical status \citep{cassano10,cassano13,cassano16,cuciti15}. We produced and analysed the \textit{Chandra} images of the remaining nine clusters (marked with $\surd$ in Tab.\ref{Tab:dynamics}).
\textit{Chandra} X-ray data were processed with CIAO 4.5 using calibration files from CALDB 4.5.8. Standard techniques to correct time-dependent issues were applied\footnote{http://cxc.harvard.edu/ciao/threads/index.html}, the screening of the events file was applied to filter out strong background flares, cosmic rays and soft protons. We identified point sources with an automatic algorithm and we removed them from the images. Images were normalised for the exposure map of the observation.

Following \citet{cassano10,cassano13} and \citet{cuciti15}, we produced \textit{Chandra} images in the 0.5-2 keV band and we analysed the X-ray surface brightness inside an aperture radius $R_{ap}=500$ kpc, centred on the cluster X-ray peak. 500 kpc is the typical radius of radio halos, in this way we evaluate the morphological properties of clusters within the region where energy is most likely dissipated and generates synchrotron emission. To provide a quantitative measure of the level of dynamical disturbance of the clusters, we used three methods, widely used in the literature to investigate the dynamics of cluster: the power ratios, $P_3/P_0$ \citep[e.g.][]{buotetsai95,jeltema05,ventimiglia08,bohringer10,cassano16,lovisari17}, the centroid shift, $w$ \citep[e.g.][]{mohr93,poole06,ohara06,ventimiglia08,maughan08,bohringer10,cassano16,lovisari17} and the concentration parameter, $c$ \citep[e.g.][]{santos08,parekh15, cassano16,rossetti17}. The power ratios represent the multipole decomposition of the mass distribution within the aperture radius. We only used the third moment, $P_3/P_0$, which is sensitive to the presence of substructures indicating ongoing dynamical activity \citep{bohringer10}. The centroid shift, $w$, is defined as the standard deviation of the projected separation between peak and the centroid of the X-ray emission. The concentration parameter is the ratio between the surface brightness inside the central region (100 kpc radius) and the `ambient' surface brightness (inside a radius of 500 kpc).

We refer to \citet{cassano10} (and references therein), for a detailed description of the morphological parameters. Here we just mention that, in general, relaxed clusters have high values of $c$ and low values of $w$ and $P_3/P_0$. Conversely, dynamically disturbed clusters have low values of $c$ and high values of $w$ and $P_3/P_0$.

\section{Dynamical properties}
\label{Sec:dynamics}

\longtab{
\begin{center}
\begin{small}
\begin{longtable}{l c c c c c}
\caption{Dynamical properties of clusters}\\
\label{Tab:dynamics}

name	  &	 c     &	 w 	   &  P3/P0	    & dynamics  & visual inspection		\\
          			&		   &  ($10^{-2}$)  &  ($10^{-7}$)   &   &	 		\\
\hline
\hline
A2163	  			&  0.116   &  5.970    &  14.850    &   M$^{1}$   &   M    	\\          
A2219	  			&  0.134   &  2.127    &  1.681     &   M$^{1}$   &   M    	\\         
A2744	  			&  0.101   &  2.637    &  10.500    &   M$^{1}$   &   M    	\\          
A1758a	  			&  0.109   &  8.217    &  2.515     &   M$^{1}$   &   M    	\\           
RXCJ2003.5-2323	  	&  0.062   &  1.824    &  4.602     &   M$^{1}$   &   M    	\\             
A1300     			&  0.191   &  4.442    &  6.847     &   M$^{1}$   &   M    	\\          
A773	  			&  0.184   &  2.403    &  1.445     &   M$^{1}$   &   M    	\\          
A209	  			&  0.176   &  1.321    &  0.518     &   M$^{1}$   &   M    	\\           
A520	  			&  0.097   &  10.050   &  5.259     &   M$^{1}$   &   M    	\\              
A521	  			&  0.108   &  2.204    &  5.090     &   M$^{1}$   &   M    	\\             
A697	  			&  0.153   &  0.731    &  1.668     &   M$^{1}$   &   M    	\\             
A1351     			&  0.083   &  4.272    &  3.506     &   M$\surd$   &   M    \\             
A665      			&  0.164   &  5.826    &  6.311     &   M$\surd$   &   M    \\             
A1689      			&  0.363   &  0.463    &  0.076     &   R$\surd$   &   M*	\\                 
A1914      			&  0.221   &  5.432    &  1.646     &   M$\surd$   &   M    \\         
A2142     			&  0.234   &  1.451    &  0.674     &   R$\surd$   &   int   \\            
A2218     			&  0.184   &  0.858    &  0.474     &   M$\surd$   &   int   \\              
A1443     			&  0.108   &  3.530    &  12.890    &   M$\surd$   &   M    \\          
A3411     			&  0.092   &  1.949    &  2.647     &   M$\surd$   &   M    \\           
RXC J1514.9-1523    &  0.064   &  1.301    &  1.411     &   M$\surd$   &   M    \\           
A2390	  			&  0.304   &  1.171    &  0.694     &   R$^{1}$   &   int      \\         
A1132 	  			&  0.111   &  3.386    &  3.059     &   M$\surd$   &   M    \\         
Zwcl 0634.1+4750    &  0.139   &  0.988    &  5.375     &   M$\surd$   &   M    \\            
Zwcl 0104.9+5350	&  0.088   &  5.693    &  0.604     &   M$\surd$   &   M    \\               
A3888	  			&  0.163   &  2.447    &  0.877     &   M$\surd$   &   M    \\           
A2261	  			&  0.334   &  0.494    &  1.026     &   R$^{1}$   &   int      \\          
PSZ1 G171.96-40.64  &  0.144   &  2.318    &  1.086     &   M$\surd$     &   M  \\         
PSZ1 G139.61+24.20  &  0.362   &  1.348    &  0.193     &   R$\surd$   &   int   \\           
RXC J1504.1-0248	&  0.624   &  0.459    &  0.147     &   R$^{1}$   &   R       \\            
A1835	  			&  0.486   &  0.996    &  0.458     &   R$\surd$   &   R    \\             
A478      			&  0.328   &  0.529    &  0.012     &   R$\surd$   &   R    \\              
A1413     			&  0.265   &  0.183    &  0.084     &   R$\surd$   &   int   \\              
S780	  			&  0.473   &  0.827    &  0.480     &   R$^{1}$   &   R       \\           
A2204     			&  0.537   &  0.125    &  0.022     &   R$\surd$   &   R    \\              
RXJ1720.1+2638  	&  0.489   &  0.279    &  0.117     &   R$\surd$   &   R    \\             
A3444	  			&  0.465   &  0.745    &  0.433     &   R$\surd$   &   R    \\             
A2667	  			&  0.406   &  0.926    &  1.395     &   R$^{1}$   &   int      \\         
A402      			&  0.323   &  1.249    &  1.350     &   R$\surd$   &   R    \\         
A2104 	  			&  0.123   &  2.198    &  2.082     &   M$\surd$   &   M    \\         
A1733	  			&  0.133   &  4.219    &  2.674     &   M$\surd$   &   M    \\         
A2355	  			&  0.075   &  4.879    &  7.495     &   M$\surd$   &   M    \\            
A2631     			&  0.121   &  1.574    &  1.550     &   M$^{1}$   &   M       \\         
A781      			&  0.111   &  6.374    &  3.143     &   M$^{1}$   &   M       \\         
RXC J0142.0+2131    &  0.186   &  0.738    &  6.625     &   M$^{2}$   &   int      \\             
A1423     			&  0.331   &  0.562    &  1.413     &   R$^{1}$   &   M       \\            
A2537     			&  0.278   &  0.561    &  0.351     &   R$^{1}$   &   M       \\              
A3088     			&  0.339   &  0.284    &  0.833     &   R$^{2}$   &   R       \\              
A1576     			&  0.235   &  1.271    &  5.950     &   R$^{2}$   &   M       \\           
A1763     			&  0.139   &  1.885    &  1.222     &   M$\surd$   &   M    \\           
A68	  	  			&  0.149   &  1.004    &  3.199     &   M$\surd$   &   M    \\           
A1437	  			&  0.085   &  7.450    &  9.505     &   M$\surd$   &   M    \\            
RXC J0616.3-2156	&  0.115   &  3.042    &  0.614     &   M$\surd$   &   M    \\             
A2895	  			&  0.161   &  4.271    &  4.851     &   M$\surd$   &   M    \\           
RXC J0510.7-0801  	&  0.134   &  2.346    &  2.171     &   M$\surd$   &   M    \\           
MACS J2135-010  	&  0.138   &  1.188    &  4.073     &   M$\surd$   &   M    \\           
A2813	  			&  0.172   &  0.311    &  1.230     &   R$\surd$   &   int   \\             
A115      			&  0.236   &  6.305    &  13.140    &   M$\surd$   &   M    \\             
A2345     			&  0.112   &  3.932    &  19.090    &   M$\surd$   &   M    \\            
A2552	  			&  0.218   &  0.639    &  0.222     &   R$\surd$   &   M    \\               
Zwcl 2120.1+2256    &  0.197   &  1.189    &  3.961     &   M$\surd$   &   int   \\           
Z5247	  			&  0.158   &  3.362    &  3.061     &   M$\surd$     &   M  \\           
A1682     			&  0.126   &  2.054    &  15.320    &   M$^{1}$   &   M       \\           
A3041     			&  0.099   &  3.342    &  13.620    &   M$\surd$     &   M  \\

\hline  \\

A56 	  			&  -	   &  -		   &  -  		&   -  	&   M       	\\			
A2697 	  			&  -	   &  -  	   &  -   		&   -   &   R 	 \\			
RXC J1314.4-2515	&  -       &  -  	   &  -   		&   -   &   M 	 \\			 
A1451	  			&  -	   &  -  	   &  -   		&   -   &   M 	 \\			 
RXC J2051.1+0216	&  -	   &  -  	   &  -   		&   -   &   M 	 \\			 
RXC J1322.8+3138 	&  -	   &  -  	   &  -   		&   -   &   int	 \\			
A384	  			&  -	   &  -  	   &  -   		&   -   &   R 	 \\			 
PSZ1 G019.12+3123	&  -	   &  -  	   &  -   		&   -   &   int	 \\			 
A2472 	  			&  -	   &  -  	   &  -   		&   -   &   int	 \\			
PSZ1 G205.07-6294   &  -	   &  -  	   &  -   		&   -   &   M 	 \\

\hline			         
\hline\\
\end{longtable}
\end{small}
\end{center}
\tablefoot{Upper panel: clusters with available X-ray \textit{Chandra} data. $\surd$ this work; $^1$ \citet{cassano10}; $^2$ \citet{cassano13}; $^3$ \citet{cassano16}; $^4$ \citet{cuciti15}; $^5$ \citet{bonafede15} * based on optical studies performed by \citet{andersson04}. Lower panel: clusters with available \textit{XMM-Newton} data. }
		         
}

\begin{figure}
\centering
\includegraphics[scale=0.4]{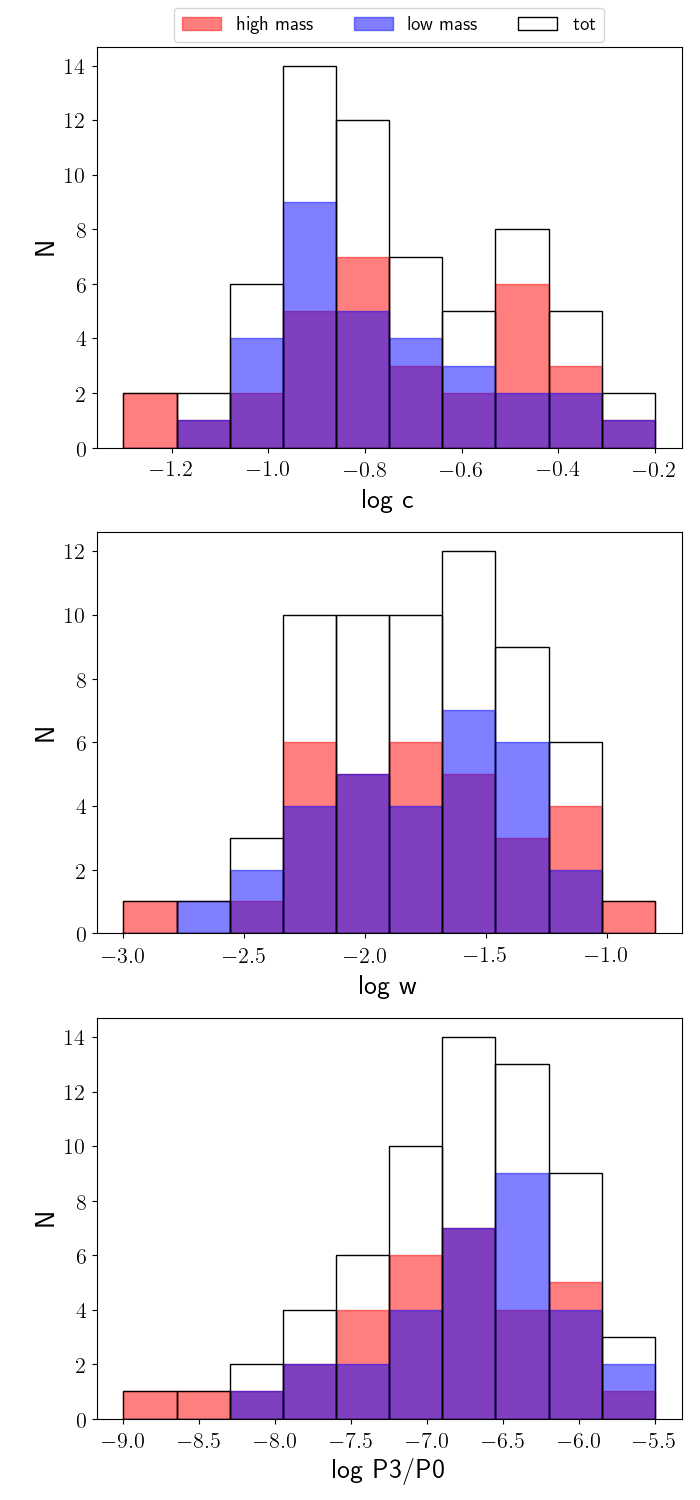}
\caption{Concentration parameter (\textit{top}), centroid shift (\textit{middle}) and power ratios (\textit{bottom}) distribution of the clusters with available X-ray \textit{Chandra} data. The red histogram refers to high mass clusters($M_{500}\geq7\times10^{14}M_\odot$), while the blue histogram refers to low mass ($M_{500}<7\times10^{14}M_\odot$) clusters.}
\label{Fig:histo_morphology}
\end{figure}

In this section we characterise the dynamical properties of the clusters of our sample on the basis of their X-ray emission. The distributions of the morphological parameters are shown in Fig \ref{Fig:histo_morphology}, both for all the clusters (black) and for two mass bins containing the same number of objects (the mass separating the two bins is $M_{500}=7\times10^{14}M_\odot$). We do not find a significant dependence of the morphological parameters on the cluster mass, in line with other recent results \citep{rossetti17,lovisari17}. 

\begin{figure}
\centering
\includegraphics[width=\columnwidth]{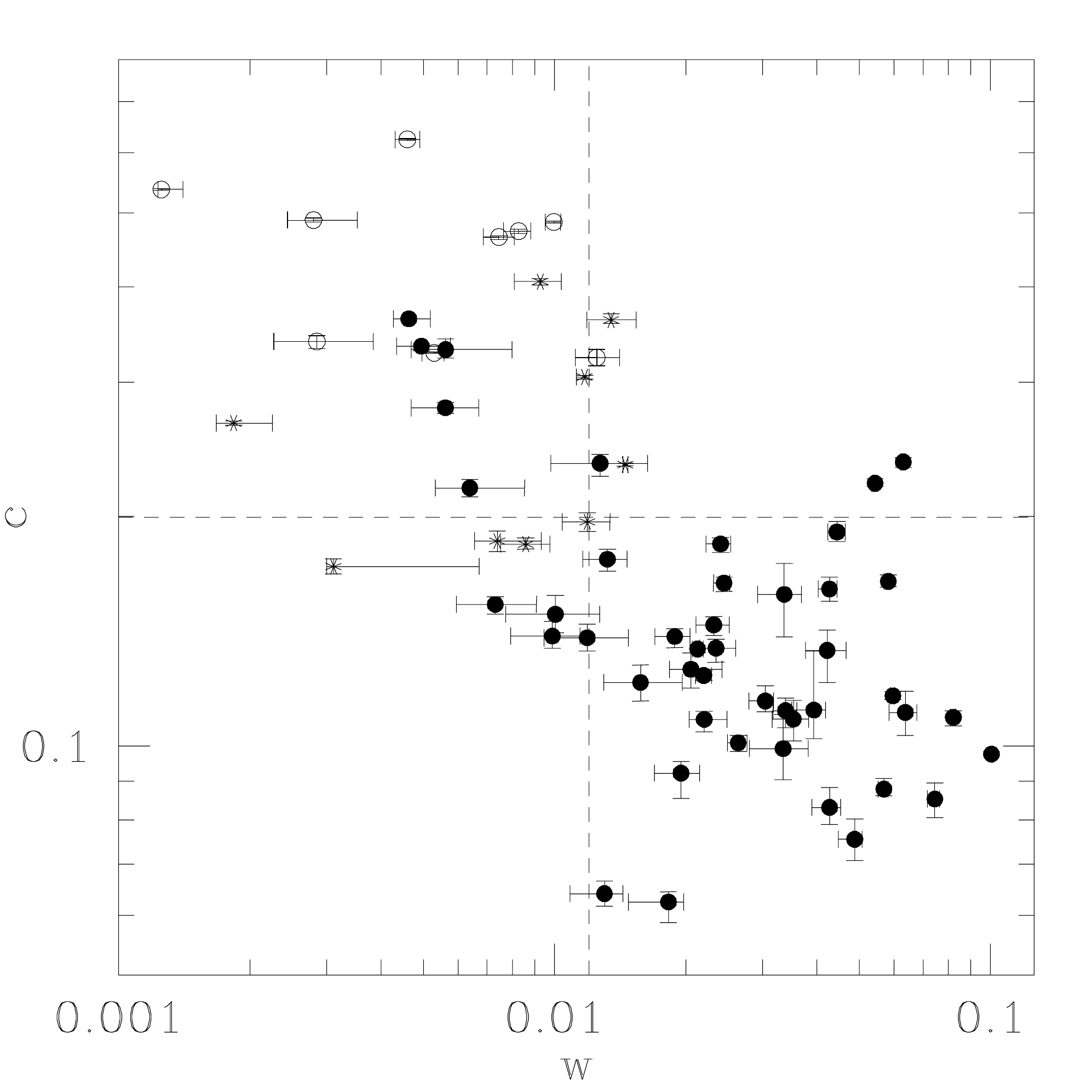}
\caption{$c-w$ morphological diagram. Filled dots are clusters classified as merging from the visual inspection, empty dots are visually classified as relaxed and asterisks are visually classified as intermediate. Black lines are adapted from \citet{cassano10} and are $c=0.2$ and $w=0.012$.}
\label{Fig:c-w}
\end{figure}

The most efficient way to characterise the dynamical properties of clusters is the combination of two morphological parameters, at least. In this Section we focus on the combination between $c$ and $w$, which has been shown to be a robust approach to distinguish between merger and relaxed clusters \citep{lovisari17}. Fig. \ref{Fig:c-w} shows the distribution of the 63 clusters of our sample with available X-ray \textit{Chandra} data in the $c-w$ morphological diagram. As expected, $c$ and $w$ are anti-correlated and the level of dynamical disturbance increases going from the top left to the bottom right corner of the diagram.

Defining a meaningful threshold between relaxed and merging clusters in the $c-w$ morphological diagram is not trivial since projection effects also play a role, in particular for unrelaxed clusters. \citet{lovisari17} has recently established the dynamical status of 150 galaxy clusters from the ESZ Planck catalogue \citep{planck11} by visually inspecting their \textit{XMM-Newton} images. The results of the visual inspection have been then combined with a number of morphological parameters, including $c$, $w$ and $P_3/P_0$, to determine the threshold values between merging and relaxed clusters. They found that among the clusters with a clear dynamical classification (excluding `intermediate' cases) $\sim40$\% are relaxed and $\sim60$\% are merging. Similar fractions have been obtained for mass-selected samples using cool core versus non-cool core clusters as indication of relaxed systems and non relaxed systems \citep{rossetti17,andrade-santos17}. Unfortunately, we cannot simply adopt the threshold found in \citet{lovisari17} because the parameters are measured within different apertures. Still, we note that the classification of merging and relaxed clusters based on the lines derived by \citet{cassano10} to separate radio halo and non-radio halo clusters provides similar results. Indeed $\sim$40\% of our clusters lie on the upper left panel of Fig. \ref{Fig:c-w}. This suggests that, although those lines were not derived to distinguish merging from relaxed clusters, they still give reasonable statistical information on the dynamical status of the clusters. Such classification of all the clusters of our sample, based on these lines is reported in Table \ref{Tab:dynamics}. 

In addition, following \citet{lovisari17}, we visually classified each cluster on the basis of the X-ray \textit{Chandra} image (Table \ref{Tab:dynamics}, column 6). In particular, we identified three classes of clusters marked with `M', `int' or `R' in Table \ref{Tab:dynamics}. `M' represents clusters where the distribution of the X-ray surface brightness and the presence of pronounced substructures clearly indicates merging activity, `R' represents relaxed clusters with circular X-ray morphology and peaked cores, while `int' indicates intermediate cases, where the morphology is fairly regular, sometimes with a peaked core, but with substructures or features in the X-ray distribution indicating a more complex situation. The two methods of classification are consistent for clusters lying at the opposite corners of the $c-w$ morphological diagram. Although the visual classification is subjective, we noted that, among fairly relaxed clusters, there are some systems showing minor disturbances that may not be caught by morphological parameters. These might be minor mergers, to which the cool core survives, but the X-ray distribution shows edges or irregularities. Alternatively, there may be signatures of disturbance on larger scales with respect to the radius adopted to measure the parameters, such as sub-clumps in the cluster outskirts.

Among the 12 clusters of our sample without available \textit{Chandra} data, ten have archival \textit{XMM-Newton} data. We did not derive morphological parameters for the clusters observed only with \textit{XMM-Newton}, to avoid introducing biases due to the different PSF and effective area compared to \textit{Chandra}. We classified the dynamical status of these clusters by visually inspecting their \textit{XMM-Newton} images and we report these classifications in the bottom panel of Table \ref{Tab:dynamics}. We could not infer the dynamical status of the two clusters without pointed X-ray observations.

\section{The full sample: Summary and conclusions}
\label{Sec:conclusions}
In this paper we built the largest mass-selected sample of galaxy clusters with deep radio observations available to date. It is made of 75 clusters with $z=0.08-0.33$ and $M_{500}\gtrsim6\times10^{14} \,M_\odot$ selected from the Planck SZ catalogue \citep{planck14}. Thanks to the radio data analysis described in Sections \ref{Sec:radio}, now each cluster of the sample has at least one pointed observation from which we have information about the possible presence of radio diffuse emission. 

Beyond the statistical value of this work, the large amount of radio data analysed led to the discovery of new radio halos, mini halos and candidate diffuse sources in clusters (Section \ref{Sec:results}). In particular, A1451 and Z0634 host radio halos \citep{cuciti18} and PSZG139 host a mini halo surrounded by steep spectrum large scale emission \citep{savini18}. We found candidate radio halos in Z2120, A3041 and a conservatively claimed a candidate mini halo in A402. Furthermore, we followed up the known radio halos in A3888 and A1443 at different frequencies. 

For clusters without any hint of diffuse emission in our observations we used the injection technique to derive upper limits to their diffuse flux (Section \ref{Sec:UL}). 

The great majority of the clusters of this sample have X-ray \textit{Chandra} data. We analysed those data to derive information about the dynamical state of clusters (Section \ref{Sec:X-ray analysis}). According to the morphological classification based on the combination of centroid shift and concentration parameter, $\sim 40$\% of the clusters of our sample are relaxed and $\sim 60$\% are merging, in line with other recent results \citep{lovisari17, rossetti17, andrade-santos17}. In addition, we visually inspected all the \textit{Chandra} images and the XMM-newton images of clusters without \textit{Chandra} data to give an independent classification of the dynamical status of clusters.

The properties (coordinates, redshift, $M_{500}$, $R_{500}$, radio classification and radio power at 1.4 GHz) of the clusters of the sample are listed in Table \ref{tab:completesample}. Radio powers in Table \ref{tab:completesample} are calculated at 1.4 GHz and are $k$-corrected. When $\alpha$ is unknown, we assume $\alpha=-1.3$. 



We summarise below the diffuse sources present in our sample (references are given in Table \ref{tab:completesample}):
\begin{itemize}
\item 28 ($\sim$37\%) clusters host radio halos, ten of which are USSRHs or candidate USSRHs
\item seven ($\sim$10\%) clusters have radio relics, five of which also have radio halos (and have been also counted among radio halos above), and two of which are without radio halos
\item 11 ($\sim$15\%) clusters have mini halos \item 31 ($\sim$41\%) clusters do not show any hint of central diffuse emission at the sensitivity of current observations \footnote{The two cluster with radio relics but without radio halos (A115 and A2345) are among these 31 clusters.}
\end{itemize}

Moreover, we found candidate diffuse emission in six clusters, one is a candidate mini halo and five are candidate radio halos. Combining our results (Section \ref{Sec:UL}) with the work done by \citet{venturi08}, \citet{kale13} and \citet{kale15} for the GMRT Radio Halo Survey and \citet{bonafede17} we have upper limits for 22 clusters. Being obtained with similar techniques (Section \ref{Sec:UL}), the upper limits from the literature are comparable with those derived in this work. The statistical analysis of the properties of radio halos in this sample, including the radio power-mass diagram, the radio emissivity-mass diagram, the radio halo-merger connection, and the occurrence of radio halos will be presented in paper II.

\begin{acknowledgements}
The authors thank the anonymous referee for the comments that improved the presentation of the paper. VC acknowledges support from the Alexander von Humboldt Foundation. AB acknowledges support from the ERC through the grant ERC-Stg Dranoel n. 714245 and from the MIUR FARE grant SMS. RJvW acknowledges support from the VIDI research programme with project number 639.042.729, which is financed by the Netherlands Organisation for Scientific Research (NWO). RK acknowledges the support of the Department of Atomic Energy, Government of India, under project no. 12-R\&D-TFR-5.02-0700. Basic research in radio astronomy at the Naval Research Laboratory is supported by 6.1 Base funding. SE acknowledges financial contribution from the contracts ASI-INAF Athena 2019-27-HH.0,
``Attivit\`a di Studio per la comunit\`a scientifica di Astrofisica delle Alte Energie e Fisica Astroparticellare''
(Accordo Attuativo ASI-INAF n. 2017-14-H.0), INAF mainstream project 1.05.01.86.10, and
from the European Union’s Horizon 2020 Programme under the AHEAD2020 project (grant agreement n. 871158). GWP acknowledges the support of the French space agency, CNES.
The National Radio Astronomy Observatory is a facility of the National Science Foundation operated under cooperative agreement by Associated Universities, Inc. We thank the staff of the GMRT that made these observations possible. GMRT is run by the National Centre for Radio Astrophysics of the Tata Institute of Fundamental Research. The scientific results reported in this article are based in part on data obtained from the \textit{Chandra} Data Archive. This research has made use of the NASA/IPAC Extragalactic Database (NED) which is operated by the Jet Propulsion Laboratory, California Institute of Technology, under contract with the National Aeronautics and Space Administration.
\end{acknowledgements}

\bibliographystyle{aa} 
\bibliography{biblio_virgi} 

\begin{thebibliography}{119}
\expandafter\ifx\csname natexlab\endcsname\relax\def\natexlab#1{#1}\fi

\bibitem[{{Andernach} {et~al.}(1986){Andernach}, {Sievers}, {Kus}, \&
  {Schnaubelt}}]{andernach86}
{Andernach}, H., {Sievers}, A., {Kus}, A., \& {Schnaubelt}, J. 1986, \aaps, 65,
  561

\bibitem[{{Andersson} \& {Madejski}(2004)}]{andersson04}
{Andersson}, K.~E. \& {Madejski}, G.~M. 2004, \apj, 607, 190

\bibitem[{{Andrade-Santos} {et~al.}(2017){Andrade-Santos}, {Jones}, {Forman},
  {Lovisari}, {Vikhlinin}, {van Weeren}, {Murray}, {Arnaud}, {Pratt},
  {D{\'e}mocl{\`e}s}, {Kraft}, {Mazzotta}, {B{\"o}hringer}, {Chon},
  {Giacintucci}, {Clarke}, {Borgani}, {David}, {Douspis}, {Pointecouteau},
  {Dahle}, {Brown}, {Aghanim}, \& {Rasia}}]{andrade-santos17}
{Andrade-Santos}, F., {Jones}, C., {Forman}, W.~R., {et~al.} 2017, \apj, 843,
  76

\bibitem[{{Bacchi} {et~al.}(2003){Bacchi}, {Feretti}, {Giovannini}, \&
  {Govoni}}]{bacchi03}
{Bacchi}, M., {Feretti}, L., {Giovannini}, G., \& {Govoni}, F. 2003, \aap, 400,
  465

\bibitem[{{Basu}(2012)}]{basu12}
{Basu}, K. 2012, \mnras, 421, L112

\bibitem[{{B{\"o}hringer} {et~al.}(2010){B{\"o}hringer}, {Pratt}, {Arnaud},
  {Borgani}, {Croston}, {Ponman}, {Ameglio}, {Temple}, \&
  {Dolag}}]{bohringer10}
{B{\"o}hringer}, H., {Pratt}, G.~W., {Arnaud}, M., {et~al.} 2010, \aap, 514,
  A32

\bibitem[{{Bonafede} {et~al.}(2017){Bonafede}, {Cassano}, {Br{\"u}ggen},
  {Ogrean}, {Riseley}, {Cuciti}, {de Gasperin}, {Golovich}, {Kale}, {Venturi},
  {van Weeren}, {Wik}, \& {Wittman}}]{bonafede17}
{Bonafede}, A., {Cassano}, R., {Br{\"u}ggen}, M., {et~al.} 2017, \mnras, 470,
  3465

\bibitem[{{Bonafede} {et~al.}(2015){Bonafede}, {Intema}, {Br{\"u}ggen},
  {Vazza}, {Basu}, {Sommer}, {Ebeling}, {de Gasperin}, {R{\"o}ttgering}, {van
  Weeren}, \& {Cassano}}]{bonafede15}
{Bonafede}, A., {Intema}, H., {Br{\"u}ggen}, M., {et~al.} 2015, \mnras, 454,
  3391

\bibitem[{{Botteon} {et~al.}(2018){Botteon}, {Shimwell}, {Bonafede},
  {Dallacasa}, {Brunetti}, {Mandal}, {van Weeren}, {Br{\"u}ggen}, {Cassano},
  {de Gasperin}, {Hoang}, {Hoeft}, {R{\"o}ttgering}, {Savini}, {White},
  {Wilber}, \& {Venturi}}]{botteon18}
{Botteon}, A., {Shimwell}, T.~W., {Bonafede}, A., {et~al.} 2018, \mnras, 478,
  885

\bibitem[{{Briggs}(1995)}]{briggs95}
{Briggs}, D.~S. 1995, in American Astronomical Society Meeting Abstracts, Vol.
  187, 112.02

\bibitem[{{Brunetti} {et~al.}(2009){Brunetti}, {Cassano}, {Dolag}, \&
  {Setti}}]{brunetti09}
{Brunetti}, G., {Cassano}, R., {Dolag}, K., \& {Setti}, G. 2009, \aap, 507, 661

\bibitem[{{Brunetti} {et~al.}(2008){Brunetti}, {Giacintucci}, {Cassano},
  {Lane}, {Dallacasa}, {Venturi}, {Kassim}, {Setti}, {Cotton}, \&
  {Markevitch}}]{brunetti08nature}
{Brunetti}, G., {Giacintucci}, S., {Cassano}, R., {et~al.} 2008, \nat, 455, 944

\bibitem[{{Brunetti} \& {Jones}(2014)}]{brunettijones14}
{Brunetti}, G. \& {Jones}, T.~W. 2014, International Journal of Modern Physics
  D, 23, 1430007

\bibitem[{{Brunetti} \& {Lazarian}(2007)}]{brunettilazarian07}
{Brunetti}, G. \& {Lazarian}, A. 2007, \mnras, 378, 245

\bibitem[{{Brunetti} \& {Lazarian}(2011)}]{brunettilazarian11}
{Brunetti}, G. \& {Lazarian}, A. 2011, \mnras, 410, 127

\bibitem[{{Brunetti} \& {Lazarian}(2016)}]{brunettilazarian16}
{Brunetti}, G. \& {Lazarian}, A. 2016, \mnras, 458, 2584

\bibitem[{{Brunetti} {et~al.}(2001){Brunetti}, {Setti}, {Feretti}, \&
  {Giovannini}}]{brunetti01}
{Brunetti}, G., {Setti}, G., {Feretti}, L., \& {Giovannini}, G. 2001, \mnras,
  320, 365

\bibitem[{{Brunetti} {et~al.}(2007){Brunetti}, {Venturi}, {Dallacasa},
  {Cassano}, {Dolag}, {Giacintucci}, \& {Setti}}]{brunetti07}
{Brunetti}, G., {Venturi}, T., {Dallacasa}, D., {et~al.} 2007, \apjl, 670, L5

\bibitem[{{Buote} \& {Tsai}(1995)}]{buotetsai95}
{Buote}, D.~A. \& {Tsai}, J.~C. 1995, \apj, 452, 522

\bibitem[{{Cassano} {et~al.}(2016){Cassano}, {Brunetti}, {Giocoli}, \&
  {Ettori}}]{cassano16}
{Cassano}, R., {Brunetti}, G., {Giocoli}, C., \& {Ettori}, S. 2016, \aap, 593,
  A81

\bibitem[{{Cassano} {et~al.}(2007){Cassano}, {Brunetti}, {Setti}, {Govoni}, \&
  {Dolag}}]{cassano07}
{Cassano}, R., {Brunetti}, G., {Setti}, G., {Govoni}, F., \& {Dolag}, K. 2007,
  \mnras, 378, 1565

\bibitem[{{Cassano} {et~al.}(2013){Cassano}, {Ettori}, {Brunetti},
  {Giacintucci}, {Pratt}, {Venturi}, {Kale}, {Dolag}, \&
  {Markevitch}}]{cassano13}
{Cassano}, R., {Ettori}, S., {Brunetti}, G., {et~al.} 2013, \apj, 777, 141

\bibitem[{{Cassano} {et~al.}(2010){Cassano}, {Ettori}, {Giacintucci},
  {Brunetti}, {Markevitch}, {Venturi}, \& {Gitti}}]{cassano10}
{Cassano}, R., {Ettori}, S., {Giacintucci}, S., {et~al.} 2010, \apjl, 721, L82

\bibitem[{{Cavagnolo} {et~al.}(2008){Cavagnolo}, {Donahue}, {Voit}, \&
  {Sun}}]{cavagnolo08}
{Cavagnolo}, K.~W., {Donahue}, M., {Voit}, G.~M., \& {Sun}, M. 2008, \apj, 682,
  821

\bibitem[{{Cavagnolo} {et~al.}(2009){Cavagnolo}, {Donahue}, {Voit}, \&
  {Sun}}]{cavagnolo09}
{Cavagnolo}, K.~W., {Donahue}, M., {Voit}, G.~M., \& {Sun}, M. 2009, \apjs,
  182, 12

\bibitem[{{Chandra} {et~al.}(2004){Chandra}, {Ray}, \& {Bhatnagar}}]{chandra04}
{Chandra}, P., {Ray}, A., \& {Bhatnagar}, S. 2004, \apj, 612, 974

\bibitem[{{Chon} {et~al.}(2012){Chon}, {B{\"o}hringer}, \& {Smith}}]{chon12}
{Chon}, G., {B{\"o}hringer}, H., \& {Smith}, G.~P. 2012, \aap, 548, A59

\bibitem[{{Colless} {et~al.}(2003){Colless}, {Peterson}, {Jackson}, {Peacock},
  {Cole}, {Norberg}, {Baldry}, {Baugh}, {Bland-Hawthorn}, {Bridges}, {Cannon},
  {Collins}, {Couch}, {Cross}, {Dalton}, {De Propris}, {Driver}, {Efstathiou},
  {Ellis}, {Frenk}, {Glazebrook}, {Lahav}, {Lewis}, {Lumsden}, {Maddox},
  {Madgwick}, {Sutherland}, \& {Taylor}}]{colless03}
{Colless}, M., {Peterson}, B.~A., {Jackson}, C., {et~al.} 2003, ArXiv
  Astrophysics e-prints [\eprint{astro-ph/0306581}]

\bibitem[{{Condon} {et~al.}(1998){Condon}, {Cotton}, {Greisen}, {Yin},
  {Perley}, {Taylor}, \& {Broderick}}]{condon98}
{Condon}, J.~J., {Cotton}, W.~D., {Greisen}, E.~W., {et~al.} 1998, \aj, 115,
  1693

\bibitem[{{Cornwell} {et~al.}(2005){Cornwell}, {Golap}, \&
  {Bhatnagar}}]{cornwell05}
{Cornwell}, T.~J., {Golap}, K., \& {Bhatnagar}, S. 2005, in Astronomical
  Society of the Pacific Conference Series, Vol. 347, Astronomical Data
  Analysis Software and Systems XIV, ed. P.~{Shopbell}, M.~{Britton}, \&
  R.~{Ebert}, 86

\bibitem[{{Cornwell} {et~al.}(2008){Cornwell}, {Golap}, \&
  {Bhatnagar}}]{cornwell08}
{Cornwell}, T.~J., {Golap}, K., \& {Bhatnagar}, S. 2008, IEEE Journal of
  Selected Topics in Signal Processing, 2, 647

\bibitem[{{Cuciti} {et~al.}(2018){Cuciti}, {Brunetti}, {van Weeren},
  {Bonafede}, {Dallacasa}, {Cassano}, {Venturi}, \& {Kale}}]{cuciti18}
{Cuciti}, V., {Brunetti}, G., {van Weeren}, R., {et~al.} 2018, \aap, 609, A61

\bibitem[{{Cuciti} {et~al.}(2015){Cuciti}, {Cassano}, {Brunetti}, {Dallacasa},
  {Kale}, {Ettori}, \& {Venturi}}]{cuciti15}
{Cuciti}, V., {Cassano}, R., {Brunetti}, G., {et~al.} 2015, \aap, 580, A97

\bibitem[{{Dallacasa} {et~al.}(2009){Dallacasa}, {Brunetti}, {Giacintucci},
  {Cassano}, {Venturi}, {Macario}, {Kassim}, {Lane}, \& {Setti}}]{dallacasa09}
{Dallacasa}, D., {Brunetti}, G., {Giacintucci}, S., {et~al.} 2009, \apj, 699,
  1288

\bibitem[{{Ensslin} {et~al.}(1998){Ensslin}, {Biermann}, {Klein}, \&
  {Kohle}}]{ensslin98}
{Ensslin}, T.~A., {Biermann}, P.~L., {Klein}, U., \& {Kohle}, S. 1998, \aap,
  332, 395

\bibitem[{{Farnsworth} {et~al.}(2013){Farnsworth}, {Rudnick}, {Brown}, \&
  {Brunetti}}]{farnsworth13}
{Farnsworth}, D., {Rudnick}, L., {Brown}, S., \& {Brunetti}, G. 2013, \apj,
  779, 189

\bibitem[{{Feretti} {et~al.}(2001){Feretti}, {Fusco-Femiano}, {Giovannini}, \&
  {Govoni}}]{feretti01}
{Feretti}, L., {Fusco-Femiano}, R., {Giovannini}, G., \& {Govoni}, F. 2001,
  \aap, 373, 106

\bibitem[{{Giacintucci} {et~al.}(2011{\natexlab{a}}){Giacintucci}, {Dallacasa},
  {Venturi}, {Brunetti}, {Cassano}, {Markevitch}, \& {Athreya}}]{giacintucci11}
{Giacintucci}, S., {Dallacasa}, D., {Venturi}, T., {et~al.} 2011{\natexlab{a}},
  \aap, 534, A57

\bibitem[{{Giacintucci} {et~al.}(2013){Giacintucci}, {Kale}, {Wik}, {Venturi},
  \& {Markevitch}}]{giacintucci13}
{Giacintucci}, S., {Kale}, R., {Wik}, D.~R., {Venturi}, T., \& {Markevitch}, M.
  2013, \apj, 766, 18

\bibitem[{{Giacintucci} {et~al.}(2011{\natexlab{b}}){Giacintucci},
  {Markevitch}, {Brunetti}, {Cassano}, \& {Venturi}}]{giacintucci11_1504}
{Giacintucci}, S., {Markevitch}, M., {Brunetti}, G., {Cassano}, R., \&
  {Venturi}, T. 2011{\natexlab{b}}, \aap, 525, L10

\bibitem[{{Giacintucci} {et~al.}(2014{\natexlab{a}}){Giacintucci},
  {Markevitch}, {Brunetti}, {ZuHone}, {Venturi}, {Mazzotta}, \&
  {Bourdin}}]{giacintucci14_1720}
{Giacintucci}, S., {Markevitch}, M., {Brunetti}, G., {et~al.}
  2014{\natexlab{a}}, \apj, 795, 73

\bibitem[{{Giacintucci} {et~al.}(2017){Giacintucci}, {Markevitch}, {Cassano},
  {Venturi}, {Clarke}, \& {Brunetti}}]{giacintucci17}
{Giacintucci}, S., {Markevitch}, M., {Cassano}, R., {et~al.} 2017, \apj, 841,
  71

\bibitem[{{Giacintucci} {et~al.}(2019){Giacintucci}, {Markevitch}, {Cassano},
  {Venturi}, {Clarke}, {Kale}, \& {Cuciti}}]{giacintucci19}
{Giacintucci}, S., {Markevitch}, M., {Cassano}, R., {et~al.} 2019, \apj, 880,
  70

\bibitem[{{Giacintucci} {et~al.}(2014{\natexlab{b}}){Giacintucci},
  {Markevitch}, {Venturi}, {Clarke}, {Cassano}, \& {Mazzotta}}]{giacintucci14}
{Giacintucci}, S., {Markevitch}, M., {Venturi}, T., {et~al.}
  2014{\natexlab{b}}, \apj, 781, 9

\bibitem[{{Giacintucci} \& {Venturi}(2009)}]{giacintucci09}
{Giacintucci}, S. \& {Venturi}, T. 2009, \aap, 505, 55

\bibitem[{{Giacintucci} {et~al.}(2008){Giacintucci}, {Venturi}, {Macario},
  {Dallacasa}, {Brunetti}, {Markevitch}, {Cassano}, {Bardelli}, \&
  {Athreya}}]{giacintucci08}
{Giacintucci}, S., {Venturi}, T., {Macario}, G., {et~al.} 2008, \aap, 486, 347

\bibitem[{{Giovannini} {et~al.}(2020){Giovannini}, {Cau}, {Bonafede},
  {Ebeling}, {Feretti}, {Girardi}, {Gitti}, {Govoni}, {Ignesti}, {Murgia},
  {Taylor}, \& {Vacca}}]{giovannini20}
{Giovannini}, G., {Cau}, M., {Bonafede}, A., {et~al.} 2020, arXiv e-prints,
  arXiv:2006.08494

\bibitem[{{Giovannini} \& {Feretti}(2000)}]{giovannini00}
{Giovannini}, G. \& {Feretti}, L. 2000, \na, 5, 335

\bibitem[{{Giovannini} {et~al.}(2006){Giovannini}, {Feretti}, {Govoni},
  {Murgia}, \& {Pizzo}}]{giovannini06}
{Giovannini}, G., {Feretti}, L., {Govoni}, F., {Murgia}, M., \& {Pizzo}, R.
  2006, Astronomische Nachrichten, 327, 563

\bibitem[{{Giovannini} {et~al.}(1999){Giovannini}, {Tordi}, \&
  {Feretti}}]{giovannini99}
{Giovannini}, G., {Tordi}, M., \& {Feretti}, L. 1999, \na, 4, 141

\bibitem[{{Gitti} {et~al.}(2002){Gitti}, {Brunetti}, \& {Setti}}]{gitti02}
{Gitti}, M., {Brunetti}, G., \& {Setti}, G. 2002, \aap, 386, 456

\bibitem[{{Govoni} {et~al.}(2001){Govoni}, {Feretti}, {Giovannini},
  {B{\"o}hringer}, {Reiprich}, \& {Murgia}}]{govoni01}
{Govoni}, F., {Feretti}, L., {Giovannini}, G., {et~al.} 2001, \aap, 376, 803

\bibitem[{{Govoni} {et~al.}(2009){Govoni}, {Murgia}, {Markevitch}, {Feretti},
  {Giovannini}, {Taylor}, \& {Carretti}}]{govoni09}
{Govoni}, F., {Murgia}, M., {Markevitch}, M., {et~al.} 2009, \aap, 499, 371

\bibitem[{{Haarsma} {et~al.}(2010){Haarsma}, {Leisman}, {Donahue}, {Bruch},
  {B{\"o}hringer}, {Croston}, {Pratt}, {Voit}, {Arnaud}, \&
  {Pierini}}]{haarsma10}
{Haarsma}, D.~B., {Leisman}, L., {Donahue}, M., {et~al.} 2010, \apj, 713, 1037

\bibitem[{{Hanisch}(1982)}]{hanisch82}
{Hanisch}, R.~J. 1982, \aap, 111, 97

\bibitem[{{Intema}(2014)}]{intema14}
{Intema}, H.~T. 2014, {SPAM: Source Peeling and Atmospheric Modeling},
  Astrophysics Source Code Library

\bibitem[{{Intema} {et~al.}(2017){Intema}, {Jagannathan}, {Mooley}, \&
  {Frail}}]{intema17}
{Intema}, H.~T., {Jagannathan}, P., {Mooley}, K.~P., \& {Frail}, D.~A. 2017,
  \aap, 598, A78

\bibitem[{{Intema} {et~al.}(2009){Intema}, {van der Tol}, {Cotton}, {Cohen},
  {van Bemmel}, \& {R{\"o}ttgering}}]{intema09}
{Intema}, H.~T., {van der Tol}, S., {Cotton}, W.~D., {et~al.} 2009, \aap, 501,
  1185

\bibitem[{{Jacob} \& {Pfrommer}(2017)}]{jacob17}
{Jacob}, S. \& {Pfrommer}, C. 2017, \mnras, 467, 1478

\bibitem[{{Jeltema} {et~al.}(2005){Jeltema}, {Canizares}, {Bautz}, \&
  {Buote}}]{jeltema05}
{Jeltema}, T.~E., {Canizares}, C.~R., {Bautz}, M.~W., \& {Buote}, D.~A. 2005,
  \apj, 624, 606

\bibitem[{{Johnston-Hollitt} \& {Pratley}(2017)}]{johnston-hollitt17}
{Johnston-Hollitt}, M. \& {Pratley}, L. 2017, ArXiv e-prints
  [\eprint{1706.04930}]

\bibitem[{{Kale} {et~al.}(2015){Kale}, {Venturi}, {Giacintucci}, {Dallacasa},
  {Cassano}, {Brunetti}, {Cuciti}, {Macario}, \& {Athreya}}]{kale15}
{Kale}, R., {Venturi}, T., {Giacintucci}, S., {et~al.} 2015, \aap, 579, A92

\bibitem[{{Kale} {et~al.}(2013){Kale}, {Venturi}, {Giacintucci}, {Dallacasa},
  {Cassano}, {Brunetti}, {Macario}, \& {Athreya}}]{kale13}
{Kale}, R., {Venturi}, T., {Giacintucci}, S., {et~al.} 2013, \aap, 557, A99

\bibitem[{{Kang} {et~al.}(2012){Kang}, {Ryu}, \& {Jones}}]{kang12}
{Kang}, H., {Ryu}, D., \& {Jones}, T.~W. 2012, \apj, 756, 97

\bibitem[{{Kempner} \& {Sarazin}(2001)}]{kempner01}
{Kempner}, J.~C. \& {Sarazin}, C.~L. 2001, \apj, 548, 639

\bibitem[{{Knowles} {et~al.}(2017){Knowles}, {Baker}, {Basu}, {Bharadwaj},
  {Deane}, {Devlin}, {Dicker}, {de Gasperin}, {Ferrari}, {Hilton}, {Hughes},
  {Intema}, {Makhathini}, {Moodley}, {Oozeer}, {Pfrommer}, {Sievers},
  {Sikhosana}, {Smirnov}, {Sommer}, {Stanchfield}, {van der Heyden}, \&
  {Zwart}}]{knowles17}
{Knowles}, K., {Baker}, A., {Basu}, K., {et~al.} 2017, arXiv e-prints,
  arXiv:1709.03318

\bibitem[{{Knowles} {et~al.}(2019){Knowles}, {Baker}, {Bond}, {Gallardo},
  {Gupta}, {Hilton}, {Hughes}, {Intema}, {L{\'o}pez-Caraballo}, {Moodley},
  {Schmitt}, {Sievers}, {Sif{\'o}n}, \& {Wollack}}]{knowles19}
{Knowles}, K., {Baker}, A.~J., {Bond}, J.~R., {et~al.} 2019, \mnras, 486, 1332

\bibitem[{{Liang} {et~al.}(2000){Liang}, {Hunstead}, {Birkinshaw}, \&
  {Andreani}}]{liang00}
{Liang}, H., {Hunstead}, R.~W., {Birkinshaw}, M., \& {Andreani}, P. 2000, \apj,
  544, 686

\bibitem[{{Lovisari} {et~al.}(2017){Lovisari}, {Forman}, {Jones}, {Ettori},
  {Andrade-Santos}, {Arnaud}, {D{\'e}mocl{\`e}s}, {Pratt}, {Randall}, \&
  {Kraft}}]{lovisari17}
{Lovisari}, L., {Forman}, W.~R., {Jones}, C., {et~al.} 2017, \apj, 846, 51

\bibitem[{{Macario} {et~al.}(2010){Macario}, {Venturi}, {Brunetti},
  {Dallacasa}, {Giacintucci}, {Cassano}, {Bardelli}, \& {Athreya}}]{macario10}
{Macario}, G., {Venturi}, T., {Brunetti}, G., {et~al.} 2010, \aap, 517, A43

\bibitem[{{Mandal} {et~al.}(2019){Mandal}, {Intema}, {Shimwell}, {van Weeren},
  {Botteon}, {R{\"o}ttgering}, {Hoang}, {Brunetti}, {de Gasperin},
  {Giacintucci}, {Hoekstra}, {Stroe}, {Br{\"u}ggen}, {Cassano}, {Shulevski},
  {Drabent}, \& {Rafferty}}]{mandal19}
{Mandal}, S., {Intema}, H.~T., {Shimwell}, T.~W., {et~al.} 2019, \aap, 622, A22

\bibitem[{{Markevitch} {et~al.}(2005){Markevitch}, {Govoni}, {Brunetti}, \&
  {Jerius}}]{markevitch05}
{Markevitch}, M., {Govoni}, F., {Brunetti}, G., \& {Jerius}, D. 2005, \apj,
  627, 733

\bibitem[{{Maughan} {et~al.}(2008){Maughan}, {Jones}, {Pierre}, {Andreon},
  {Birkinshaw}, {Bremer}, {Pacaud}, {Ponman}, {Valtchanov}, \&
  {Willis}}]{maughan08}
{Maughan}, B.~J., {Jones}, L.~R., {Pierre}, M., {et~al.} 2008, \mnras, 387, 998

\bibitem[{{Mazzotta} \& {Giacintucci}(2008)}]{mazzotta08}
{Mazzotta}, P. \& {Giacintucci}, S. 2008, \apjl, 675, L9

\bibitem[{{Mohan} \& {Rafferty}(2015)}]{pybdsm}
{Mohan}, N. \& {Rafferty}, D. 2015, {PyBDSM: Python Blob Detection and Source
  Measurement}, Astrophysics Source Code Library

\bibitem[{{Mohr} {et~al.}(1993){Mohr}, {Fabricant}, \& {Geller}}]{mohr93}
{Mohr}, J.~J., {Fabricant}, D.~G., \& {Geller}, M.~J. 1993, \apj, 413, 492

\bibitem[{{Motl} {et~al.}(2005){Motl}, {Hallman}, {Burns}, \&
  {Norman}}]{motl05}
{Motl}, P.~M., {Hallman}, E.~J., {Burns}, J.~O., \& {Norman}, M.~L. 2005,
  \apjl, 623, L63

\bibitem[{{Murgia} {et~al.}(2009){Murgia}, {Govoni}, {Markevitch}, {Feretti},
  {Giovannini}, {Taylor}, \& {Carretti}}]{murgia09}
{Murgia}, M., {Govoni}, F., {Markevitch}, M., {et~al.} 2009, \aap, 499, 679

\bibitem[{{Nagai}(2006)}]{nagai06}
{Nagai}, D. 2006, \apj, 650, 538

\bibitem[{{O'Hara} {et~al.}(2006){O'Hara}, {Mohr}, {Bialek}, \&
  {Evrard}}]{ohara06}
{O'Hara}, T.~B., {Mohr}, J.~J., {Bialek}, J.~J., \& {Evrard}, A.~E. 2006, \apj,
  639, 64

\bibitem[{{Orr{\'u}} {et~al.}(2007){Orr{\'u}}, {Murgia}, {Feretti}, {Govoni},
  {Brunetti}, {Giovannini}, {Girardi}, \& {Setti}}]{orru07}
{Orr{\'u}}, E., {Murgia}, M., {Feretti}, L., {et~al.} 2007, \aap, 467, 943

\bibitem[{{Parekh} {et~al.}(2015){Parekh}, {van der Heyden}, {Ferrari},
  {Angus}, \& {Holwerda}}]{parekh15}
{Parekh}, V., {van der Heyden}, K., {Ferrari}, C., {Angus}, G., \& {Holwerda},
  B. 2015, \aap, 575, A127

\bibitem[{{Perley} \& {Butler}(2013)}]{perleybutler13}
{Perley}, R.~A. \& {Butler}, B.~J. 2013, \apjs, 204, 19

\bibitem[{{Petrosian}(2001)}]{petrosian01}
{Petrosian}, V. 2001, \apj, 557, 560

\bibitem[{{Pfrommer} \& {En{\ss}lin}(2004)}]{pfrommer04}
{Pfrommer}, C. \& {En{\ss}lin}, T.~A. 2004, \aap, 413, 17

\bibitem[{{Pinzke} {et~al.}(2013){Pinzke}, {Oh}, \& {Pfrommer}}]{pinzke13}
{Pinzke}, A., {Oh}, S.~P., \& {Pfrommer}, C. 2013, \mnras, 435, 1061

\bibitem[{{Pinzke} {et~al.}(2017){Pinzke}, {Oh}, \& {Pfrommer}}]{pinzke17}
{Pinzke}, A., {Oh}, S.~P., \& {Pfrommer}, C. 2017, \mnras, 465, 4800

\bibitem[{{Planck Collaboration} {et~al.}(2014){Planck Collaboration}, {Ade},
  {Aghanim}, {Armitage-Caplan}, {Arnaud}, {Ashdown}, {Atrio-Barandela},
  {Aumont}, {Aussel}, {Baccigalupi}, \& et~al.}]{planck14}
{Planck Collaboration}, {Ade}, P.~A.~R., {Aghanim}, N., {et~al.} 2014, \aap,
  571, A29

\bibitem[{{Planck Collaboration} {et~al.}(2011){Planck Collaboration},
  {Aghanim}, {Arnaud}, {Ashdown}, {Aumont}, {Baccigalupi}, {Balbi}, {Banday},
  {Barreiro}, {Bartelmann}, {Bartlett}, {Battaner}, {Benabed}, {Beno{\^i}t},
  {Bernard}, {Bersanelli}, {Bhatia}, {Bock}, {Bonaldi}, {Bond}, {Borrill},
  {Bouchet}, {Brown}, {Bucher}, {Burigana}, {Cabella}, {Cardoso}, {Catalano},
  {Cay{\'o}n}, {Challinor}, {Chamballu}, {Chary}, {Chiang}, {Chiang}, {Chon},
  {Christensen}, {Churazov}, {Clements}, {Colafrancesco}, {Colombi}, {Couchot},
  {Coulais}, {Crill}, {Cuttaia}, {da Silva}, {Dahle}, {Danese}, {de Bernardis},
  {de Gasperis}, {de Rosa}, {de Zotti}, {Delabrouille}, {Delouis},
  {D{\'e}sert}, {Diego}, {Dolag}, {Donzelli}, {Dor{\'e}}, {D{\"o}rl},
  {Douspis}, {Dupac}, {Efstathiou}, {En{\ss}lin}, {Finelli}, {Flores-Cacho},
  {Forni}, {Frailis}, {Franceschi}, {Fromenteau}, {Galeotta}, {Ganga},
  {G{\'e}nova-Santos}, {Giard}, {Giardino}, {Giraud-H{\'e}raud},
  {Gonz{\'a}lez-Nuevo}, {G{\'o}rski}, {Gratton}, {Gregorio}, {Gruppuso},
  {Harrison}, {Henrot-Versill{\'e}}, {Hern{\'a}ndez-Monteagudo}, {Herranz},
  {Hildebrandt}, {Hivon}, {Hobson}, {Holmes}, {Hovest}, {Hoyland},
  {Huffenberger}, {Jaffe}, {Jones}, {Juvela}, {Keih{\"a}nen}, {Keskitalo},
  {Kisner}, {Kneissl}, {Knox}, {Kurki-Suonio}, {Lagache}, {Lamarre}, {Lasenby},
  {Laureijs}, {Lawrence}, {Leach}, {Leonardi}, {Linden-V{\o}rnle},
  {L{\'o}pez-Caniego}, {Lubin}, {Mac{\'{\i}}as-P{\'e}rez}, {MacTavish},
  {Maffei}, {Maino}, {Mandolesi}, {Mann}, {Maris}, {Marleau},
  {Mart{\'{\i}}nez-Gonz{\'a}lez}, {Masi}, {Matarrese}, {Matthai}, {Mazzotta},
  {Melchiorri}, {Melin}, {Mendes}, {Mennella}, {Mitra},
  {Miville-Desch{\^e}nes}, {Moneti}, {Montier}, {Morgante}, {Mortlock},
  {Munshi}, {Murphy}, {Naselsky}, {Natoli}, {Netterfield},
  {N{\o}rgaard-Nielsen}, {Noviello}, {Novikov}, {Novikov}, {Osborne}, {Pajot},
  {Pasian}, {Patanchon}, {Perdereau}, {Perotto}, {Perrotta}, {Piacentini},
  {Piat}, {Pierpaoli}, {Piffaretti}, {Plaszczynski}, {Pointecouteau},
  {Polenta}, {Ponthieu}, {Poutanen}, {Pratt}, {Pr{\'e}zeau}, {Prunet}, {Puget},
  {Rebolo}, {Reinecke}, {Renault}, {Ricciardi}, {Riller}, {Ristorcelli},
  {Rocha}, {Rosset}, {Rubi{\~n}o-Mart{\'{\i}}n}, {Rusholme}, {Sandri},
  {Santos}, {Schaefer}, {Scott}, {Seiffert}, {Smoot}, {Starck}, {Stivoli},
  {Stolyarov}, {Sunyaev}, {Sygnet}, {Tauber}, {Terenzi}, {Toffolatti},
  {Tomasi}, {Tristram}, {Tuovinen}, {Valenziano}, {Vibert}, {Vielva}, {Villa},
  {Vittorio}, {Wandelt}, {White}, {White}, {Yvon}, {Zacchei}, \&
  {Zonca}}]{planck11}
{Planck Collaboration}, {Aghanim}, N., {Arnaud}, M., {et~al.} 2011, \aap, 536,
  A10

\bibitem[{{Poole} {et~al.}(2006){Poole}, {Fardal}, {Babul}, {McCarthy},
  {Quinn}, \& {Wadsley}}]{poole06}
{Poole}, G.~B., {Fardal}, M.~A., {Babul}, A., {et~al.} 2006, \mnras, 373, 881

\bibitem[{{Pratt} {et~al.}(2009){Pratt}, {Croston}, {Arnaud}, \&
  {B{\"o}hringer}}]{pratt09}
{Pratt}, G.~W., {Croston}, J.~H., {Arnaud}, M., \& {B{\"o}hringer}, H. 2009,
  \aap, 498, 361

\bibitem[{{Press} \& {Schechter}(1974)}]{press74}
{Press}, W.~H. \& {Schechter}, P. 1974, \apj, 187, 425

\bibitem[{{Reid} {et~al.}(1999){Reid}, {Hunstead}, {Lemonon}, \&
  {Pierre}}]{reid99}
{Reid}, A.~D., {Hunstead}, R.~W., {Lemonon}, L., \& {Pierre}, M.~M. 1999,
  \mnras, 302, 571

\bibitem[{{Rossetti} {et~al.}(2017){Rossetti}, {Gastaldello}, {Eckert}, {Della
  Torre}, {Pantiri}, {Cazzoletti}, \& {Molendi}}]{rossetti17}
{Rossetti}, M., {Gastaldello}, F., {Eckert}, D., {et~al.} 2017, \mnras, 468,
  1917

\bibitem[{{Rudnick} {et~al.}(2006){Rudnick}, {Delain}, \&
  {Lemmerman}}]{rudnick06}
{Rudnick}, L., {Delain}, K.~M., \& {Lemmerman}, J.~A. 2006, Astronomische
  Nachrichten, 327, 549

\bibitem[{{Russell} {et~al.}(2011){Russell}, {van Weeren}, {Edge}, {McNamara},
  {Sanders}, {Fabian}, {Baum}, {Canning}, {Donahue}, \& {O'Dea}}]{russell11}
{Russell}, H.~R., {van Weeren}, R.~J., {Edge}, A.~C., {et~al.} 2011, \mnras,
  417, L1

\bibitem[{{Santos} {et~al.}(2008){Santos}, {Rosati}, {Tozzi}, {B{\"o}hringer},
  {Ettori}, \& {Bignamini}}]{santos08}
{Santos}, J.~S., {Rosati}, P., {Tozzi}, P., {et~al.} 2008, \aap, 483, 35

\bibitem[{{Savini} {et~al.}(2019){Savini}, {Bonafede}, {Br{\"u}ggen},
  {Rafferty}, {Shimwell}, {Botteon}, {Brunetti}, {Intema}, {Wilber}, {Cassano},
  {Vazza}, {van Weeren}, {Cuciti}, {De Gasperin}, {R{\"o}ttgering}, {Sommer},
  {B{\^\i}rzan}, \& {Drabent}}]{savini19}
{Savini}, F., {Bonafede}, A., {Br{\"u}ggen}, M., {et~al.} 2019, \aap, 622, A24

\bibitem[{{Savini} {et~al.}(2018){Savini}, {Bonafede}, {Br{\"u}ggen}, {van
  Weeren}, {Brunetti}, {Intema}, {Botteon}, {Shimwell}, {Wilber}, {Rafferty},
  {Giacintucci}, {Cassano}, {Cuciti}, {de Gasperin}, {R{\"o}ttgering}, {Hoeft},
  \& {White}}]{savini18}
{Savini}, F., {Bonafede}, A., {Br{\"u}ggen}, M., {et~al.} 2018, \mnras, 478,
  2234

\bibitem[{{Scaife} \& {Heald}(2012)}]{scaifeheald12}
{Scaife}, A.~M.~M. \& {Heald}, G.~H. 2012, \mnras, 423, L30

\bibitem[{{Shakouri} {et~al.}(2016{\natexlab{a}}){Shakouri},
  {Johnston-Hollitt}, \& {Dehghan}}]{shakouri16opt}
{Shakouri}, S., {Johnston-Hollitt}, M., \& {Dehghan}, S. 2016{\natexlab{a}},
  \mnras, 458, 3083

\bibitem[{{Shakouri} {et~al.}(2016{\natexlab{b}}){Shakouri},
  {Johnston-Hollitt}, \& {Pratt}}]{shakouri16}
{Shakouri}, S., {Johnston-Hollitt}, M., \& {Pratt}, G.~W. 2016{\natexlab{b}},
  \mnras, 459, 2525

\bibitem[{{Shimwell} {et~al.}(2019){Shimwell}, {Tasse}, {Hardcastle}, {Mechev},
  {Williams}, {Best}, {R{\"o}ttgering}, {Callingham}, {Dijkema}, {de Gasperin},
  {Hoang}, {Hugo}, {Mirmont}, {Oonk}, {Prandoni}, {Rafferty}, {Sabater},
  {Smirnov}, {van Weeren}, {White}, {Atemkeng}, {Bester}, {Bonnassieux},
  {Br{\"u}ggen}, {Brunetti}, {Chy{\.z}y}, {Cochrane}, {Conway}, {Croston},
  {Danezi}, {Duncan}, {Haverkorn}, {Heald}, {Iacobelli}, {Intema}, {Jackson},
  {Jamrozy}, {Jarvis}, {Lakhoo}, {Mevius}, {Miley}, {Morabito}, {Morganti},
  {Nisbet}, {Orr{\'u}}, {Perkins}, {Pizzo}, {Schrijvers}, {Smith}, {Vermeulen},
  {Wise}, {Alegre}, {Bacon}, {van Bemmel}, {Beswick}, {Bonafede}, {Botteon},
  {Bourke}, {Brienza}, {Calistro Rivera}, {Cassano}, {Clarke}, {Conselice},
  {Dettmar}, {Drabent}, {Dumba}, {Emig}, {En{\ss}lin}, {Ferrari}, {Garrett},
  {G{\'e}nova-Santos}, {Goyal}, {G{\"u}rkan}, {Hale}, {Harwood}, {Heesen},
  {Hoeft}, {Horellou}, {Jackson}, {Kokotanekov}, {Kondapally},
  {Kunert-Bajraszewska}, {Mahatma}, {Mahony}, {Mandal}, {McKean}, {Merloni},
  {Mingo}, {Miskolczi}, {Mooney}, {Nikiel-Wroczy{\'n}ski}, {O'Sullivan},
  {Quinn}, {Reich}, {Roskowi{\'n}ski}, {Rowlinson}, {Savini}, {Saxena},
  {Schwarz}, {Shulevski}, {Sridhar}, {Stacey}, {Urquhart}, {van der Wiel},
  {Varenius}, {Webster}, \& {Wilber}}]{shimwell19}
{Shimwell}, T.~W., {Tasse}, C., {Hardcastle}, M.~J., {et~al.} 2019, \aap, 622,
  A1

\bibitem[{{Skrutskie} {et~al.}(2006){Skrutskie}, {Cutri}, {Stiening},
  {Weinberg}, {Schneider}, {Carpenter}, {Beichman}, {Capps}, {Chester},
  {Elias}, {Huchra}, {Liebert}, {Lonsdale}, {Monet}, {Price}, {Seitzer},
  {Jarrett}, {Kirkpatrick}, {Gizis}, {Howard}, {Evans}, {Fowler}, {Fullmer},
  {Hurt}, {Light}, {Kopan}, {Marsh}, {McCallon}, {Tam}, {Van Dyk}, \&
  {Wheelock}}]{2mass}
{Skrutskie}, M.~F., {Cutri}, R.~M., {Stiening}, R., {et~al.} 2006, \aj, 131,
  1163

\bibitem[{{Sommer} \& {Basu}(2014)}]{sommerbasu14}
{Sommer}, M.~W. \& {Basu}, K. 2014, \mnras, 437, 2163

\bibitem[{{Sommer} {et~al.}(2017){Sommer}, {Basu}, {Intema}, {Pacaud},
  {Bonafede}, {Babul}, \& {Bertoldi}}]{sommer17}
{Sommer}, M.~W., {Basu}, K., {Intema}, H., {et~al.} 2017, \mnras, 466, 996

\bibitem[{{Vacca} {et~al.}(2011){Vacca}, {Govoni}, {Murgia}, {Giovannini},
  {Feretti}, {Tugnoli}, {Verheijen}, \& {Taylor}}]{vacca11}
{Vacca}, V., {Govoni}, F., {Murgia}, M., {et~al.} 2011, \aap, 535, A82

\bibitem[{{van Weeren} {et~al.}(2011){van Weeren}, {Br{\"u}ggen},
  {R{\"o}ttgering}, {Hoeft}, {Nuza}, \& {Intema}}]{vanweeren11}
{van Weeren}, R.~J., {Br{\"u}ggen}, M., {R{\"o}ttgering}, H.~J.~A., {et~al.}
  2011, \aap, 533, A35

\bibitem[{{van Weeren} {et~al.}(2019){van Weeren}, {de Gasperin}, {Akamatsu},
  {Br{\"u}ggen}, {Feretti}, {Kang}, {Stroe}, \& {Zandanel}}]{vanweeren19}
{van Weeren}, R.~J., {de Gasperin}, F., {Akamatsu}, H., {et~al.} 2019, \ssr,
  215, 16

\bibitem[{{van Weeren} {et~al.}(2013){van Weeren}, {Fogarty}, {Jones},
  {Forman}, {Clarke}, {Br{\"u}ggen}, {Kraft}, {Lal}, {Murray}, \&
  {R{\"o}ttgering}}]{vanweeren13}
{van Weeren}, R.~J., {Fogarty}, K., {Jones}, C., {et~al.} 2013, \apj, 769, 101

\bibitem[{{Ventimiglia} {et~al.}(2008){Ventimiglia}, {Voit}, {Donahue}, \&
  {Ameglio}}]{ventimiglia08}
{Ventimiglia}, D.~A., {Voit}, G.~M., {Donahue}, M., \& {Ameglio}, S. 2008,
  \apj, 685, 118

\bibitem[{{Venturi} {et~al.}(2007){Venturi}, {Giacintucci}, {Brunetti},
  {Cassano}, {Bardelli}, {Dallacasa}, \& {Setti}}]{venturi07}
{Venturi}, T., {Giacintucci}, S., {Brunetti}, G., {et~al.} 2007, \aap, 463, 937

\bibitem[{{Venturi} {et~al.}(2008){Venturi}, {Giacintucci}, {Dallacasa},
  {Cassano}, {Brunetti}, {Bardelli}, \& {Setti}}]{venturi08}
{Venturi}, T., {Giacintucci}, S., {Dallacasa}, D., {et~al.} 2008, \aap, 484,
  327

\bibitem[{{Venturi} {et~al.}(2017){Venturi}, {Rossetti}, {Brunetti},
  {Farnsworth}, {Gastaldello}, {Giacintucci}, {Lal}, {Rudnick}, {Shimwell},
  {Eckert}, {Molendi}, \& {Owers}}]{venturi17}
{Venturi}, T., {Rossetti}, M., {Brunetti}, G., {et~al.} 2017, \aap, 603, A125

\bibitem[{{Wei{\ss}mann} {et~al.}(2013){Wei{\ss}mann}, {B{\"o}hringer}, {{\v
  S}uhada}, \& {Ameglio}}]{weissmann13}
{Wei{\ss}mann}, A., {B{\"o}hringer}, H., {{\v S}uhada}, R., \& {Ameglio}, S.
  2013, \aap, 549, A19

\bibitem[{{Wilber} {et~al.}(2018){Wilber}, {Br{\"u}ggen}, {Bonafede}, {Savini},
  {Shimwell}, {van Weeren}, {Rafferty}, {Mechev}, {Intema}, {Andrade-Santos},
  {Clarke}, {Mahony}, {Morganti}, {Prand oni}, {Brunetti}, {R{\"o}ttgering},
  {Mandal}, {de Gasperin}, \& {Hoeft}}]{wilber18}
{Wilber}, A., {Br{\"u}ggen}, M., {Bonafede}, A., {et~al.} 2018, \mnras, 473,
  3536

\bibitem[{{Willott} {et~al.}(2003){Willott}, {Rawlings}, {Jarvis}, \&
  {Blundell}}]{willott03}
{Willott}, C.~J., {Rawlings}, S., {Jarvis}, M.~J., \& {Blundell}, K.~M. 2003,
  \mnras, 339, 173

\bibitem[{{ZuHone} {et~al.}(2015){ZuHone}, {Brunetti}, {Giacintucci}, \&
  {Markevitch}}]{zuhone15}
{ZuHone}, J.~A., {Brunetti}, G., {Giacintucci}, S., \& {Markevitch}, M. 2015,
  \apj, 801, 146

\bibitem[{{ZuHone} {et~al.}(2013){ZuHone}, {Markevitch}, {Brunetti}, \&
  {Giacintucci}}]{zuhone13}
{ZuHone}, J.~A., {Markevitch}, M., {Brunetti}, G., \& {Giacintucci}, S. 2013,
  \apj, 762, 78

\end{thebibliography}

\clearpage
\onecolumn

\begin{appendix}
\pagebreak
\clearpage
\section{Images of clusters without diffuse emission}
\label{app:UL}
These are images of clusters without diffuse emission.
\begin{figure*}[h]
   \centering
   \includegraphics[scale=0.5]{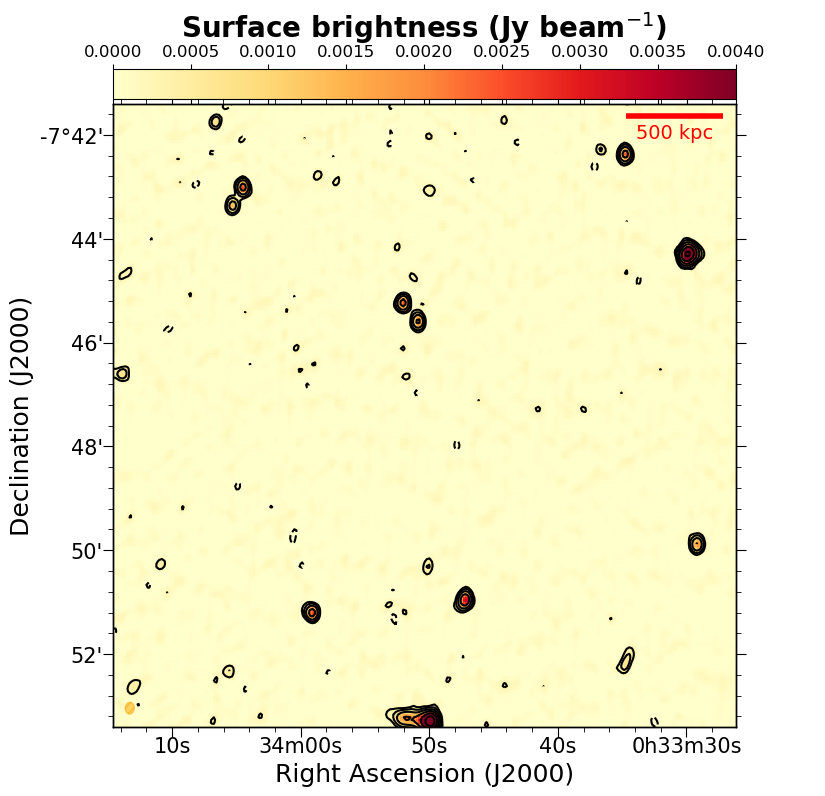}
 \caption{A56 JVLA C array 1.5 GHz image. Contours start at 3$-\sigma_{rms}$ and are spaced by a factor of two. 1$-\sigma_{rms}=0.08$ mJy/beam with beam=$13.''\times10.4''$.}
         \label{Fig:A56}
   \end{figure*}
   
\begin{figure*}[h]
   \centering
   \includegraphics[scale=0.5]{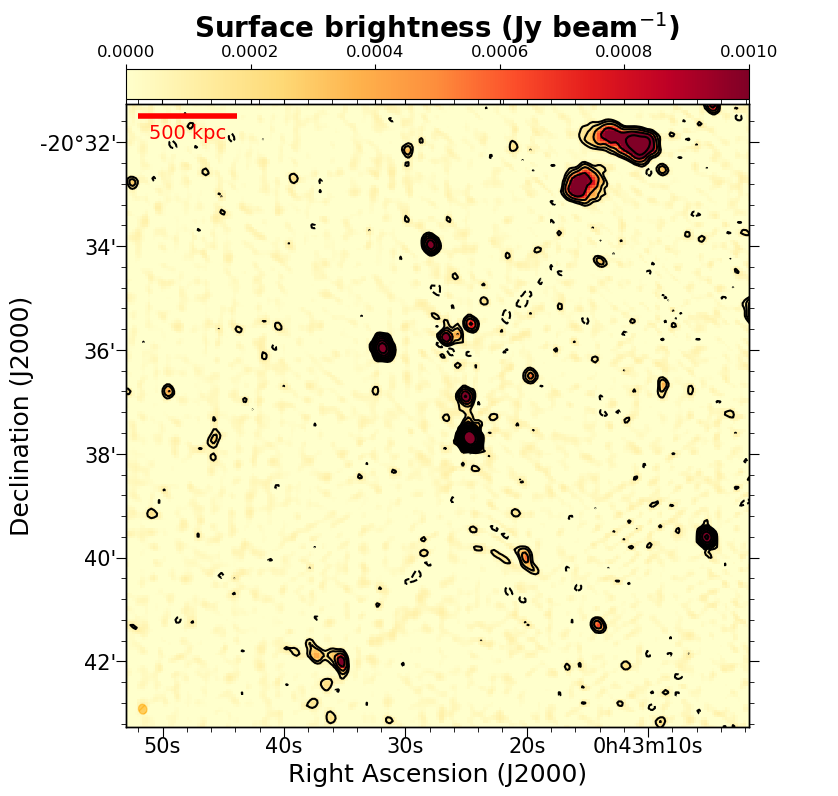}
 \caption{A2813 JVLA C array 1.5 GHz image. Contours start at 3$-\sigma_{rms}$ and are spaced by a factor of two. 1$-\sigma_{rms}=0.035$ mJy/beam with beam=$11.4''\times9.5''$.}
         \label{Fig:A2813}
   \end{figure*}   
   
\begin{figure*}[h]
   \centering
   \includegraphics[scale=0.4]{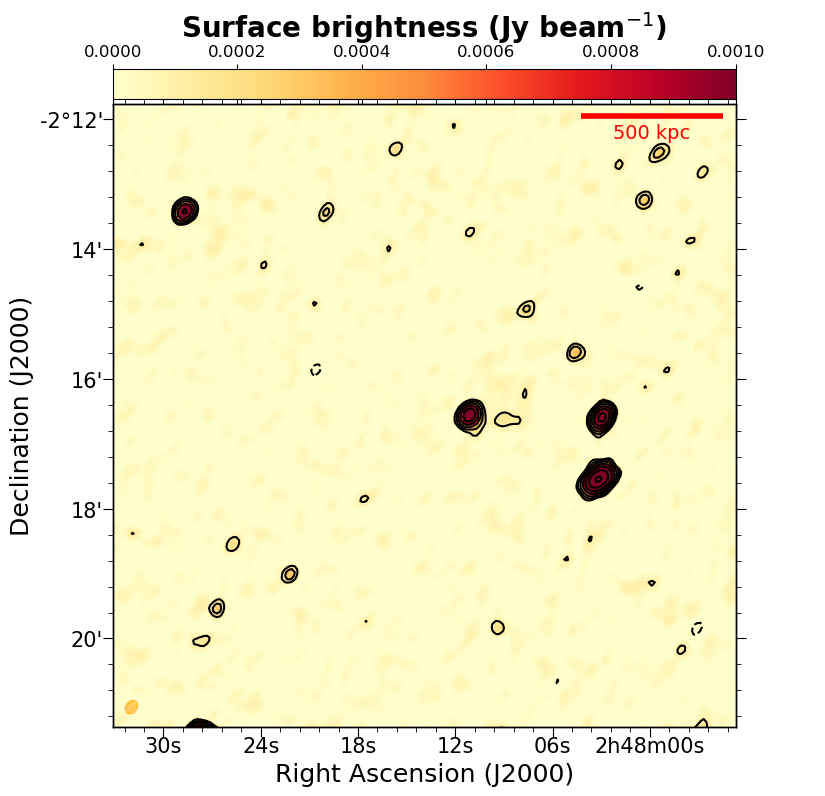}
   \includegraphics[scale=0.4]{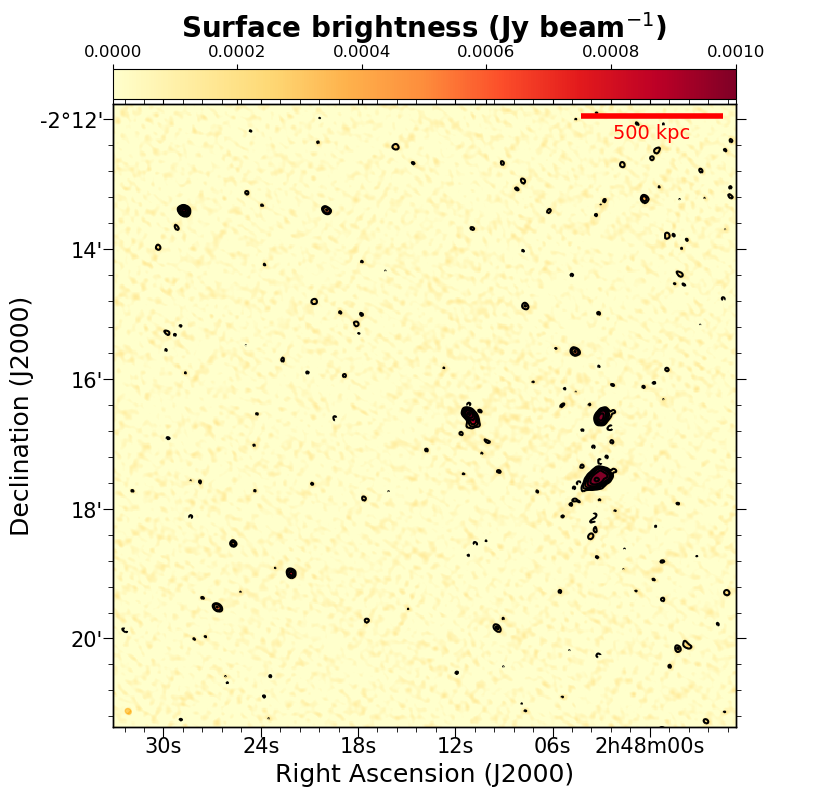}
 \caption{Images of the cluster A384. \textit{Left panel}: JVLA C array 1.5 GHz image. Contours start at 3$-\sigma_{rms}$ and are spaced by a factor of two. 1$-\sigma_{rms}=0.035$ mJy/beam with beam=$13.5''\times10.4''$. \textit{Right panel}: GMRT 610 MHz image. Contours start at 3$-\sigma_{rms}$ and are spaced by a factor of two. 1$-\sigma_{rms}=0.05$ mJy/beam with beam=$5.9''\times4.8''$.}
         \label{Fig:A384}
   \end{figure*}

 \begin{figure*}[h]
   \centering
   \includegraphics[scale=0.5]{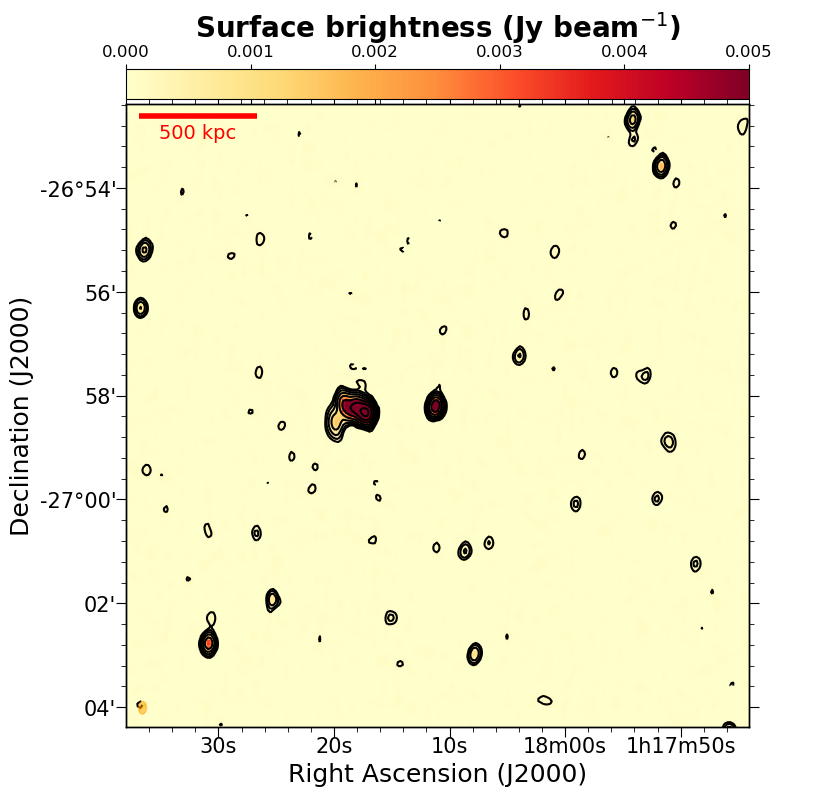}
 \caption{A2895 JVLA C array 1.5 GHz image. Contours start at 3$-\sigma_{rms}$ and are spaced by a factor of 2. 1$-\sigma_{rms}=0.04$ mJy/beam with beam=$14.6''\times9.0''$.}
         \label{Fig:A2895}
   \end{figure*}  
   
\begin{figure*}[h]
   \centering
   \includegraphics[scale=0.5]{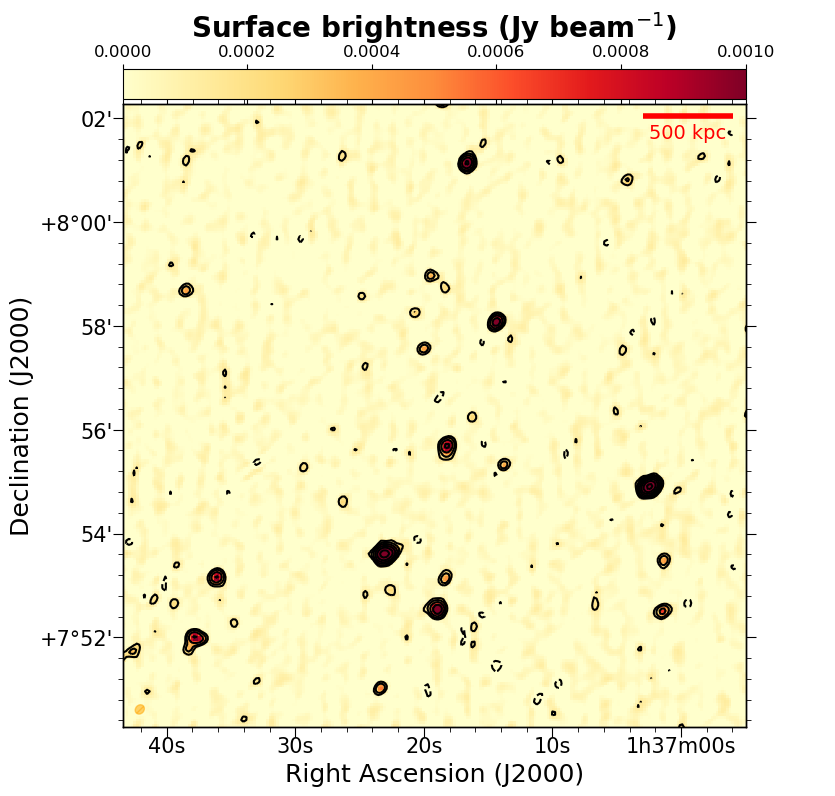}
 \caption{A220 JVLA C array 1.5 GHz image. Contours start at 3$-\sigma_{rms}$ and are spaced by a factor of two. 1$-\sigma_{rms}=0.045$ mJy/beam with beam=$11.6''\times9.8''$.}
         \label{Fig:A220}
   \end{figure*}  
   
\begin{figure*}[h]
   \centering
   \includegraphics[scale=0.5]{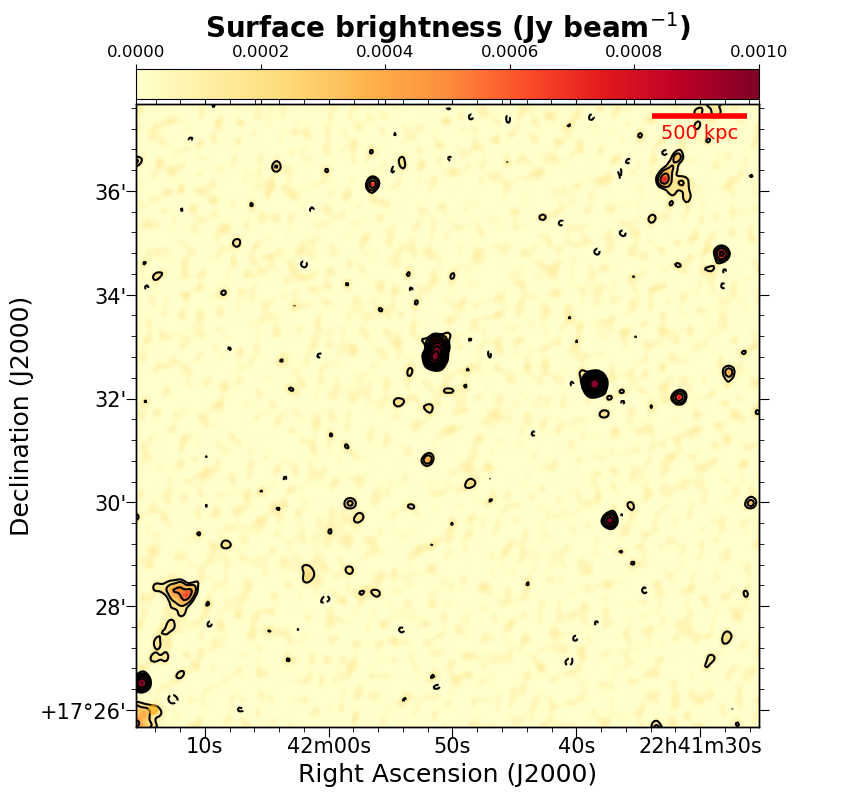}
 \caption{A2472 JVLA C array 1.5 GHz image. Contours start at 3$-\sigma_{rms}$ and are spaced by a factor of two. 1$-\sigma_{rms}=0.04$ mJy/beam with beam=$10.5''\times10.1''$.}
         \label{Fig:A2472}
   \end{figure*}

\begin{figure*}[h]
   \centering
   \includegraphics[scale=0.4]{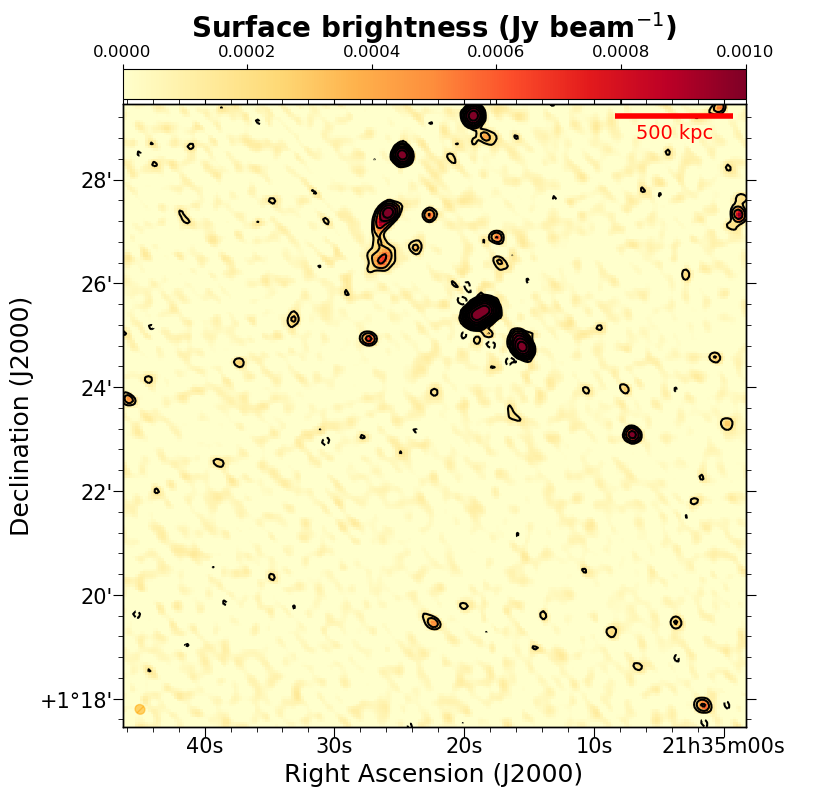}
   \includegraphics[scale=0.4]{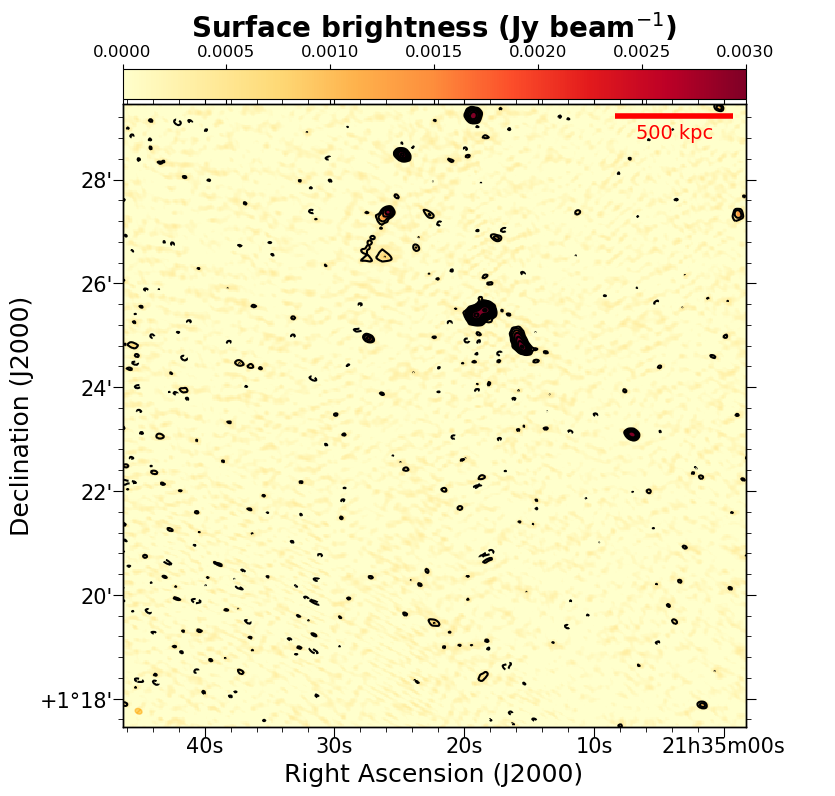}
 \caption{Images of the cluster A2355. \textit{Left panel}: JVLA C array 1.5 GHz image. Contours start at 3$-\sigma_{rms}$ and are spaced by a factor of two. 1$-\sigma_{rms}=0.04$ mJy/beam with beam=$11.6''\times10.8''$. \textit{Right panel}: GMRT 610 MHz image. Contours start at 3$-\sigma_{rms}$ and are spaced by a factor of two. 1$-\sigma_{rms}=0.13$ mJy/beam with beam=$8.3''\times6.1''$.}
         \label{Fig:A2355}
   \end{figure*}

\begin{figure*}[h]
   \centering
   \includegraphics[scale=0.4]{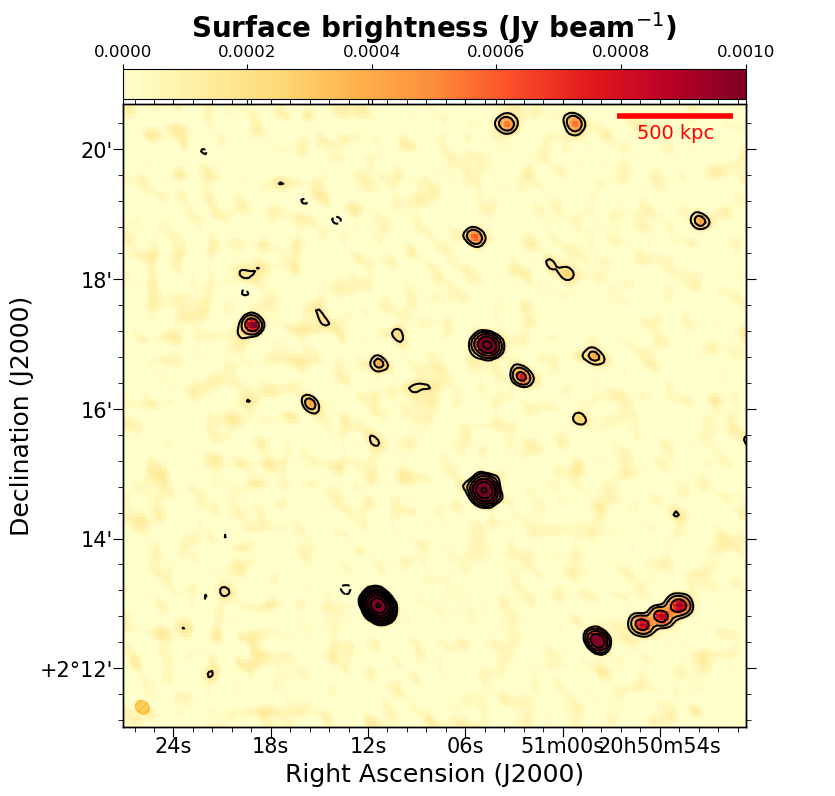}
   \includegraphics[scale=0.4]{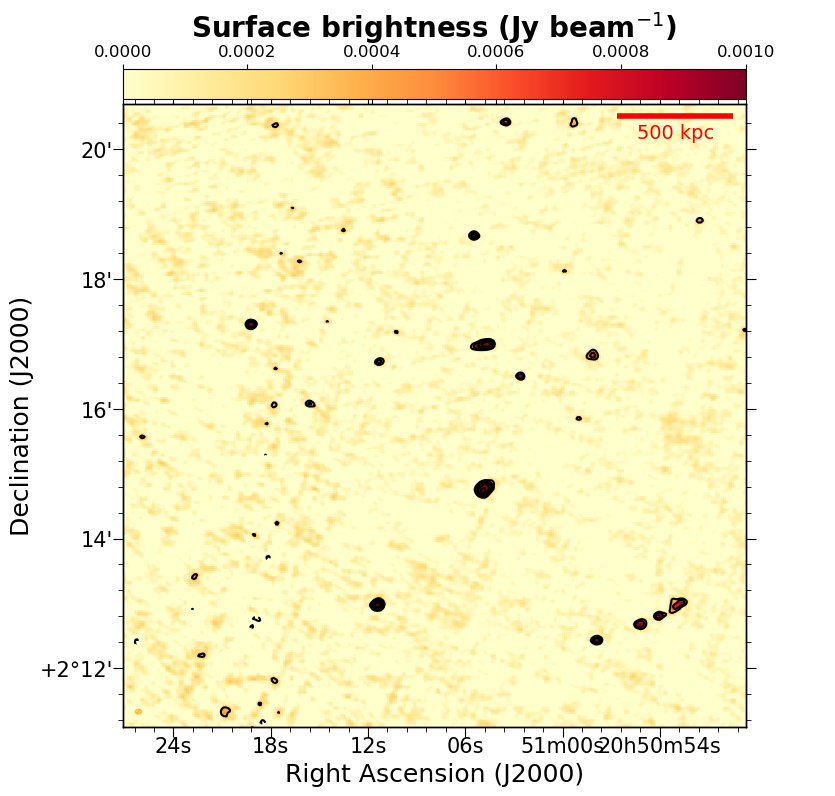}
 \caption{Images of the cluster RXC J2051.1+0216. \textit{Left panel}: JVLA C array 1.5 GHz image. Contours start at 3$-\sigma_{rms}$ and are spaced by a factor of two. 1$-\sigma_{rms}=0.05$ mJy/beam with beam=$14.1''\times11.5''$. \textit{Right panel}: GMRT 610 MHz image. Contours start at 3$-\sigma_{rms}$ and are spaced by a factor of two. 1$-\sigma_{rms}=0.1$ mJy/beam with beam=$6.0''\times4.8''$.}
         \label{Fig:R2051}
   \end{figure*} 

\begin{figure*}[h]
   \centering
   \includegraphics[scale=0.5]{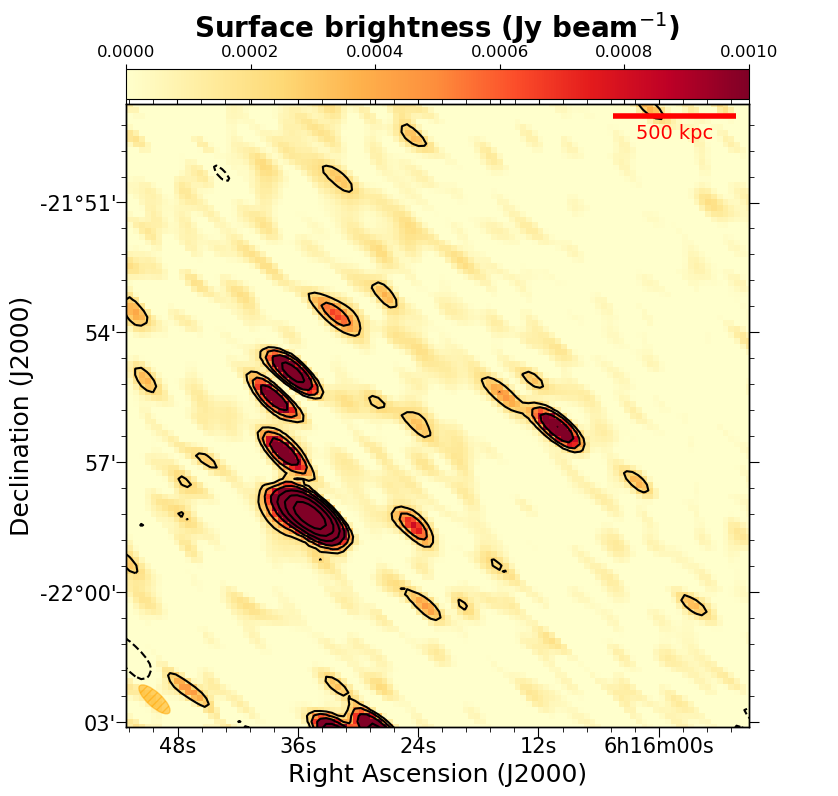}
 \caption{RXC J0616.3-2156 JVLA DnC array 1.5 GHz image. Contours start at 3$-\sigma_{rms}$ and are spaced by a factor of 2. 1$-\sigma_{rms}=0.08$ mJy/beam with beam=$55.4''\times19.9''$.}
         \label{Fig:R0616}
   \end{figure*}

\begin{figure*}[h]
   \centering
   \includegraphics[scale=0.5]{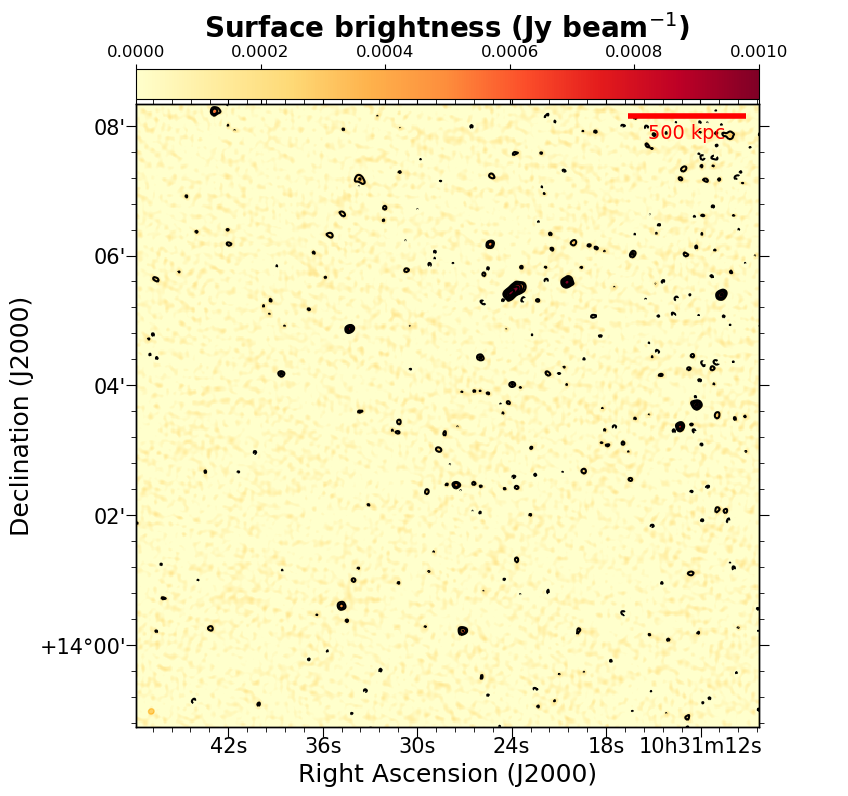}
 \caption{Zwcl 1028.8+1419 GMRT 610 MHz image. Contours start at 3$-\sigma_{rms}$ and are spaced by a factor of two. 1$-\sigma_{rms}=0.056$ mJy/beam with beam=$5.3''\times4.8''$.}
         \label{Fig:Z1028}
   \end{figure*}

\begin{figure*}[h]
   \centering
   \includegraphics[scale=0.4]{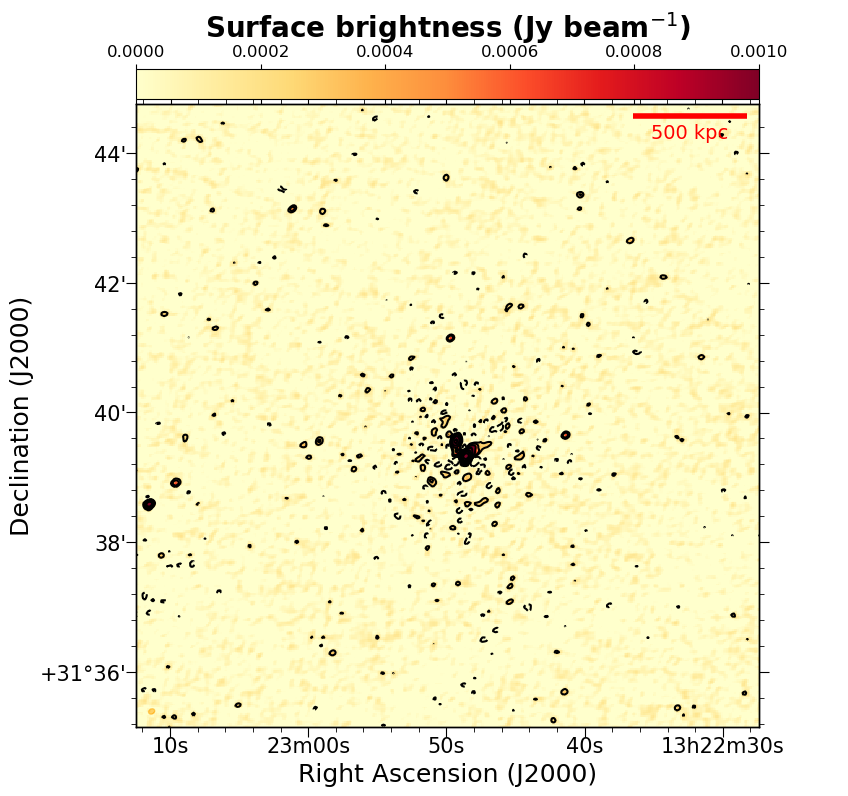}
 \caption{RXC J1322.8+3138 GMRT 610 MHz image. Contours start at 3-$\sigma_{rms}$ and are spaced by a factor of two. 1$-\sigma_{rms}=0.06$ mJy/beam with beam=$5.7''\times4.4''$. }
         \label{Fig:R1322}
   \end{figure*}

\begin{figure*}[h]
   \centering
   \includegraphics[scale=0.5]{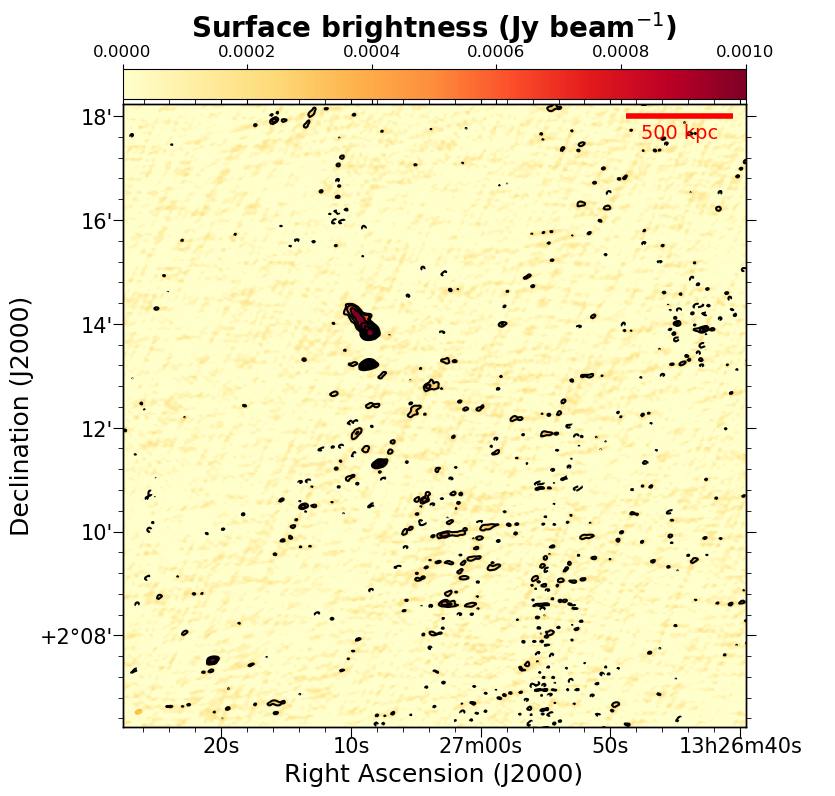}
 \caption{A1733 GMRT 610 MHz image. Contours start at 3$-\sigma_{rms}$ and are spaced by a factor of two. 1$-\sigma_{rms}=0.06$ mJy/beam with beam=$7.9''\times5.0''$.}
         \label{Fig:A1733}
   \end{figure*}

\begin{figure*}[h]
   \centering
   \includegraphics[scale=0.5]{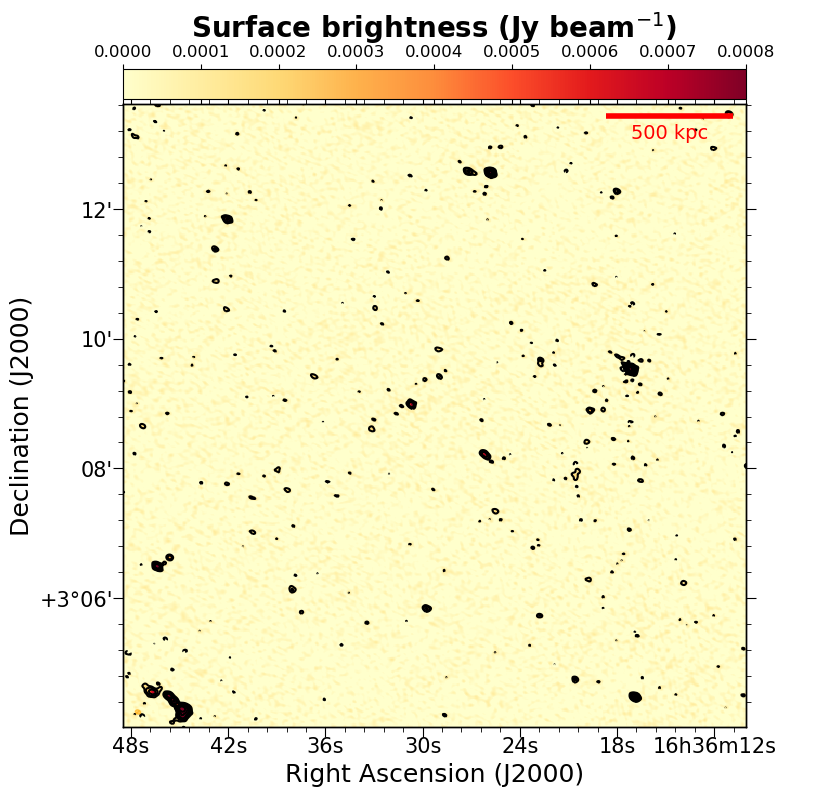}
 \caption{PSZ1 G019.12+3123 GMRT 610 MHz image. Contours start at 3$-\sigma_{rms}$ and are spaced by a factor of two. 1$-\sigma_{rms}=0.035$ mJy/beam with beam=$5.0''\times3.8''$.}
         \label{Fig:P019}
   \end{figure*}

\begin{figure*}[h]
   \centering
   \includegraphics[scale=0.5]{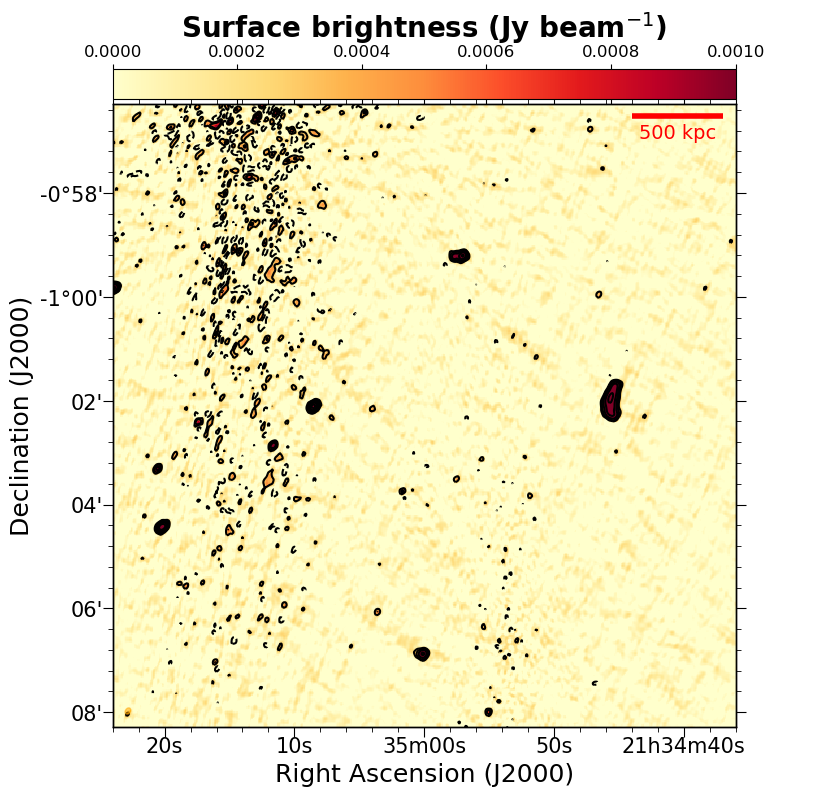}
 \caption{MACS J2135-010 GMRT 610 MHz image. Contours start at 3$-\sigma_{rms}$ and are spaced by a factor of two. 1$-\sigma_{rms}=0.08$ mJy/beam with beam=$7.8''\times5.8''$.}
         \label{Fig:M2135}
   \end{figure*}

\begin{figure*}[h]
   \centering
   \includegraphics[scale=0.4]{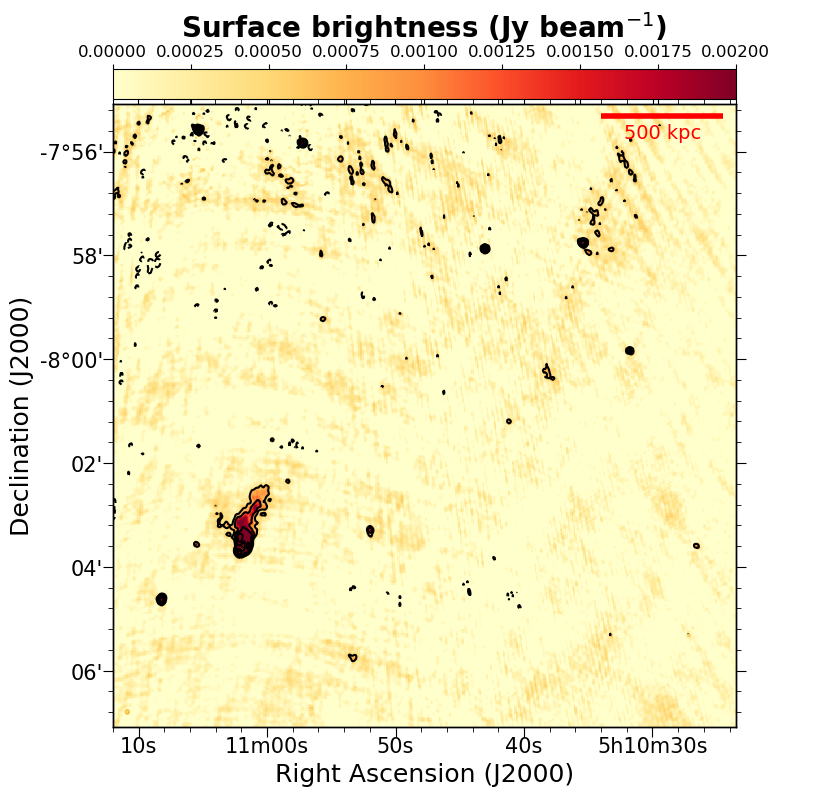}
   \includegraphics[scale=0.4]{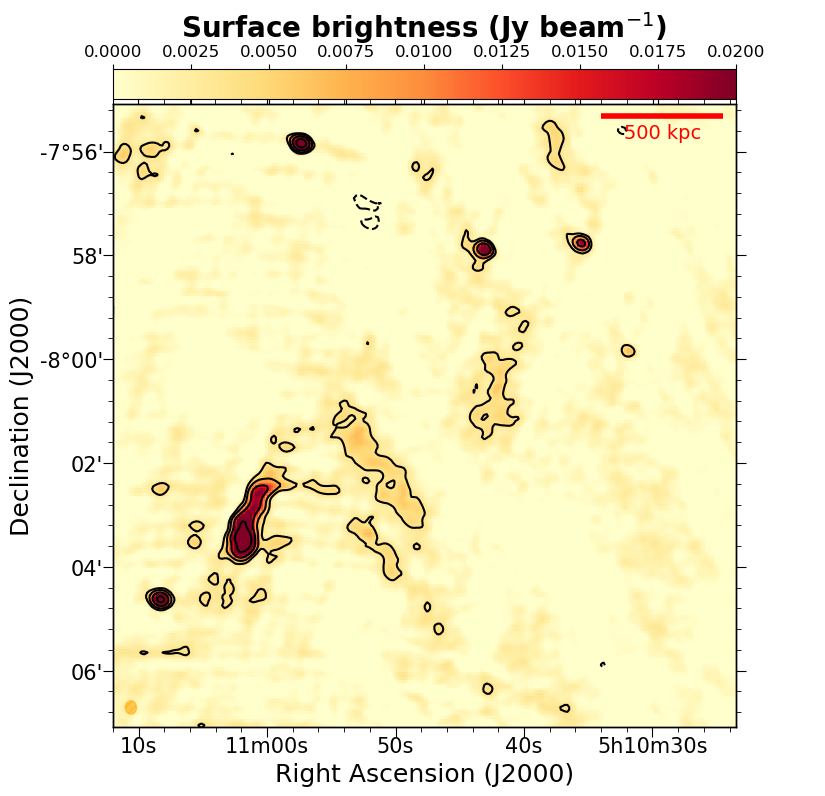}
 \caption{Images of the cluster RXC J0510.7-0801. \textit{Left panel}: GMRT 610 MHz image. Contours start at 3$-\sigma_{rms}$ and are spaced by a factor of two. 1$-\sigma_{rms}=0.2$ mJy/beam with beam=$5.4\times4.8$. \textit{Right panel}: RXC J0510.7-0801 GMRT 240 MHz image. Contours start at 3$-\sigma_{rms}$ and are spaced by a factor of two. 1$-\sigma_{rms}=1.2$mJy/beam with beam=$15.7''\times13.1''$.}
         \label{Fig:R0510}
   \end{figure*}

\begin{figure*}[h]
   \centering
   \includegraphics[scale=0.5]{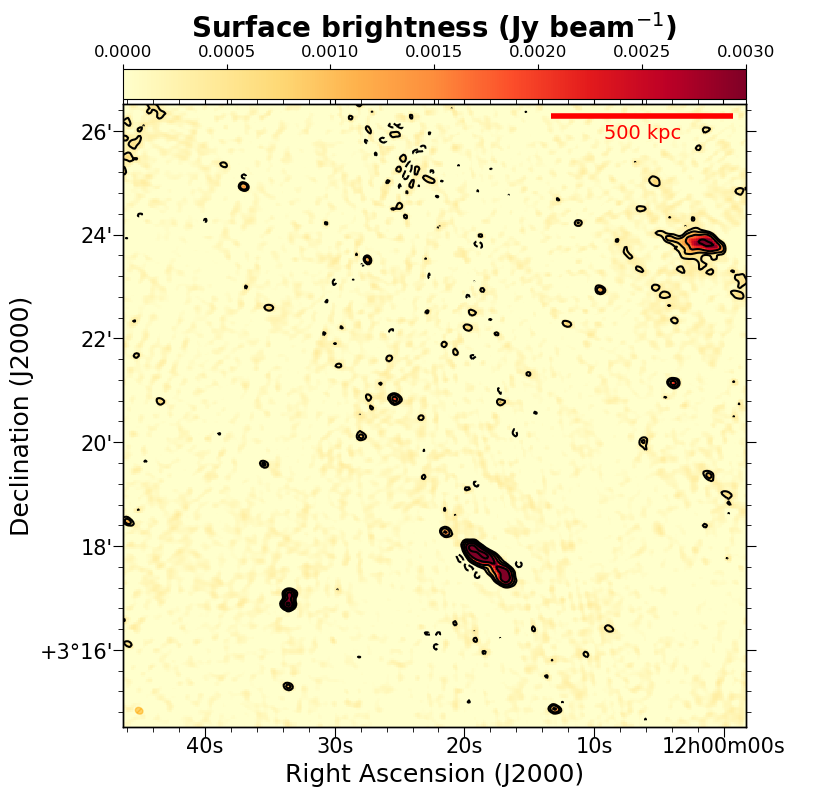}
 \caption{A1437 GMRT 330 MHz image. Contours start at 3$-\sigma_{rms}$ and are spaced by a factor of two. 1$-\sigma_{rms}=0.4$ mJy/beam with beam=$9.0''\times7.4''$.}
         \label{Fig:A1437}
   \end{figure*}
   
\begin{figure*}[h]
   \centering
   \includegraphics[scale=0.5]{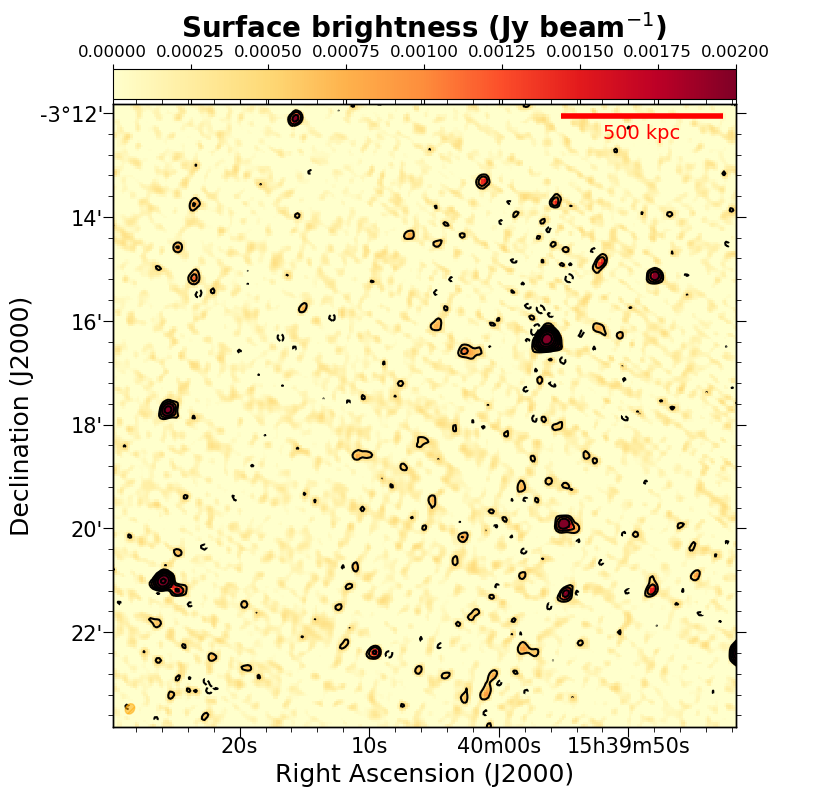}
 \caption{A2104 GMRT 240 MHz image. Contours start at 3$-\sigma_{rms}$ and are spaced by a factor of two. 1$-\sigma_{rms}=0.13$ mJy/beam with beam=$13.3''\times10.4''$.}
         \label{Fig:A2104}
   \end{figure*}

\clearpage   
\pagebreak   
\section{Radio halo surface brightness radial profiles}
\label{app:profile}

\begin{figure*}[h]
   \centering
   \includegraphics[height=8cm]{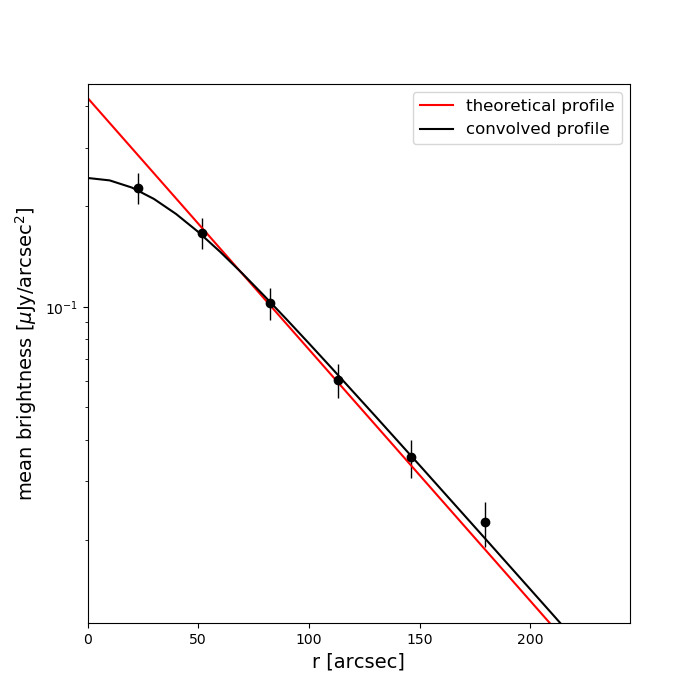}
   \includegraphics[height=8cm]{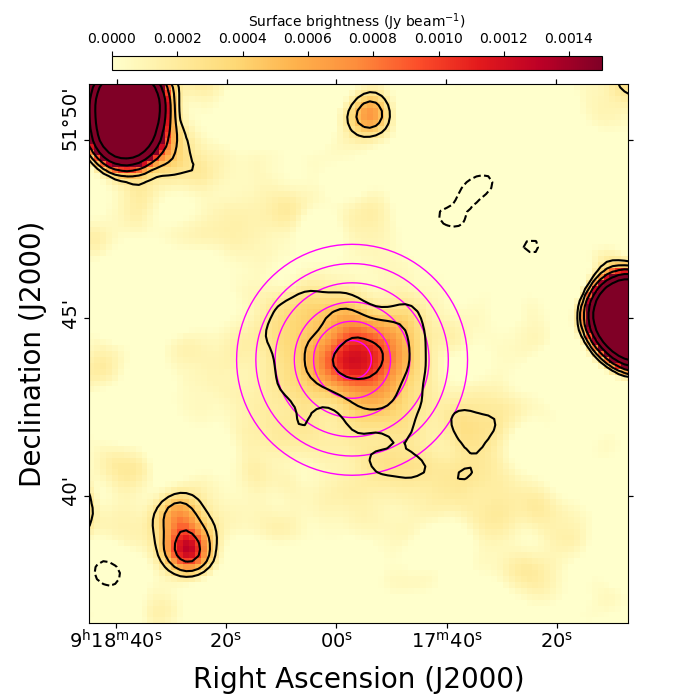}
   
 \caption{Radial surface brightness profile of the radio halo in A773. \textit{Left}: Data points represents the averaged surface brightness measured in the magenta annuli shown in the right panel. The red curve is the theoretical exponential profile and the black curve is the profile convolved with the beam of the image. \textit{Right}: Radio image of A773 (see Tab. \ref{tab:profiles}). Contours start at 3$-\sigma$ rms noise and are spaced by a factor of two. The orange circle in the bottom left corner shows the size of the beam.}
         \label{Fig:profile_A773}
   \end{figure*}

\begin{figure*}[h]
   \centering
   \includegraphics[height=8cm]{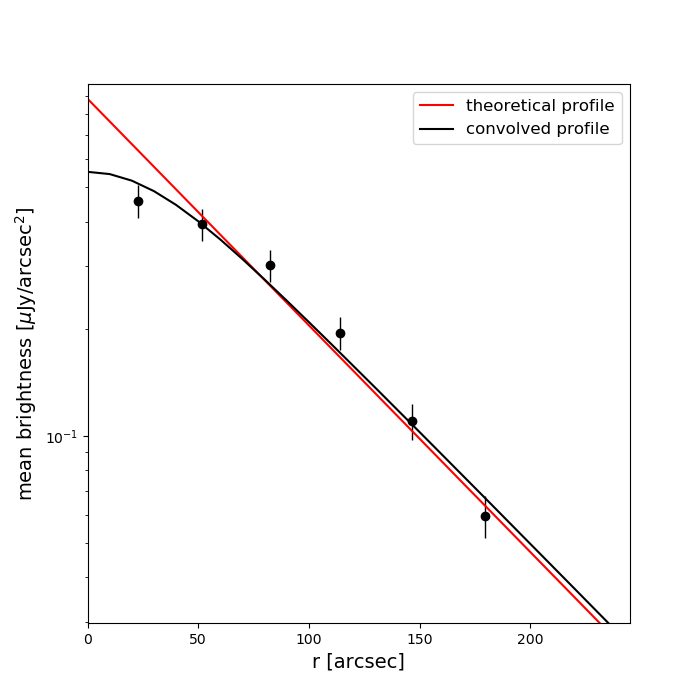}
   \includegraphics[height=8cm]{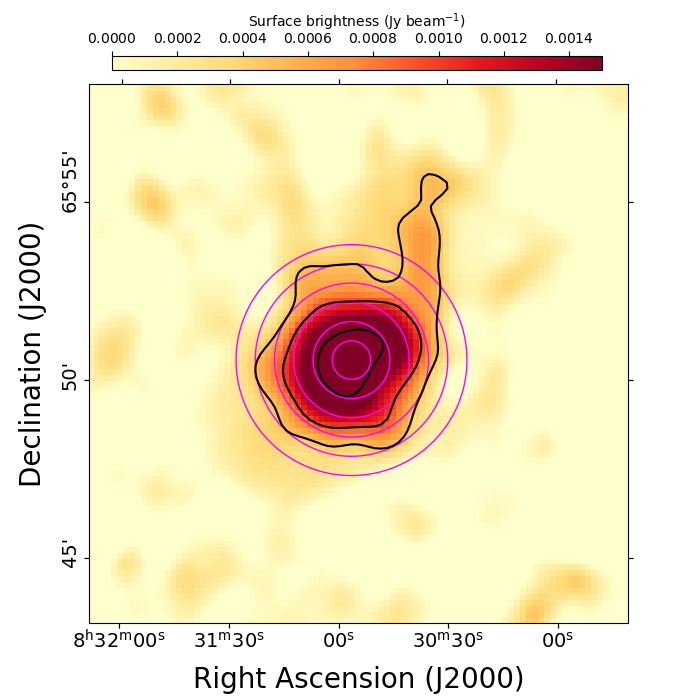}
   
 \caption{Same as Fig. \ref{Fig:profile_A773} for A665.}
         \label{Fig:profile_A665}
   \end{figure*}

\begin{figure*}[h]
   \centering
   \includegraphics[height=8cm]{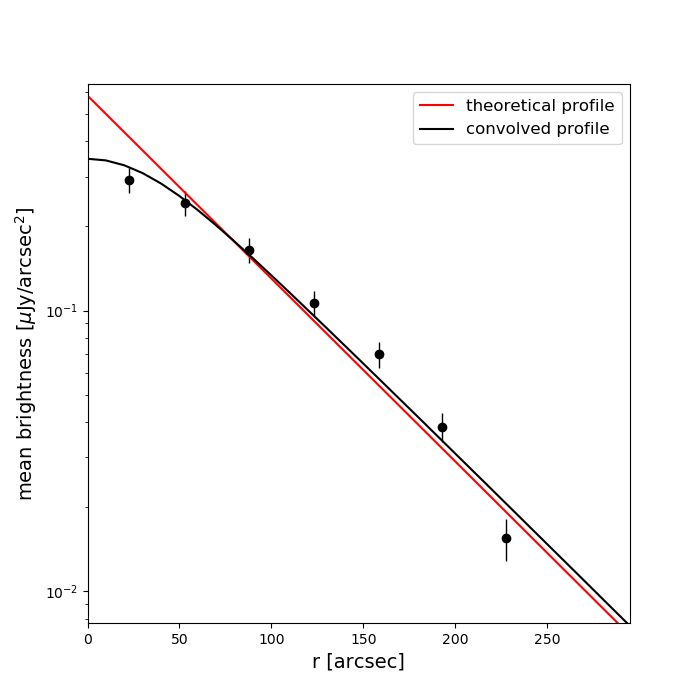}
   \includegraphics[height=8cm]{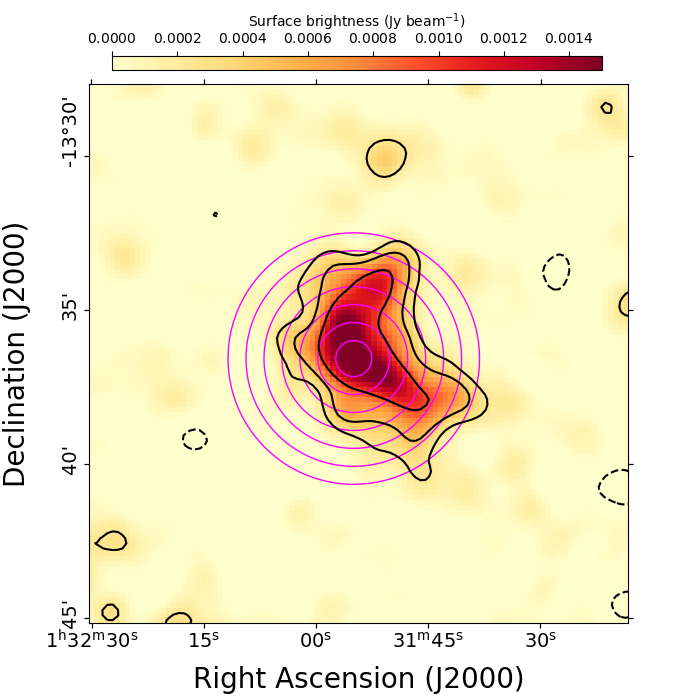}
   
 \caption{Same as Fig. \ref{Fig:profile_A773} for A209.}
         \label{Fig:profile_A209}
   \end{figure*}

\begin{figure*}[h]
   \centering
   \includegraphics[height=8cm]{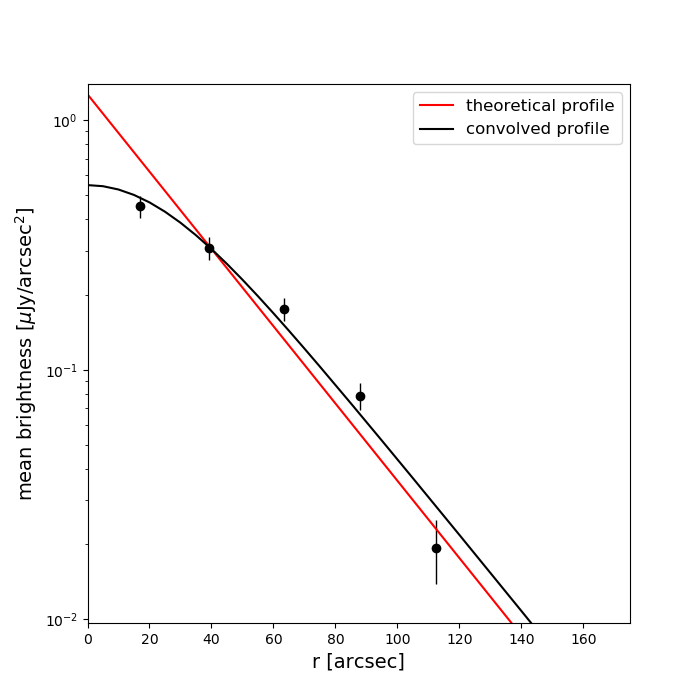}
   \includegraphics[height=8cm]{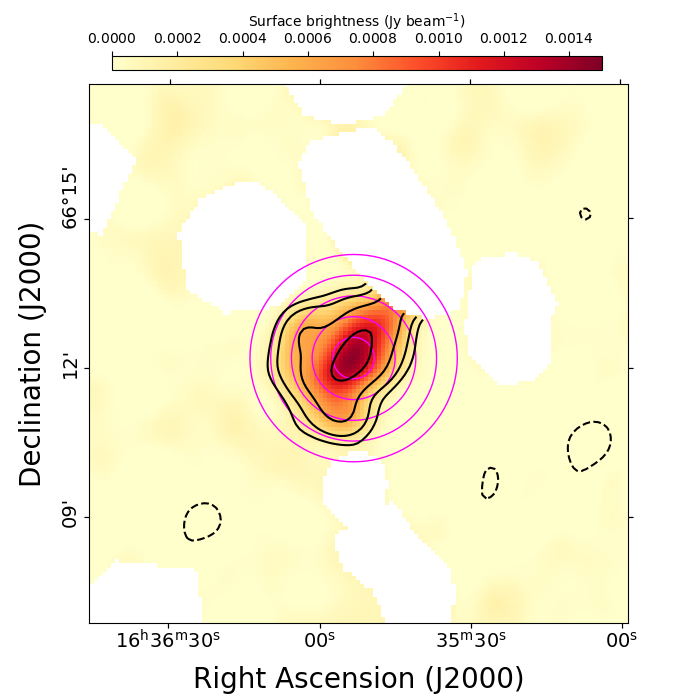}
   
 \caption{Same as Fig. \ref{Fig:profile_A773} for A2218.}
         \label{Fig:profile_A2218}
   \end{figure*}

\begin{figure*}[h]
   \centering
   \includegraphics[height=8cm]{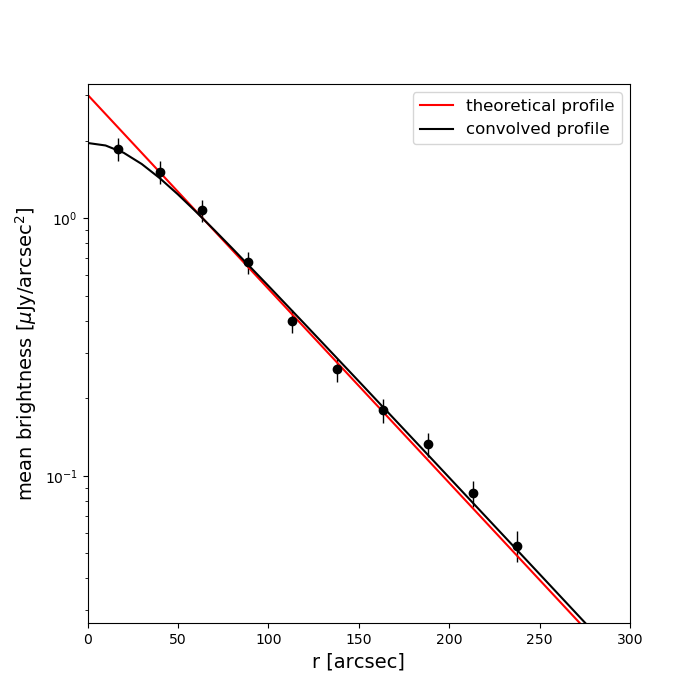}
   \includegraphics[height=8cm]{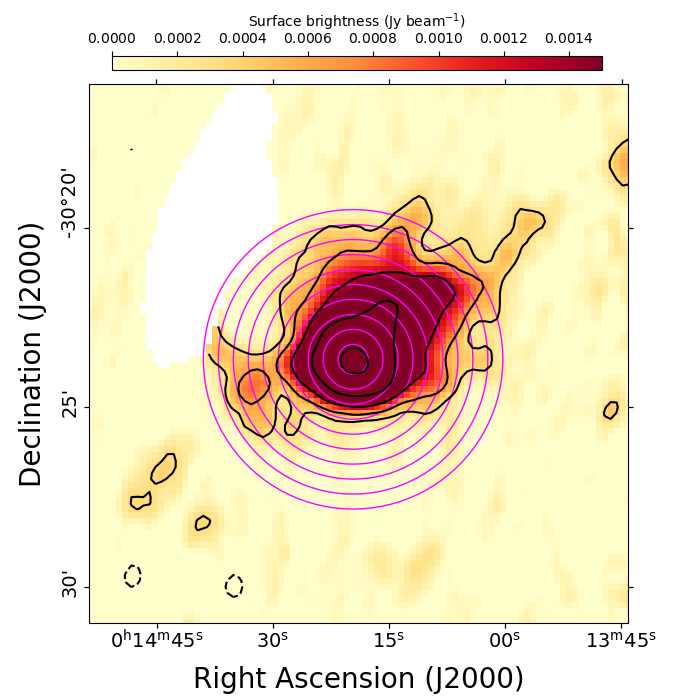}
   
 \caption{Same as Fig. \ref{Fig:profile_A773} for A2744.}
         \label{Fig:profile_A2744}
   \end{figure*}

\begin{figure*}[h]
   \centering
   \includegraphics[height=8cm]{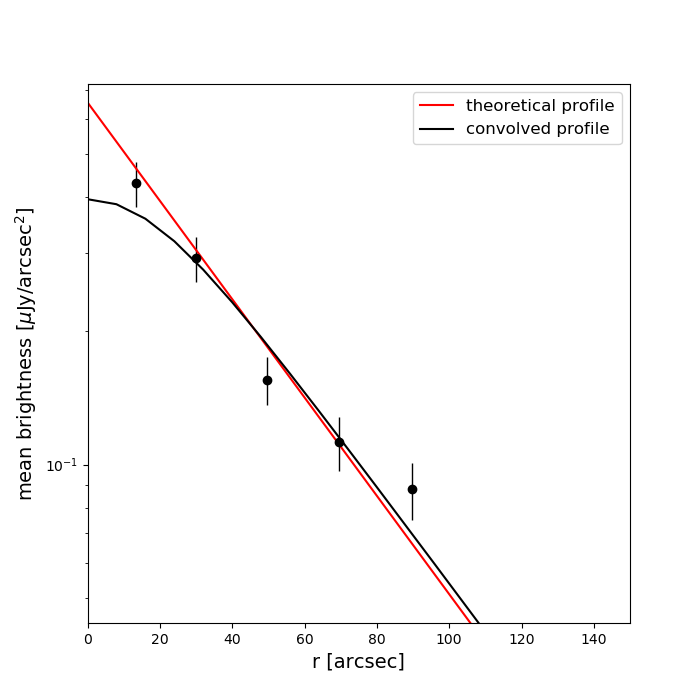}
   \includegraphics[height=8cm]{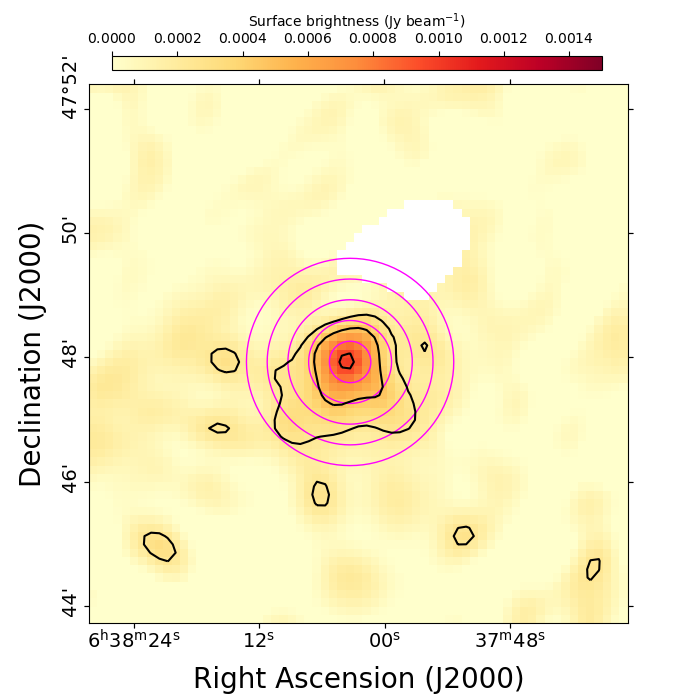}
   
 \caption{Same as Fig. \ref{Fig:profile_A773} for Z0634.}
         \label{Fig:profile_Z0634}
   \end{figure*}

\begin{figure*}[h]
   \centering
   \includegraphics[height=8cm]{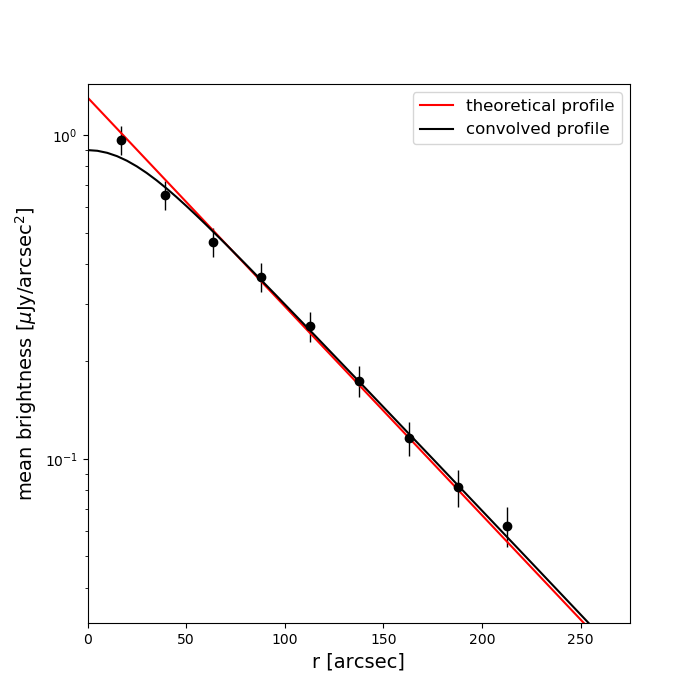}
   \includegraphics[height=8cm]{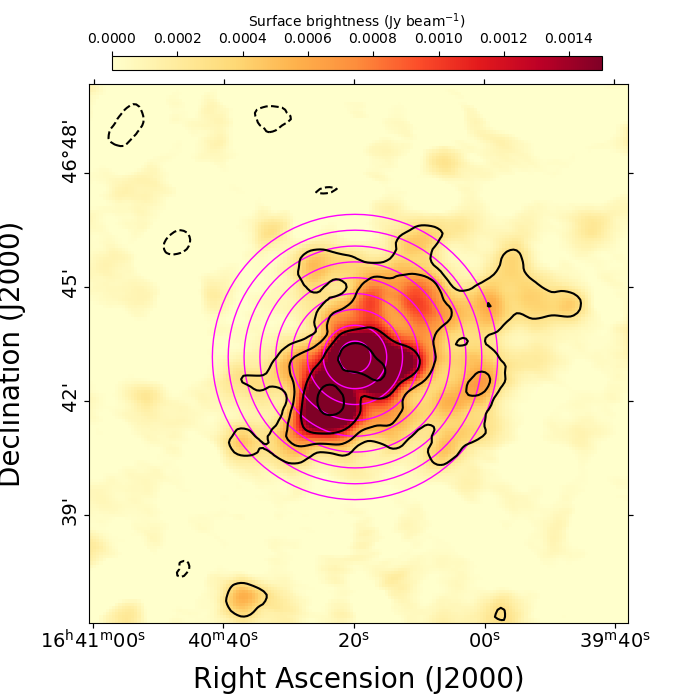}
   
 \caption{Same as Fig. \ref{Fig:profile_A773} for A2219.}
         \label{Fig:profile_A2219}
   \end{figure*}

\begin{figure*}[h]
   \centering
   \includegraphics[height=8cm]{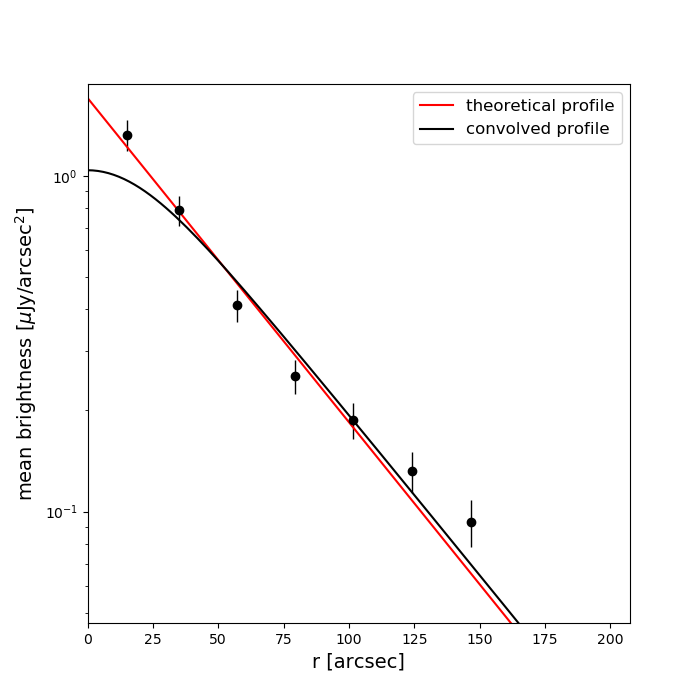}
   \includegraphics[height=8cm]{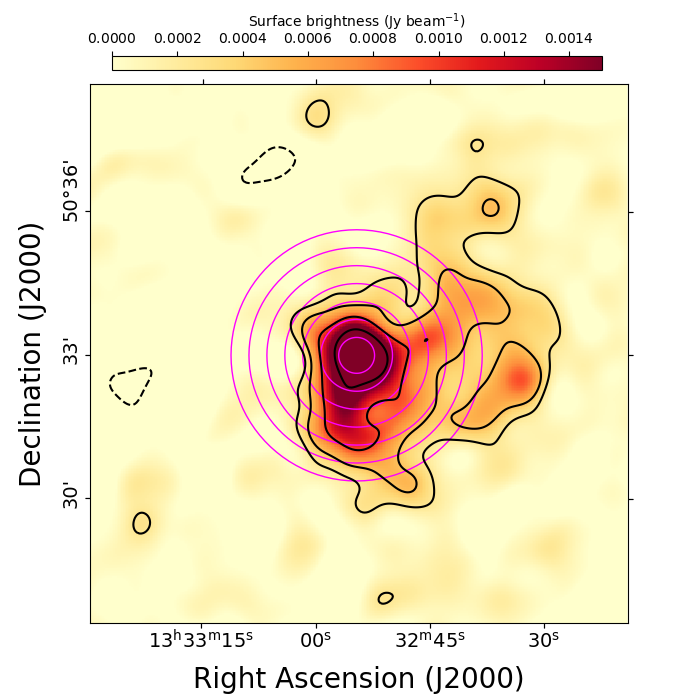}
   
 \caption{Same as Fig. \ref{Fig:profile_A773} for A1758.}
         \label{Fig:profile_A1758}
   \end{figure*}

\begin{figure*}[h]
   \centering
   \includegraphics[height=8cm]{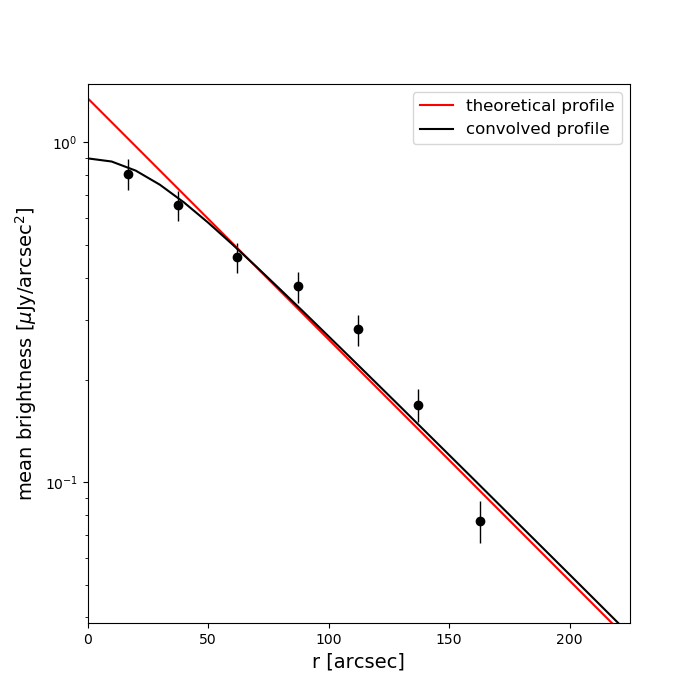}
   \includegraphics[height=8cm]{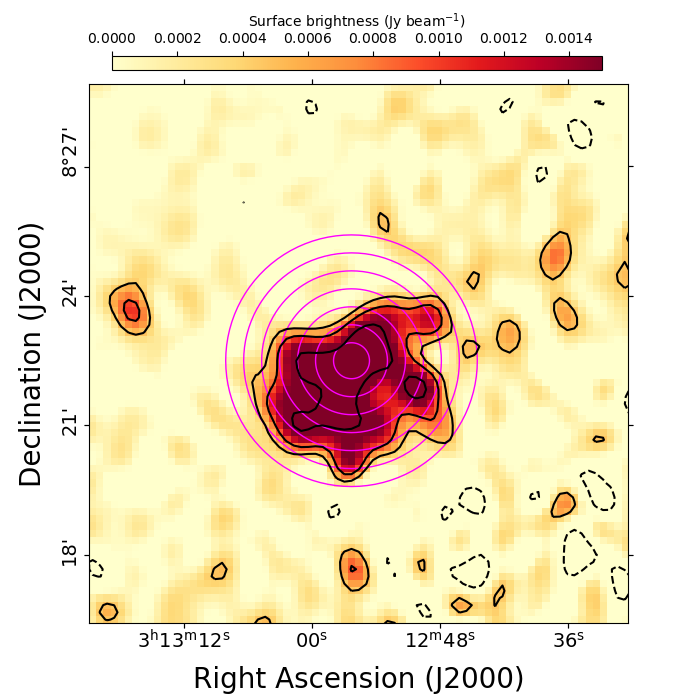}
   
 \caption{Same as Fig. \ref{Fig:profile_A773} for PSZ1 G171.96-40.64.}
         \label{Fig:profile_P171}
   \end{figure*}

\begin{figure*}[h]
   \centering
   \includegraphics[height=8cm]{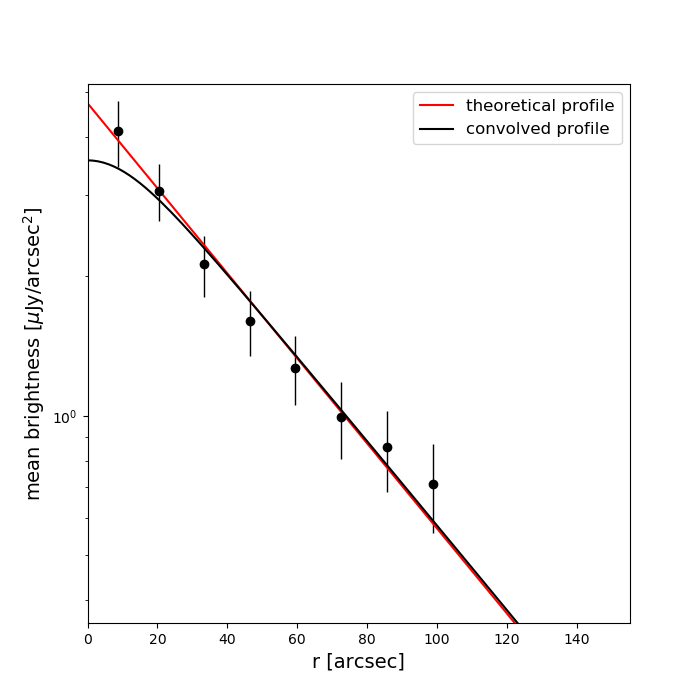}
   \includegraphics[height=8cm]{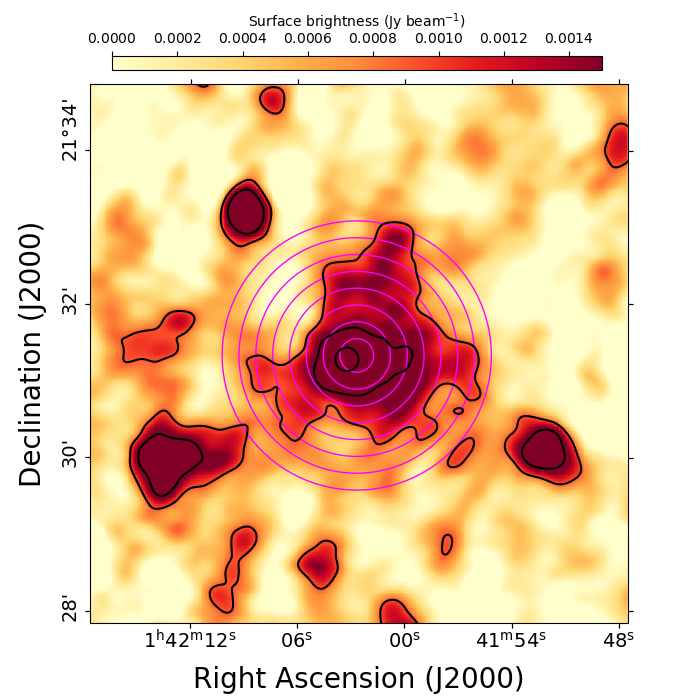}
   
 \caption{Same as Fig. \ref{Fig:profile_A773} for RXC J0142.0+2131.}
         \label{Fig:profile_R0142}
   \end{figure*}

\begin{figure*}[h]
   \centering
   \includegraphics[height=8cm]{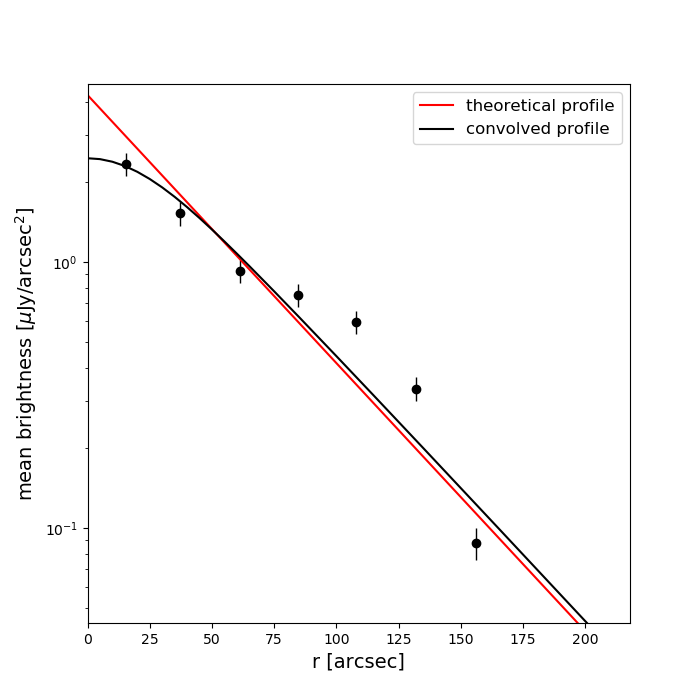}
   \includegraphics[height=8cm]{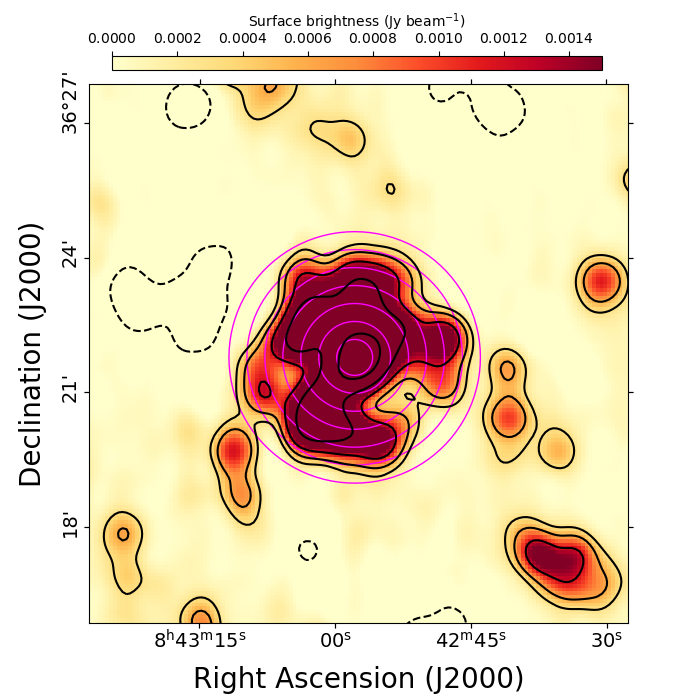}
   
 \caption{Same as Fig. \ref{Fig:profile_A773} for A697.}
         \label{Fig:profile_A697}
   \end{figure*}

\begin{figure*}[h]
   \centering
   \includegraphics[height=8cm]{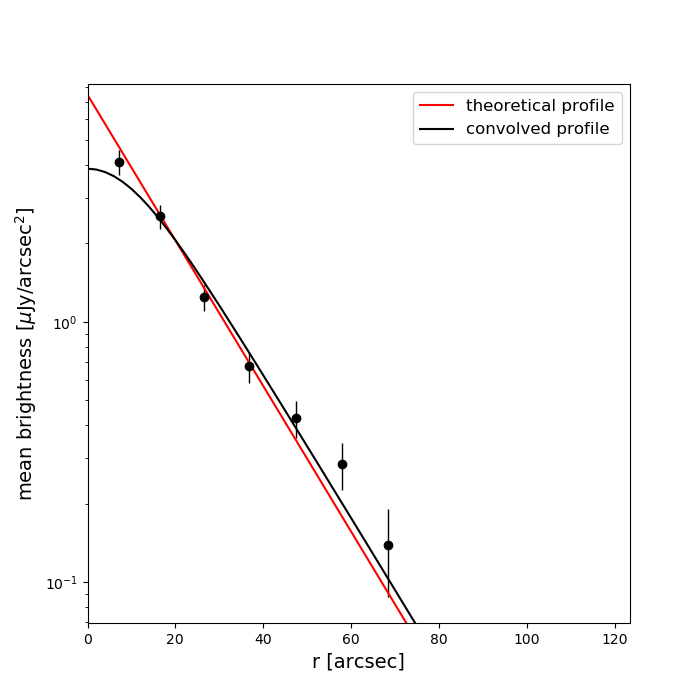}
   \includegraphics[height=8cm]{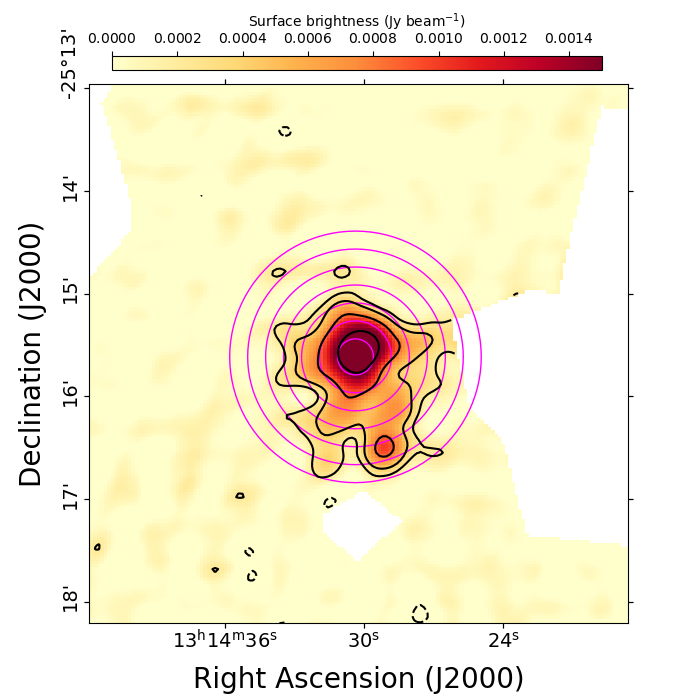}
   
 \caption{Same as Fig. \ref{Fig:profile_A773} for RXC J1314.4-2515.}
         \label{Fig:profile_R1314}
   \end{figure*}

The properties of the radio images used to derive the surface brightness radial profiles shown in Fig. \ref{Fig:profile_A773}--\ref{Fig:profile_R1314} are listed in Table \ref{tab:profiles}. 
We analysed 1.4 GHz data from the VLA archive for the clusters marked with an asterisk in the column 'Reference' of Table \ref{tab:profiles}, that is to say, A773 (project AF349), A665 (AG690, AF304), A209 (AG639), A2218 (AG344), A2744 (AF349) and A2219 (AF367, AF372). Each field was observed with both the C-array and D-array configurations. 
We calibrated each configuration dataset separately using AIPS, following the standard calibration scheme, with amplitude and phase calibration carried out using the primary and secondary calibration sources. The flux density scale was set using the Perley \& Butler (2013) coefficients. We applied phase-only self-calibration to each dataset and produced the final images using the multi-scale CLEAN algorithm in the IMAGR task. After self-calibration, we combined the C- and D-configuration data into a single dataset for each cluster. A final cycle of phase-only self-calibration was applied to the combined datasets to improve the quality of the final images. 

For each cluster, we first identified the discrete radio sources in (or projected onto) and around the cluster region using the higher-resolution images from the C-array datasets alone. We subtracted the discrete sources from the \textit{uv}-data using the same procedure outlined in Section \ref{Sec:GMRT_data} and used the resulting datasets to obtain images of the diffuse radio emission at low-resolution using the multi-scale CLEAN.
The angular resolution and noise levels of our final radio halo images are listed in Table \ref{tab:profiles}.

\begin{table*}[h]
\begin{center}
\caption{Images for surface brightness radial profiles}
\begin{small}
\begin{tabular}{l c c c c c}
\midrule
\midrule
Name		&	telescope	&	$\nu$		&	beam FWHM		&	rms	&	Reference\\	
		&				&	(MHz)	&	(arcsec)	&	(mJy/beam)	&	\\
\midrule

A773 &	VLA		&	1400	& 65	& 0.08	&  \large{*}	\\
A665 &	VLA		&	1400	& 65	& 0.16	&	\large{*}\\
A209 &	VLA		&	1400	& 70	& 0.08	&	\large{*}\\
A2163 &	VLA		&	1400	& 60	& 0.10	&   Rojas et al. 2020, to be submitted\\
A2218 &	VLA		&	1400	& 50 	& 0.05	&   \large{*}	\\
A2744 &	VLA		&	1400	& 50	& 0.11	&	\large{*}\\
Z0634 &	JVLA	&	1400	& 40	& 0.08	&   \citet{cuciti18}	\\
A2219 &	VLA		&	1400	& 50	& 0.10	&	\large{*} \\
A1758 &	VLA		&	1400	& 45	& 0.08	&	\citet{botteon18}\\
A697 &	GMRT	&	327		& 48	& 1.10	&   \citet{macario10}\\
RXC J1314.4-2515 &	GMRT	&	610	& 21	&	0.07	& \citet{venturi07}	\\
A521 &	GMRT	&	235		& 40	& 0.17  &	\citet{brunetti08nature}\\
PSZ1 G171.96-40.64 &	VLA	&	1400& 50	&	0.17    &	\citet{giacintucci13}\\
RXC J0142.0+2131 &	LOFAR	&	144	& 25	&	0.30	&	\citet{savini19}\\

\midrule
\midrule
\end{tabular}
\label{tab:profiles}
\end{small}
\end{center}
\end{table*}

\begin{figure*}[h]
   \centering
   \includegraphics[height=8cm]{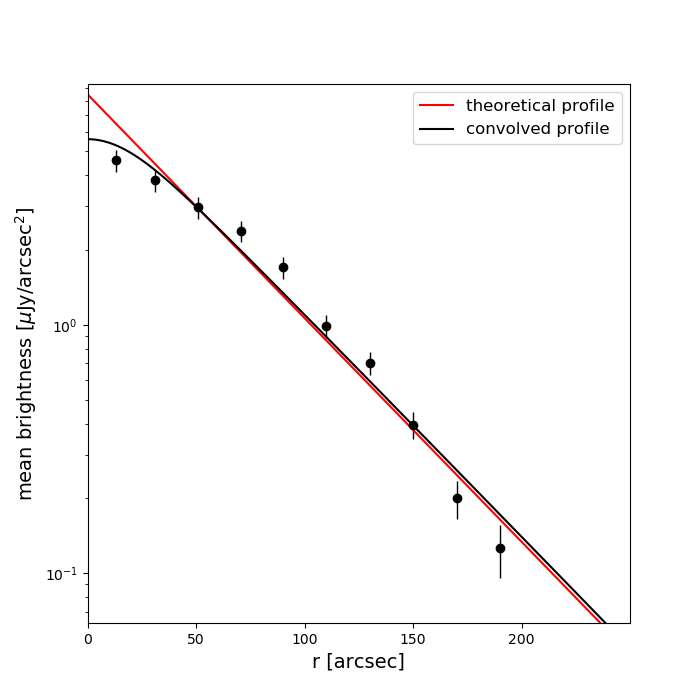}
   \includegraphics[height=8cm]{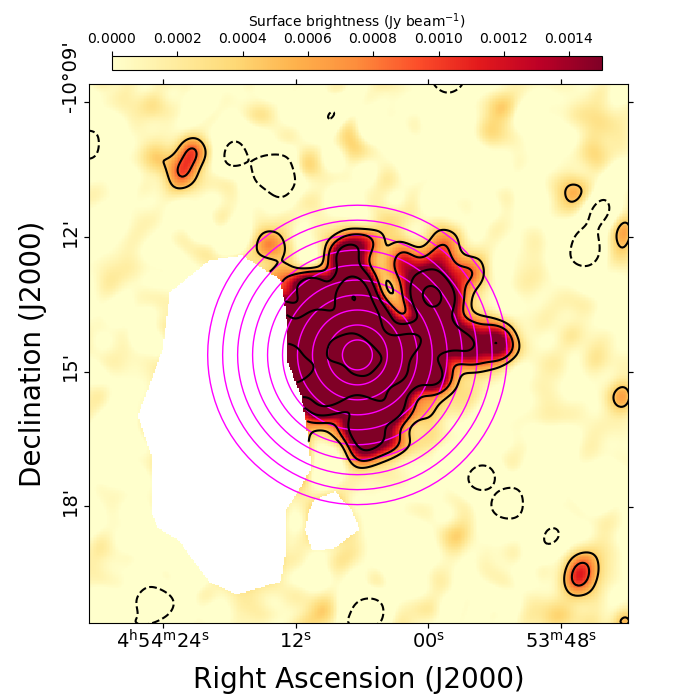}
   
 \caption{Same as Fig. \ref{Fig:profile_A773} for A521.}
         \label{Fig:profile_A521}
   \end{figure*}      
  
\end{appendix}
\end{document}